\documentclass[preprint2]{aastex}

\usepackage{graphicx}
\usepackage{color}

\newcommand{\Bband}{B$_{435}$}
\newcommand{\Vband}{V$_{606}$}
\newcommand{\iband}{i$_{775}$}
\newcommand{\zband}{z$_{850}$}

\slugcomment{To appear in the Astronomical Journal}

\shorttitle{The HUDF}
\shortauthors{Beckwith et al.}

\begin{document}

%% LaTeX will automatically break titles if they run longer than one line. However, you may use \\ to force a line break if you desire.

\title{The Hubble Ultra Deep Field}

%% Use \author, \affil, and the \and command to format
%% author and affiliation information.
%% Note that \email has replaced the old \authoremail command
%% from AASTeX v4.0. You can use \email to mark an email address
%% anywhere in the paper, not just in the front matter.
%% As in the title, use \\ to force line breaks.

\author{Steven V. W. Beckwith\altaffilmark{1,2}, Massimo Stiavelli\altaffilmark{1}, Anton M. Koekemoer\altaffilmark{1}, John A. R. Caldwell\altaffilmark{1,3}, Henry C. Ferguson\altaffilmark{1}, Richard Hook\altaffilmark{5,6}, Ray A. Lucas\altaffilmark{1}, Louis E. Bergeron\altaffilmark{1}, Michael Corbin\altaffilmark{7}, Shardha Jogee\altaffilmark{1,8}, Nino Panagia\altaffilmark{4}, Massimo Robberto\altaffilmark{4}, Patricia Royle\altaffilmark{1}, Rachel S. Somerville\altaffilmark{1,9}, and Megan Sosey\altaffilmark{1}}
\affil{Space Telescope Science Institute, 3700 San Martin Drive, Baltimore, MD 21218, USA}
 \email{svwb@stsci.edu}

%% Notice that each of these authors has alternate affiliations, which
%% are identified by the \altaffilmark after each name.  Specify alternate
%% affiliation information with \altaffiltext, with one command per each
%% affiliation1

\altaffiltext{1}{Space Telescope Science Institute}
\altaffiltext{2}{Johns Hopkins University}
\altaffiltext{3}{McDonald Observatory, University of Texas}
\altaffiltext{4}{European Space Agency}
\altaffiltext{5}{European Southern Observatory}
\altaffiltext{6}{Space Telescope European Coordinating Facility}
\altaffiltext{7}{US Naval Observatory, Flagstaff Station}
\altaffiltext{8}{University of Texas}
\altaffiltext{9}{Max-Planck-Institut f\"ur Astronomie}

%% Mark off your abstract in the ``abstract'' environment. In the manuscript
%% style, abstract will output a Received/Accepted line after the
%% title and affiliation information. No date will appear since the author
%% does not have this information. The dates will be filled in by the
%% editorial office after submission.

% abstract
\begin{abstract}
This paper presents the Hubble Ultra Deep Field (HUDF), a one million second exposure of an 11 square minute-of-arc region in the southern sky with the Advanced Camera for Surveys on the Hubble Space Telescope using Director's Discretionary Time. The exposure time was divided among four filters, F435W (\Bband ), F606W (\Vband ), F775W (\iband ), and F850LP (\zband ), to give approximately uniform limiting magnitudes $m_{\rm AB} \sim 29$ for point sources. The image contains at least 10,000 objects presented here as a catalog, the vast majority of which are galaxies. Visual inspection of the images shows few if any galaxies at redshifts greater than $\sim 4$ that resemble present day spiral or elliptical galaxies. The image reinforces the conclusion from the original Hubble Deep Field that galaxies evolved strongly during the first few billion years in the infancy of the universe. Using the Lyman break dropout method to derive samples of galaxies at redshifts between 4 and 7, it is possible to study the apparent evolution of the galaxy luminosity function and number density.  Examination of the catalog for dropout sources yields 504 \Bband -dropouts, 204 \Vband -dropouts, and 54 \iband -dropouts. The \iband -dropouts are most likely galaxies at redshifts between 6 and 7. Using these samples that are at different redshifts but derived from the same data, we find no evidence for a change in the characteristic luminosity of galaxies but some evidence for a decrease in their number densities between redshifts of 4 and 7. Assessing the factors needed to derive the luminosity function from the data suggests there is considerable uncertainty in parameters from samples discovered with different instruments and derived using independent assumptions about the source populations. This assessment calls into question some of the strong conclusions of recently published work on distant galaxies. The ultraviolet luminosity density of these samples is dominated by galaxies fainter than the characteristic luminosity, and the HUDF reveals considerably more luminosity than shallower surveys. The apparent ultraviolet luminosity density of galaxies appears to decrease from redshifts of a few to redshifts greater than 6, although this decrease may be the result of faint-end incompleteness in the most distant samples. The highest redshift samples show that star formation was already vigorous at the earliest epochs that galaxies have been observed, less than one billion years after the Big Bang.
 \end{abstract}
 

%------------ body of article ------------------->>

\section{INTRODUCTION}

A primary motivation for deep exposures of the sky has been to detect the most distant objects allowed by the observing technology. Over the last ten years, the use of ground-based telescopes combined with the Hubble Space Telescope produced large samples of galaxies at redshifts as high as 5 to study early structure formation and the assembly of stars into present-day galaxies (Steidel et al. 1996a,b, 1999; Ellis 1998; Giavalisco 2002). These programs successfully revealed the distant populations recognized for several decades as important for understanding how the present-day universe came to be (Eggen et al. 1962; Partridge and Peebles 1967a, b; Tinsley 1972a, b). Because of the complications arising from star formation, gas dynamics, and feedback into the early intergalactic medium, theoretical predictions about the earliest galaxies are challenging, and the subject has been driven mainly by observations. 

Even though it has been possible to {\it detect} galaxies at redshifts above one, it has been difficult to determine the redshifts and thus distances to objects from images only, where large samples may be rapidly assembled. Early workers recognized that Ly$_\alpha$ radiation should be especially prominent around the first generation of galaxies, despite some uncertainty about the amount of scattering and absorption, and there should be a strong edge or break in the rest frame UV spectra at 912 \AA\ owing to absorption by hydrogen internal to the galaxies and in the intergalactic medium (e.g. Partridge 1974; Davis \& Wilkinson 1974; Koo \& Kron 1980.) Subsequently, Steidel and Hamilton (Steidel \& Hamilton 1992; Steidel 1996a,b) developed search techniques to exploit the Lyman edge using broad band colors to find galaxies with a paucity of short-wavelength flux, the so-called ``dropout'' galaxies. This technique has proven most productive in discovering large samples of high redshift galaxies in multi-band images. There are now samples of several thousand galaxies at redshifts between about 2 and 5 (Steidel et al. 1999, hereafter SAGDP99; Steidel et al. 2003; Giavalisco et al. 2004.)

When it became evident that the most distant galaxies were characterized by compact high-surface brightness features (Driver et al. 1995), the Hubble Space Telescope took a prominent role in the study of young galaxies. An important advance came from the Hubble Deep Field (HDF; Williams et al. 1996), a four-band, 0.5 million second exposure with  the Wide Field Planetary Camera 2. This seminal program using 150 orbits of Director's Discretionary time on Hubble uncovered a large number of sources at redshifts above 1 that would have been difficult to discover from the ground. The HDF revealed a population of small, irregular galaxies that often appeared in pairs or small groups. Much of the light from these objects was high surface brightness---owing to high rates of star formation---but concentrated, requiring the resolution of Hubble to identify them as distant galaxies as opposed to red stars, say. Extension of the deep field approach to the southern hemisphere (Williams et al. 2000) confirmed the main conclusions of the HDF but also showed the limitations of a pencil beam survey in drawing broad conclusions about distant populations; cosmic error within small fields can be substantial.

Several advances since the HDF suggested that even deeper observations could reveal important aspects of the way that galaxies were created. The standard cosmology holds that the atoms in the universe were neutral following recombination at a redshift, $z \sim 1100$, until somewhere around $z \sim 6-10$, at which time they were reionized by stars and black holes. The first observation of this epoch came with detection of the Gunn-Peterson hydrogen edge in the spectra of distant quasars discovered in the Sloan Digital Sky Survey (Becker et al. 2001, Fan et al. 2002), putting the redshift of complete reionization around 6. The WMAP experiment made an indirect determination of a reionization era that started as early as redshift $11$ (Kogut et al. 2003; Spergel et al. 2003; Spergel et al. 2006). Cold Dark Matter (CDM) models with a cosmological constant had some constraints that were not in accord with such an early epoch of reionization (Frenk et al. 1985), although most of these models have sufficient freedom to accommodate even the most discrepant data. Precise determination of the reionization history of the universe remains one of the important goals of observational astronomy. 

The luminosity function inferred from the HDF suggested that searching a wider area to less depth than the HDF would be efficient at picking up large populations of high redshift galaxies. An important advance since the HDF was the Great Observatories Origins Deep Survey: GOODS (Giavalisco et al. 2004.) GOODS used the Advanced Camera for Surveys (ACS; Ford et al. 2003) on Hubble to image an area thirty times larger but 1 magnitude shallower than the HDF. The GOODS sample contains more than 60,000 galaxies with photometric magnitudes in four bands, \Bband\ (F445W), \Vband\ (F606W), \iband\  (F775W), and \zband\ (F850LP), and sufficient resolution to study structures as small as 1\,kpc at redshifts approaching 6. This sample is excellent for statistical studies of bright galaxies at high redshifts.  

Deep fields have an advantage over shallow fields to study the faint end of the luminosity function and for increasing sample sizes when the slope of the luminosity function is large near the limiting magnitude of the survey. For a steep slope, the sample size will increase faster by investing additional observing time in more exposure on a single field rather than covering more area. The luminosity function at high redshifts is imprecise, but the current evidence indicates that it is consistent with a Schechter function (Schechter 1976) with a characteristic luminosity, $L_*$, somewhat brighter than the local value and a faint end slope steep enough to warrant investment in a deep field (SAGDP99, Gabasch 2004a,b.)

The redshift at which the limiting magnitude of GOODS makes a deep field preferable to a wide field can be estimated using a standard Schechter function. A deep field becomes preferable at a redshift greater than 5, where the GOODS limiting magnitude is $\sim L_*$, depending on the exact assumptions about how $L_*$ evolves with redshift. The upper limit to the redshifts of the objects in a deep optical survey is when the Lyman edge goes beyond the longest wavelength filter. A practical limit for the ACS is when intergalactic absorption shortward of Ly$_\alpha$ shifts through the \zband\  filter, $z\sim 7.4$. A deep field should produce samples of objects in the range $5 < z < 7$ that are fainter than those found in GOODS and other wide surveys and allow a good characterization of the luminosity function in the early universe.

It is most important to characterize the luminosity function below $L_*$ to see the transition from exponential to power-law form, to measure the slope, and to assess the total luminosity of faint galaxies. The GOODS survey was limited to studying galaxies at the bright end of the luminosity function for redshifts greater than about 5. Even for lower redshifts, a deep field is useful to observe galaxies fainter than the characteristic brightness, providing important information about samples in the intermediate redshift ranges $2 < z < 5$ where much of the early star formation in the universe took place. There is a strong degeneracy between derivations of object density and characteristic luminosity unless the luminosity function is well characterized below $L_*$. Such a degeneracy hampers the interpretation of shallow surveys even with large samples.

As shown in the next section, it is possible to reach well below $L_*$ out to redshifts near 7 with ACS on Hubble. This capability motivated a deep ACS field.

The appearance of high redshift galaxies in the HDF and shallower surveys indicates substantial evolution in size and structure between early times and today. This evolution was already known at the time of the HDF, and subsequent observations tend to confirm the conclusion that the galaxy populations look markedly different at high redshift compared to the present time. But the apparent morphology of high redshift galaxies is affected strongly by the loss of low-surface brightness features owing to cosmological dimming. An important way to test whether the loss of these features significantly distorts our perception of galaxies at high redshift is to make deeper observations of the sample at intermediate redshifts. Thus, an ultra deep field can provide an important complement to the pioneering observations of the HDF, GOODS, and ground-based surveys by searching for low-surface brightness components of faint galaxies.

It was evident that a deep field with the new capabilities of Hubble following the installation of the ACS could address several important issues in early galaxy formation. In addition to augmenting the samples of galaxies at redshifts greater than 2, there was also the tantalizing possibility of pushing back the observational boundaries to redshifts greater than 6 to reach the reionization epoch. With these motivations in mind and following the same philosophy pioneered by Robert Williams for the HDF, we held a series of meetings asking for advice on the scientific importance of another deep field and then assembled a Scientific Advisory Committee with a wide range of expertise to recommend specific parameters for the survey: choice of field, choice of filters, and depth needed for a meaningful advance. 

As with the original HDF, our purpose was to provide a public database using Director's Discretionary Time on the Hubble Space Telescope for community use. This paper emphasizes the parameters of the database rather than the subsequent analysis, but it provides a first-order analysis of the data to assess changes in the galaxy populations from the highest observable redshifts until the present. 

Thus, we assembled a team at the Space Telescope Science Institute to create the deepest visual-band image of the universe to date and put the observations in the public domain for community analysis. Like the HDF, this is a multi-color, pencil beam survey in a single ACS field. We call the resulting multi-color image the Hubble Ultra Deep Field (HUDF). 

\section{Observations}

\subsection{Field selection}

The field choice derived from a desire to minimize the celestial foreground radiation, maximize the accessibility to other astronomical observatories, maximize the overlap with extant or planned deep observations at x-ray, infrared, and radio wavelengths, and maximize the observing efficiency of Hubble. The original HDF was located in Hubble's continuous viewing zone (CVZ) to allow uninterrupted observations over a long period. The background light in CVZ orbits is often bright when observing in the part of the orbit grazing the bright earth limb. The HDF overcame this limitation by taking images in the ultraviolet filter, F300W, during the bright periods, because WFPC2 images in this filter are detector noise limited and  relatively unaffected by increased background. The Wide Field Camera of ACS is not sensitive at ultraviolet wavelengths, and the enhanced background of the bright CVZ orbits would seriously limit their usefulness. We, therefore, decided not to require that the target field be located in the CVZ, since it would not enhance the efficiency of the observations.

Zodiacal dust within approximately 30\degr\ of the ecliptic plane is bright for Hubble; it was desirable to locate the field as far from the ecliptic as possible. Declinations north of 35\degr\ are inaccessible from all major southern hemisphere observatories, particularly the planned Atacama Large Millimeter Array (ALMA), designed to be an important tool for observations of distant galaxies. Declinations south of -40\degr\ are inaccessible from Hawaii and all observatories northward. At the outset, we concentrated on fields between  -40\degr\ and +35\degr\ declination and more than 35\degr\ from the ecliptic plane.

Within this declination range, there are a few places with very low Galactic dust and substantial investments of observing time from other programs. The most prominent was the Chandra Deep Field South (CDF-S), a large (15\arcmin $\times 15$\arcmin ) field located in the direction $3^{\rm h} 30^{\rm m}$ -28\degr. This field has very low Galactic cirrus emission and atomic hydrogen column density (Schlegel, Finkbeiner, \& Davis 1998), it passes through the zenith at the major observatories in Chile (the VLT, CTIO, Gemini South, Magellan, and ALMA), and it is accessible from as far north as the VLA site in New Mexico. Furthermore, the CDF-S already has a substantial investment in deep x-ray observations with Chandra and XMM, and there are existing ACS observations through the Great Observatories Origins Deep Survey (GOODS, Giavalisco et al. 2004) allowing some useful comparisons for the HUDF. There are also deep infrared observations with the Spitzer Space Telescope (Dickinson et al. 2004)

CDF-S is larger than a single ACS field; several additional criteria guided the exact choice of pointing within it. The x-ray sensitivity with Chandra varies across the field, and it was desirable for the HUDF to coincide with a region of good x-ray sensitivity. There are several interesting objects identified through GOODS that deep observations would be most useful for, specifically a galaxy at redshift 5.8 and an old supernova. We centered the field such that the high redshift galaxy and old supernova were both covered, and the x-ray sensitivity was also very good. This choice produced a field centered on: RA (J2000) = $3^{\rm h}32^{\rm m}39^{\rm s}$,  Dec (J2000) = -27\degr 47\arcmin 29\arcsec.1. Table~1 lists the major characteristics of this field.

\subsection{Filters}
The filter choice was identical to that chosen by the GOODS team. This choice provides enough color information for rudimentary classification of objects and enough wavelength coverage to search for the highest redshift galaxies. It also makes possible an easy comparison of samples derived from both surveys. To detect objects at the highest possible redshifts, the observations needed to include the longest wavelength filter, F850LP (\zband ), a band that was sufficiently insensitive in WFPC2 (F814W) to limit its use for the HDF. The adjacent F775W (\iband ) filter gives minimal overlap but contiguous wavelength coverage. Together, these two bands provide excellent sensitivity to the highest redshift objects detectable with ACS, the \iband-dropout sources, and are mandatory to search for objects at redshifts approaching 7.

Four bands are desirable to provide crude spectroscopic analysis of the objects. Since the long-wavelength observations are background-limited, additional sensitivity could be gained by adding images at shorter wavelengths without loss of signal-to-noise ratio. The \Vband\ filter, F606W, is immediately adjacent to \iband , broad, and provides excellent sensitivity to all objects at redshifts less than about 4. 

The HDF incorporated an ultraviolet filter useful for identifying dropout sources at redshifts near 3. The ACS wide field camera is optimized for red wavelengths, making ultraviolet observations relatively insensitive. Since a primary goal was to identify samples at higher redshifts than the HDF, we chose the bluest filter to be the \Bband\ band, F435W, immediately adjacent to \Vband\ band. Following the conventions of the GOODS team, the bands are hereafter called \Bband, \Vband, \iband, and \zband.  

There are two other advantages to using the same filter set as GOODS. The overlap between the GOODS CDF-S field and the HUDF makes it possible to compare objects directly in both surveys to calibrate completeness estimates for the shallower survey.  Furthermore, science analyses can be carried out on the two data sets using identical methods and minimizing systematic differences. 

Figure 1 plots the total detection efficiency for the four bands used for the observations. The figure includes the spectrum of a model galaxy at a redshift of 5.8 for comparison. Figure 1 shows that this very high redshift galaxy produces a sharp drop in flux between \zband\ and \iband\ with no flux at all in the shorter wavelength bands. The relative detection efficiency also indicates the need for longer exposures in the longest wavelength bands to detect objects whose spectra are either flat or blue, typical of star forming galaxies at high redshift.

%Figure 1: Filter transmission curves
\begin{figure}[ht]
\includegraphics[width=7.5cm]{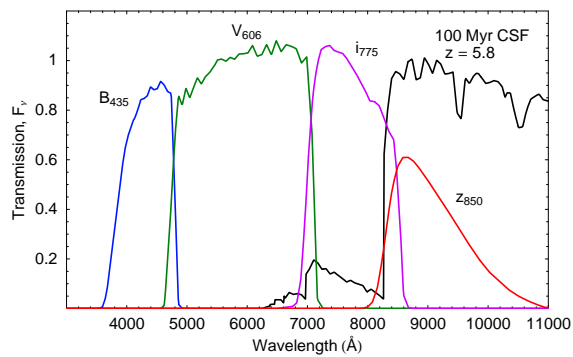}
\caption{The filter transmission curves for the four bands used in this survey (colored lines) along with the spectrum of a model star forming galaxy at $z = 5.8$ (black). The spectrum is from the model of Bruzual \& Charlot (2003) for a 0.4 solar metallicity galaxy undergoing continuous star formation for 100\,Myr using the intergalactic absorption curves of Madau (1995).}
\end{figure}

\subsection{Depth}
There are several ways to estimate the depth needed to address the different problems described in the introduction. However, we caution at the outset that so little is known about objects at the highest redshifts of interest that a conservative approach would require depths well beyond those possible even with Hubble. We recognized at the outset that our goal was to obtain as deep an observation as possible with the amount of discretionary time available to the Director, and the depth would be constrained by the available pool. The resulting sensitivity is, nevertheless, well-suited to make progress on each issue.

We adopt the concordance cosmology, $\Omega_\Lambda = 0.73$, $\Omega_M = 0.27$, $h = 0.71$, throughout this paper for all analyses.

The primary goal was to detect a statistically significant sample of galaxies at redshifts between 5 and 7. We set this goal at about 100 objects. Two additional goals are to study the luminosity function of high redshift galaxies at the faint end down to $\sim 0.1 L_*$, and to observe low surface brightness features in galaxies that are missed in shallower surveys such as GOODS and the HDF, both of which will be aided by the faintest limiting magnitude that can be achieved by Hubble. The following paragraphs estimate what might be achieved with a limiting AB magnitude of $\sim 29^{\rm m}$, say.

Assuming galaxy luminosity distributions are described by a Schechter function, it is possible to estimate the number of galaxies accessible to observation in any volume element of the universe. SAGDP99 determined a characteristic apparent magnitude at redshift 3, $m_*(3)$, of 24.5 (B-band) in the rest frame ultraviolet, corresponding to $M_* = -20.8$ corrected to the assumed cosmology; the local value is $-20.2$ (Schechter 1976.) Assuming $M_*=-20.2$, $\phi_* = 0.016$ Mpc$^{-3}$, and $\alpha = -1.6$, the number of galaxies detected as a function of redshift per unit redshift in a single field of $\Delta\Omega$ sr (11 arcmin$^2$ for ACS) is:
\begin{eqnarray}
{d N \over d z} & = & \Delta\Omega_{ACS} {d V \over d z} \int_{x(z)}^\infty \phi_* L^\alpha e^{-L} dL,
\end{eqnarray}
where ${d V \over d z}$ is the co-moving volume, and the fraction, $x$, of $L_*$ detectable with limiting magnitude, $m_{lim}$, is:
\begin{eqnarray}
x(z) & = & {D_L(z)^2 \over (1+z)} 10^{(M_* + k_b(z) - m_{lim})/2.5 },
\end{eqnarray}
with the luminosity distance, $D_L$, in units of 10\,pc and $k_b(z)$ the k-correction at redshift, z. These equations may be evaluated numerically; the results are given in Table~2 for an assumed magnitude limit between 28 and 29. We also calculated the expected number of objects appearing in the \Vband, \iband, and \zband\ bands assuming all objects had the spectrum of a source undergoing continuous star formation for 100 Myr (Bruzual and Charlot 2003) and accounting for intergalactic hydrogen absorption (Madau 1995) to illustrate where the objects drop out of each filter.

The numbers in Table~2 are certainly much larger than a real survey would see. We have not attempted to correct for many observational effects that would preclude galaxy detection in a survey or for the expected mixture of source sizes, types, colors, etc. Furthermore, the density, $\phi_*$, of distant populations appears to be several times smaller than the local value used in these calculations (SAGDP99.) The point of this estimate is to show that the expected number of sources is so large that even a few percent detection probability would yield statistically useful samples of galaxies.

It is evident from Table~2 that even a detection limit of $28^m$ in \zband\ would easily satisfy the goal of detecting $\sim 100$ objects above redshift 5. It would also reach below $L_*$ at redshifts approaching 7. There are several large uncertainties that could change these numbers in either direction: if $L_*$ continues to increase beyond redshift 3 (e.g. Gabasch et al. 2004a,b), there will be more faint objects to increase the counts and move further into the faint end of the luminosity function; however, if $\phi_*$ decreases owing either to a smaller number density of galaxies in the early universe or cosmic error, the number of detected galaxies will decrease. The two effects offset one another for number counts and introduce a difficulty of interpretation without a well-characterized luminosity function.

A \zband\ band limiting magnitude of 28 requires of order one hundred orbits of dedicated Hubble observations. The maximum available was 400, and those needed to be divided among the four bands to give adequate spectral information. We chose to allocate 144 orbits each to the \zband\ and \iband\ bands with estimated limiting magnitudes of 28.7 and 29.2, respectively. The remaining 112 orbits were split equally between \Bband\ and \Vband, whose estimated limiting magnitudes were then 29.1 and 29.3. With these sensitivities, the HUDF would, therefore, move firmly into the range needed to assemble samples of high redshift galaxies and might even see substantial evolution in the properties of galaxies when compared with later epochs.

\subsection{Schedule and field orientation}
The observations were scheduled for two periods during which the roll angles could be controlled to produce a nearly square image. Scheduling 400 orbits at the same pointing and with constrained orientations required the use of four roll angles in all: 40\degr , 44\degr , 310\degr , and 314\degr\ for the position angle of the +U3 axis on the sky to increase the target visibility and facilitate scheduling. Table~3 lists the schedule of observations in orbits for each filter.

Each observation or visit consisted of two orbits with two exposures per orbit. The exposure time was typically 1200 seconds but in a few cases the exposures had to be shortened to 850 seconds. The total exposure time is just under 1 million seconds. 

\subsection{Small Scale Pointing: Dithers}
In addition to the large rotations between different phases of the observations that were required for scheduling, smaller shifts in telescope pointing were applied to different observations at the same position angle. These small changes in pointing between exposures, referred to as dithers, were introduced on two scales.

First, small-scale dithers were applied to each of the four exposures within a two orbit visit. This dither pattern improved the sampling of the final image by introducing half-pixel offsets. The ACS/WFC detector critically samples the point spread function (PSF) in the reddest bands but significantly undersamples the PSF at shorter wavelengths. Such undersampling leads to loss of spatial information and aliasing artifacts. The introduction of sub-pixel dithering improves the sampling and allows the reconstruction of a higher-resolution final image and a reduction of artifacts. 

In the case of ACS/WFC, half-pixel dithers or small integer numbers of pixels plus a half-pixel in both X and Y directions provide adequate sampling. The integer pixel components of the dithering were chosen to create the most compact dither pattern that ensured that a bad row or column could not overlap in the combined image because the pointings were always at least 1.5 pixels away from others in both X and Y. This final four-point dither, suggested by Stefano Casertano, is given in Table~4. The most compact pattern was chosen because the exact sub-pixel shifts will be different far from the center of the detectors owing to the very large non-linear component of the ACS/WFC optical distortion. The dither pattern minimizes this effect and ensures good sampling across the field.

Additional dithers of approximately 3 and 6 arcseconds in length in the direction perpendicular to the gap between the two ACS/WFC chips were introduced between visits. These offsets ensure that the regions of sky falling in the gap between the two ACS/WFC chips had at least two thirds of the exposure of the rest of the field and hence minimized the lack of uniformity of the final exposure map.

\section{Data analysis}
 
\subsection{Basic data reduction}
Each of the ACS/WFC exposures was processed through the standard ACS data pipeline, CALACS.  The first step removed the bias level, subtracted the dark current, corrected the flat field and gain variations, eliminated known bad pixels, and calculated the photometric zeropoint. 

To achieve optimal calibration, several reference files were created specifically for these observations: improved dark current correction files (hyperdarks), improved flatfields, and bad pixel files. 

The hyperdarks were created using all the dark-current frames from the 6-month period encompassing the HUDF observations. These files provide higher signal-to-noise ratios than the typical dark reference files that are subtracted during standard calibration and provide a more accurate representation of the overall dark current structure appropriate to the HUDF exposures.

New flatfield images for each filter were produced by applying a flatfield technique that corrects only low spatial-frequency variations based on stellar photometry of 47 Tucanae. These new flatfield images (L-flats) produced a more uniform sky level across the images than the standard pipeline products. After re-calibrating the data with these L-flats, the images had residual flux of order 2\% of the sky level that we ascribe to scattered light from bright sources outside the field of view. We produced images of these residuals from the re-calibrated images and subsequently removed the scattered light by the following procedure:
 
\begin{enumerate}
\item
 All exposures from the pipeline were combined to create an image of the field.

\item
All astronomical objects were identified in this image to create a mask that eliminated all pixels with detectable light from an object; this process also provided improved cosmic ray masks.

\item
A median image was created from the original calibrated exposures using the combined object masks and cosmic ray plus bad pixel masks to exclude all pixels affected by celestial sources, cosmic rays or bad pixels.

\item
The median image was convolved with a 100 pixel-wide smoothing function corresponding to the spatial frequency of the flat-field residuals.

\item
The calibrated files were divided by this smoothed median image to produce another image of the sky.

\item
This sky image was scaled and subtracted from each individual calibrated image to remove the scattered light.
\end{enumerate}

There is a bad-pixel file specific to the HUDF data, containing a number of additional bad columns and other defects that were not present in the standard data quality arrays when the data were taken. The bad pixel file for each band was created after correcting the images with the improved sky values in the previous step, then subtracting an image made by combining all exposures registered to the same pointing. The resulting images made it easy to identify bad pixels, cosmic rays, and satellite tracks that were significantly above the noise level. A difference image was then created for each exposure to determine the root-mean-square variations of each pixel including Poisson noise from any objects. All 800 images were subsequently combined using both median and averaging to produce an image that showed the median or average deviation of each pixel. Additional bad pixels were identified as those exceeding a threshold set to five times the variance of the distribution (``sigma-clipping''). This technique was especially valuable in identifying charge traps not present in the original data quality file. 

Finally, the bias level correction performed by CALACS did not fully remove the bias levels in the four amplifier quadrants but left residual offsets of a few tenths of a photo-electron. These residuals were corrected by first reversing the multiplicative flatfield correction, solving iteratively for the residual bias differences between the quadrants, and removing these differences before re-applying the flatfield. 

The ACS/WFC geometric distortion calibration developed by King (2005) given for ACS by Anderson and King (ISR 2006-001) from large numbers of images of the globular cluster 47 Tucanae provided a means to refine the astrometric coordinates of the final images. This calibration includes filter-dependent scale variations, new fourth-order correction polynomials and additional distortion correction images to model remaining systematic effects. The latter cannot be modeled by polynomials of lower order but can introduce shifts of up to 0.2 ACS/WFC pixels. The inclusion of these corrections into the drizzle software (Fruchter \& Hook 2002) that was used for the image combination has reduced the RMS geometric correction error to significantly less than 0.1 pixels across the full field.

\subsection{Image combination}
The calibrated images were combined into a single image for each filter by means of the MultiDrizzle program (Koekemoer 2002). This program first performs sky subtraction on each input exposure, after which it uses the ``drizzle'' approach (Fruchter \& Hook 2002) with tools developed for the HUDF to correct for the geometric distortion in each exposure and remove the shifts and rotations that were introduced by the observational dither pattern, thereby producing a set of geometrically rectified output images that are all registered onto a common grid.

The MultiDrizzle program then combined all the rectified images for each band to create a clean median image after having rejected the highest and lowest values at each pixel to help minimize the effect of cosmic rays and negative-valued pixels. The clean median image was then transformed back to the original distorted frame of each input exposure to identify cosmic rays.

The algorithm for identifying cosmic rays depends upon comparing the original input image, $I_{in}$, with the clean image, $I_{cl}$, as well as the derivative of the clean image, $\Delta I_{cl}$, to reject pixels deviating from the clean image using a technique similar to that employed for the Hubble Deep Field but with parameters tuned specifically for the HUDF (cf. Williams et al. 1996). The algorithm identifies cosmic rays using the following criterion:
\begin{equation}
| I_{in} - I_{cl} |  >  S \Delta I_{cl} + {S/N\over G}  \sqrt{\sigma_{rn}^2 + G | I_{cl} + I_{sky} |} \\
 \end{equation}
where $S$ was set to 1.5, $S/N$ was set to 3.5, $\sigma_{rn}$ is the read-out noise, $G$ is the electronic A-to-D conversion ratio, and $I_{sky}$ is the sky value measured in the original input image. This procedure is essentially sigma-clipping, with the addition of the derivative image which helps to soften the rejection and prevents pixels from being incorrectly identified as cosmic rays in regions of extremely sharp gradients, such as near bright stars or galaxy cores. A second iteration using a lower level of statistical significance for the detection threshold identified additional pixels to reject surrounding those that were rejected in the first iteration. This technique ensured robust rejection of cosmic rays while at the same time avoiding over-rejection of useful but noisy pixels in bright objects.

The resulting cosmic ray masks were used as input to the final drizzle combination of all the images. Each image was weighted by the inverse variance of the exposure at the mean sky level as calculated from the noise model before combination.  This choice of weighting ensured that the inverse variance of the final image is the sum of the inverse variances of the input images. An input variance image was calculated separately for each exposure, taking into account the flatfield variation, the sky level, the read noise and dark current, and all the bad pixel information.

This step essentially performs a weighted sum of the input images and allows input pixels to be shrunk by a specific amount before being mapped onto the output plane. The final pixel scale was set to 0.6 of the input ACS pixels or 0.030\arcsec/pixel. Since this scale provides Nyquist-limited sampling of the PSF in all 4 bands, it is the optimum pixel size to use. It was also the same as used for the GOODS data products, allowing direct comparisons of the data sets without transformation.

A number of tests provided a means to optimize the choice of drizzle parameters, in particular to examine the effect of the drizzle kernel on the uniformity of the output weight maps and the extent of the noise correlation in the final drizzled image. Drizzle provides the ability to shrink the input pixels before mapping them onto the output plane, which reduces the degree of correlated noise but can lead to large variations in the output weight map if there are insufficient numbers of images. We examined the weight map statistics in detail for a wide range of kernel sizes (given by the value of the drizzle parameter $pixfrac$) aiming to ensure that $pixfrac$ values ranging from 0.3 down to 0 did not introduce too large a degree of variation in the weight map.

Since the HUDF has such a large number of pointings in each band with a good sampling of the sub-pixel space, the weight map was uniform and well behaved across the entire central region of the field, including the intersection of the chip gaps which typically had half the exposure coverage of the remainder of the field, even with $pixfrac = 0$. Therefore, the final images were drizzled using a point kernel (i.e. setting $pixfrac = 0$), corresponding to pure interlacing, so that each input pixel maps onto only a single output pixel. This choice has the advantage of minimizing the amount by which the output image has been processed (convolved with a smoothing function), thus providing the sharpest possible image and the least amount of correlated noise.

\subsection{Data quality assessments}
A series of tests confirmed the quality of the point-spread-function (PSF) in the final combined images. These aimed to quantify the extent to which the original Hubble PSF could be recovered, as well as verifying that the astrometric alignment of the full set of 800 images was sufficiently accurate not to affect the resulting PSF. A sample of isolated stars was identified across the entire HUDF image, covering a range of locations and brightnesses. PSF fits produced radial profiles of these stars as well as determining quantities such as their full width at half maximum (FWHM), enclosed flux and Moffat function fit parameters.  The mean FWHM of the stars was 0\arcsec.084 in \Bband, 0\arcsec.079 in \Vband, 0\arcsec.081 in \iband\ and 0\arcsec.089 in \zband, with a scatter of 1-2 milli-seconds of arc about these values. These values agree to within 2 sigma with the expected diffraction-limited values for Hubble, after taking into account the initial convolution due to discrete sampling by the 0\arcsec.05 ACS pixels, as well as PSF smearing by the charge diffusion kernel between adjacent pixels and subsequent convolution by the 0\arcsec.03 output pixel size. Figure 2 presents the PSF at the center and at a position near the edge of the image.

%Figure 2: Point-spread-functions
\begin{figure}[ht]
\includegraphics[width=7.5cm]{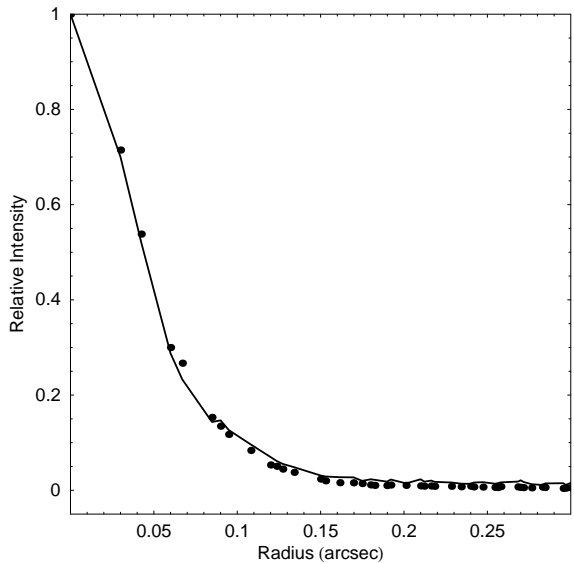}
\caption{The point-spread function derived from stars in the image shown at two positions: near the center of the image (solid line) and near the edge of the image (filled circles).}
\end{figure}

The final image noise may be compared with the sky-limited predictions in two ways. The first is to calculate the noise in small regions that are free from obvious sources and compare with the estimated values from typical sky brightness. This was done for 50-pixel areas, equivalent to a square aperture 0\arcsec.2 on a side. The predicted  background-limited variances are calculated from the sky counts in each filter as $-2.5 \log(SN\, \sqrt{50\, ( R_{sky} t_{\rm exp} +N \sigma_{\rm read}^2)}/t_{\rm exp} ) + ZP$, where $SN$ is the signal-to-noise ratio (assumed to be 10), 50 is the number of pixels in the circular aperture, $R_{sky}$ is the count rate in photo-electrons s$^{-1}$ pix$^{-1}$ from the sky, $t_{exp}$ is the exposure time, $N$ is the number of readouts in $t_{\rm exp}$ (102 for \Bband\ \& \Vband, 288 for \iband\ \& \zband), $\sigma_{\rm read} = 5$ is the read out noise, and $ZP$ is the zero point magnitude (AB.) This calculation gives the magnitude of a uniform disk of 50-pixel area whose flux would be 10 times the noise in the image across the same area. 

Table~5 compares the results ($10\sigma$/0\arcsec.2) with expected sensitivities calculated from the average sky values, readout noises, and the zero points from Table~6.1 in the ACS Handbook (http://www.stsci.edu/hst/acs/documents/hand\linebreak books/cycle14/c06\_expcalc3.html\#328554). The magnitudes corresponding to 10 times the rms noise of a 50-pixel area achieved in the HUDF are very close to those predicted for purely zodiacal-light limited performance. The small differences between the predictions and the results are likely due to the use of an average sky value for the predictions as opposed to the actual sky value in the direction of the HUDF at the time of the observations. These estimates demonstrate that the ACS continued to gain sensitivity with the square root of the integration time even in exposures of more than 340 ksec. The observations achieved the natural limits allowed by zodiacal emission and the overall transmission and quantum efficiency of the instrument. 

%Figure 3: Recovery fractions of disks and bulges
\begin{figure*}[ht]
\includegraphics[width=7.5 cm]{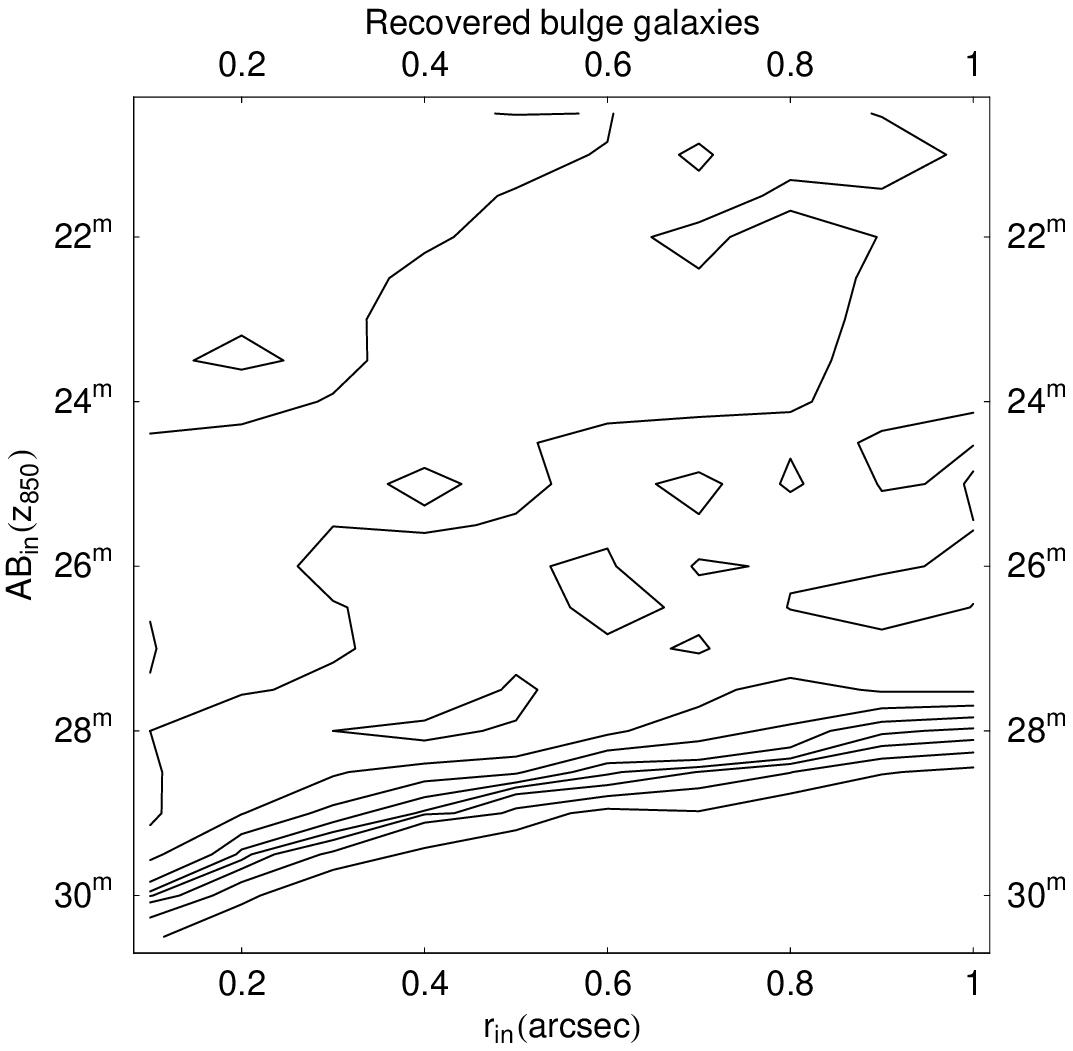} \hspace{1. cm}
\includegraphics[width=7.5cm]{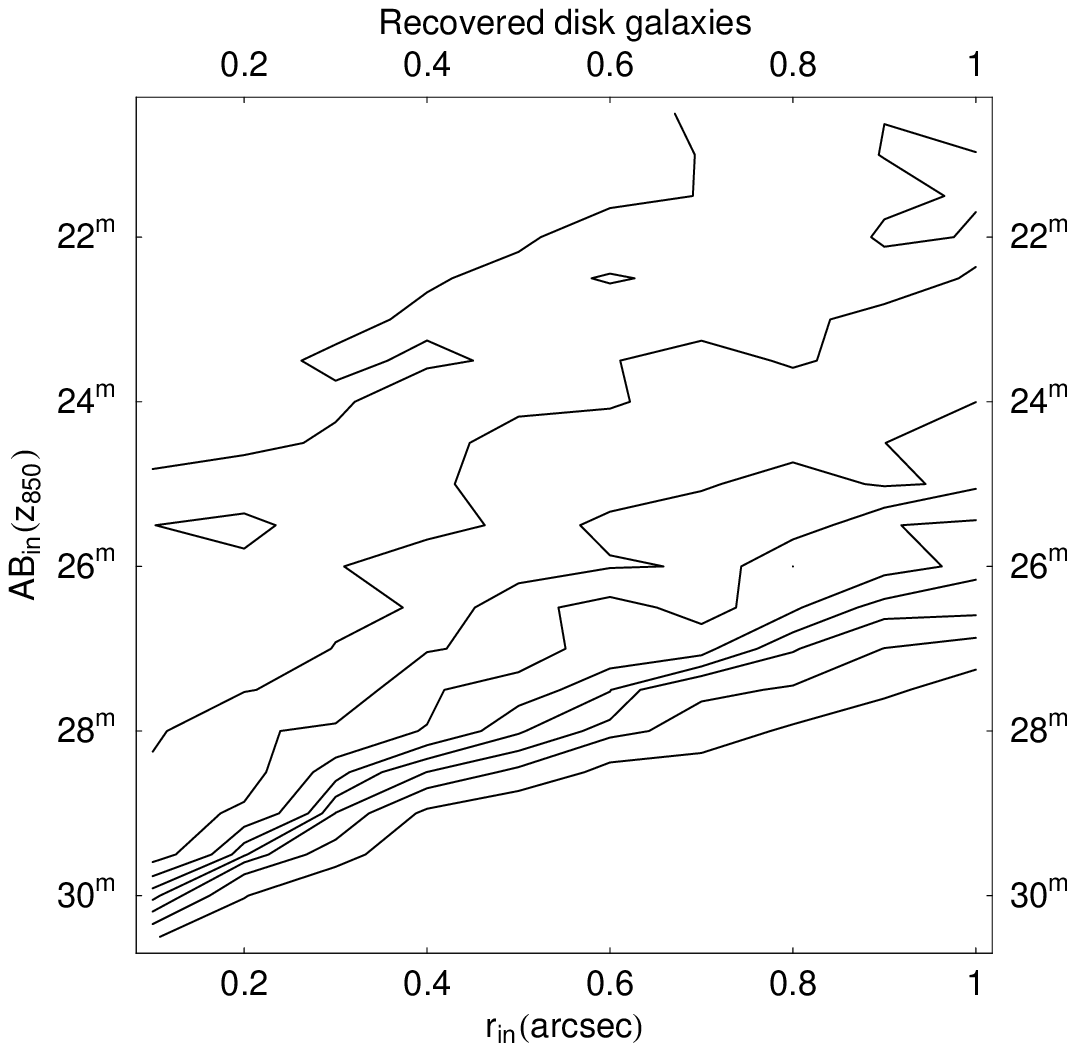}
\caption{Recovery fractions for bulge-like (a) and disk-like (b) sources are shown as a function of input magnitude and input source radius (\arcsec ) in \zband . Contour levels starting from the large magnitudes are: 0.1 through 0.9 in steps of 0.1. The recovery fractions are uncorrected for the cases where the input sources overlapped objects in the image, leading to a slight decrease in recovery fraction as both the source size and magnitude increase.}
\end{figure*}

The second method mimics the technique used to construct the source catalogs by adding artificial sources of known brightness and size to the data sets and using the SExtractor program (Bertin \& Arnouts 1996) to recover them. These Monte-Carlo tests on the \zband\ image simulated the recovery of actual galaxies by using input sources similar to elliptical (``bulge-like'') and spiral (``disk-like'') galaxies. The bulge-like sources in this simulation were oblate spheroids with a uniform distribution of minor/major axes ($b\over a$) from 0.3 to 0.9 at arbitrary projection angles on the sky. The disk-like sources included a variety of orientations for thin disks, and the results discussed below represent an average over the orientations. Figure 3 shows the results of the recovery simulations, plotted as contours with the recovery fractions as functions of input source radius (major axis for disks) and total source magnitude. 

In these simulations, there was no correction for those cases where the artificial sources overlapped with actual objects in the image, in which case SExtractor typically did not recover the artificial sources separately. The effect was to decrease the recovery fraction as the source magnitude became fainter. It is most pronounced for the extended objects that have a higher chance of overlap. The effect is apparent in the contour plots giving recovery fractions less than 1 even for objects much brighter than the limiting magnitude.

The simulations make it straightforward to derive the limiting \zband\ magnitude as a function of input source size. The contour at which 50\% of the galaxies are recovered is well described by quadratic fits for input source radii, $r$ (seconds of arc) , between 0\arcsec\ and 1\arcsec :
\begin{eqnarray}
{\rm m50_{bulge}} & = & 29.27 - 3.98 r + 1.91 r^2 \\
{\rm m50_{disk}} & = & 29.60 - 6.40 r + 1.99 r^2
\end{eqnarray}
The sensitivity to average surface brightness for actual galaxies also varies with source size, at least for sources smaller than 1\arcsec\ in radius, the ones of most interest for studying very high redshift objects.

Table~5 also shows the limiting magnitude at 50\% recovery point from the Monte-Carlo simulations for a point source in \zband\ (``50\% Recovery''). The limiting sensitivity predicted this way is most applicable to assess the completeness of catalogs derived from the images. The underlying size distribution of sources is needed to calculate a robust limiting magnitude for objects in the catalogs. Although galaxies with a variety of sizes up to 1\arcsec\ were successfully recovered from these experiments, the recovered sizes were generally smaller than the actual artificial sources for input radii greater than about 0\arcsec.2, corresponding to about 7 pixels, and the recovered magnitudes were often larger than the actual magnitudes as a result.

\section{Results}

Figure 4 displays a color rendition of the final HUDF image cropped to display an area of uniform exposure. The color rendition allows one to distinguish red and blue galaxies easily and gives the viewer a good visual impression of the colors of typical galaxies in the HUDF. The URL archive.stsci.edu/prepds/udf/udf\_hlsp.html  gives the reader access to the complete data sets as well as the catalogs described below.

%Figure 4: Full color HUDF
\begin{figure*}
\includegraphics[width=16. cm]{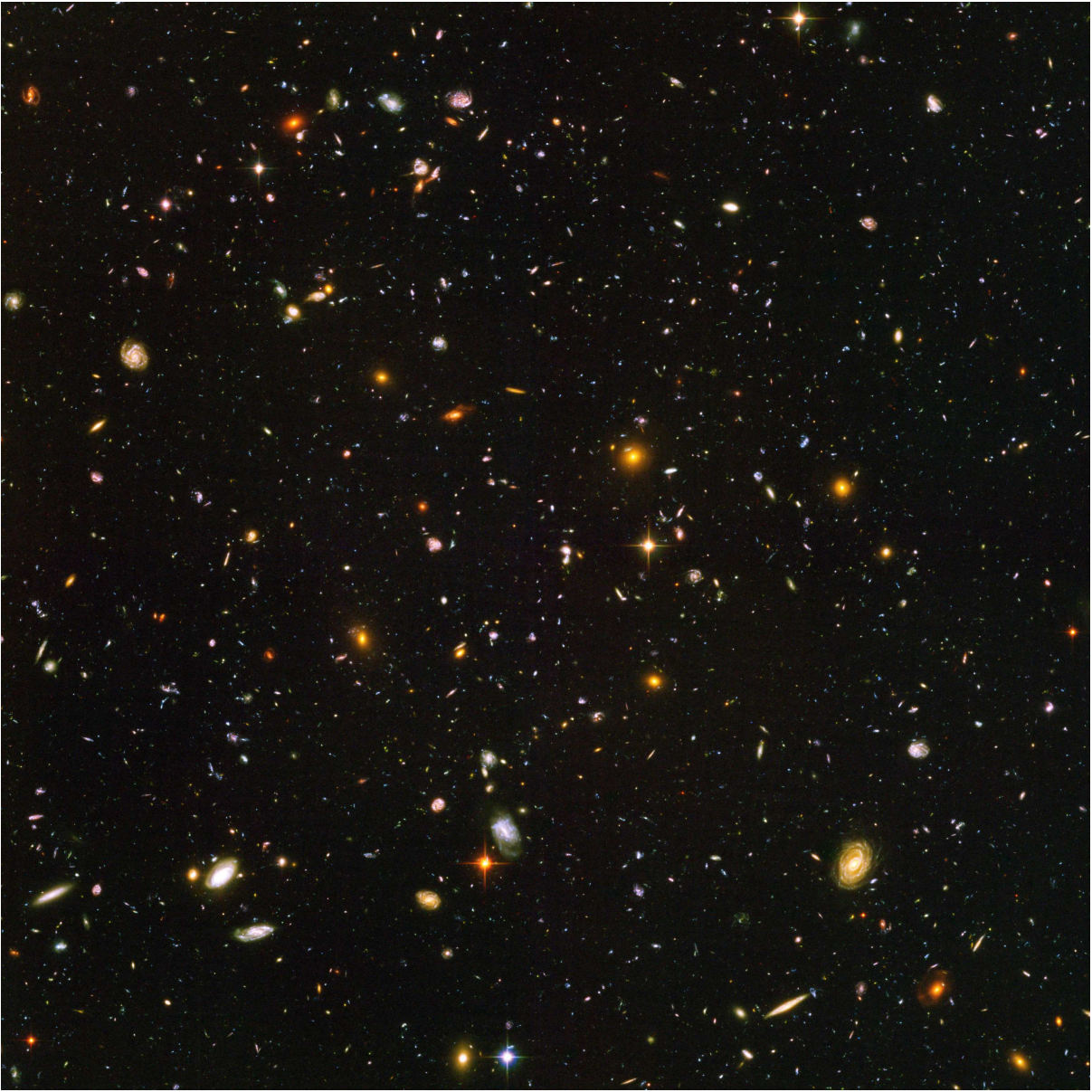} \hspace{1. cm}
\caption{The orientation is such that the left-hand edge is approximately NE and the top edge is approximately NW. The color mapping used to produce this rendition is: blue = (combination of \Bband\ \& \Vband ), green = (combination of \Vband\ \& \iband ), red = (combination of \iband\ \& \zband ). See also Wherry, Blanton, \&Hogg (2004) for a different approach to the color scheme.}
\end{figure*}

\subsection{Source Catalogs}
The SExtractor program (Bertin \& Arnouts 1996) produced catalogs from the final drizzled ACS images. We used the GOODS version of SExtractor that contains a few improvements in background determination and segmentation; thus, the results presented here may not be identical to those produced with the published version of the program. The inverse variance images were used as weights for calculating the RMS uncertainties associated with the flux measurements for each source. Two catalogs of sources were produced using the \iband\ and \zband\ images to identify the sources and the ``dual-image'' mode in SExtractors, using isophotes of each source determined by the selection band to measure the photometric brightness of each source in the other three bands, thereby producing isophotally matched magnitudes that can be directly compared to produce colors for each source.

Catalogs were compiled independently by three team members using the same software with slightly different search parameters. The results were essentially the same among these independent trials; one version was chosen for the final catalog. The SExtractor parameters for this catalog were optimized for the pixel scale and PSF. The FWHM was set to 0\arcsec.09. Source detection required a minimum of 9 contiguous pixels with a detection threshold above 0.61 with a total of 32 deblending sub-thresholds and a contrast parameter of 0.03. Additional details about the SExtractor parameters are available from the electronic HUDF distribution.

The \iband\ catalog contains 10,040 sources, at: \newline http://archive.stsci.edu/prepds/udf/udf\_hlsp.html. A visual inspection of the sources revealed a small number ($<0.1$\%) of spurious detections that are not included in the final catalog. Moreover, there are about 100 additional sources identified visually that were not picked up by SExtractor, owing to their proximity to brighter sources and the inadequacy of the deblending algorithms. These sources were formally added by doing another SExtractor run with different deblending parameters. An initial list of 208 sources was produced, which was then reduced to a total of 100 sources after visual inspection and rejection of sources that were clearly part of previously identified sources. Note that the SExtractor magnitudes of these 100 sources at bands other than \iband\ are suspect, owing to the close proximity of other bright objects. 

The \zband\ catalog contains 39 sources that are not in the \iband\ catalog. This catalog may also be obtained at the URL referenced above.

While SExtractor is useful to identify and measure the photometric properties of sources in the images, its use introduces some uncertainties that are often overlooked in subsequent analyses. Many objects consist of complicated structures that may either be lumped together as a single source or broken up into several different sources depending on the choice of deblending parameter. These choices will then affect the distribution of source sizes, for example, and could lead to false conclusions about the nature of source sizes in different populations owing to the subjective criteria involved in identifying individual sources. It could also affect the luminosity function by weighting either towards many small faint sources or fewer large brighter complexes combined as individual sources.

Similarly, the photometric magnitudes and colors depend on the approach to aperture photometry. In this paper, we select sources and measure their colors entirely with isophotal magnitudes---mag-ISO in the SExtractor output---but we subsequently use the larger Kron-like aperture photometry in the selection band---mag-AUTO---to measure total magnitudes. Thus, for example, if a source were selected for a particular sample in the \zband -band, the selection would be done on mag-ISO(\zband ), the colors would be set by mag-ISO(band 1)-mag-ISO(band 2), but the final magnitudes would be scaled to make the \zband\ magnitude equal to mag-AUTO(\zband ). 

The GOODS survey produced an image of the same field as the HUDF through the same filter set but with less exposure time. One way to check the completeness of catalogs produced with SExtractor is to compare the sources recovered from the GOODS data with the HUDF and examine differences in the recovered magnitudes. Figure 5 shows the differences between the GOODS and HUDF catalog \zband-magnitudes as a function of HUDF \zband\ for common sources that were detected by GOODS with at least $5\sigma$ confidence. 

%Figure 5: Plot comparing GOODS & HUDF
\begin{figure}[ht]
\includegraphics[width=7.5cm]{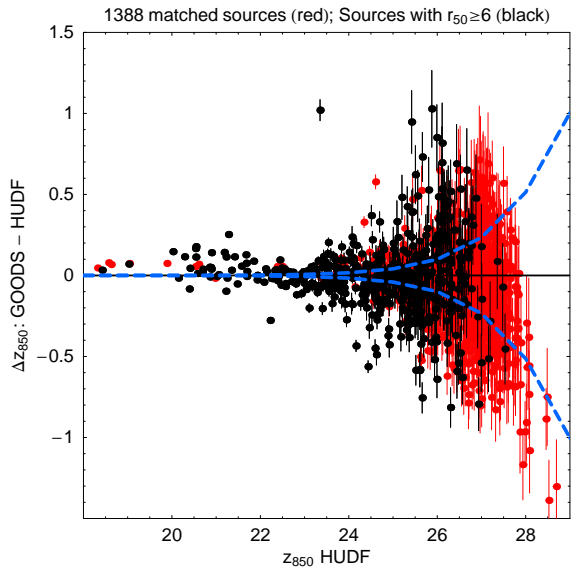}
\caption{The \zband -magnitude differences between GOODS and the HUDF as a function of the HUDF \zband - magnitudes for 1388 sources appearing in both catalogs. The red points are sources with 50\% flux radii less than 6 pixels (0\arcsec.18); the black points are sources with larger 50\% flux radii. The dashed blue lines show the expected 1$\sigma$ scatter from photon, background, and detector noise only in a 112 pixel area corresponding to $r_{50} = 6$.}
\end{figure}

There is generally good agreement between the two surveys. In particular, we find no systematic differences between GOODS and the UDF and no trend of galaxies becoming systematically brighter in the UDF as expected if most galaxies had low-surface brightness regions lost at the GOODS depth and recovered by the HUDF. The direct comparison confirms also that the GOODS \zband\  selected catalog is complete down to $\sim 27^{\rm m}$. 

The HUDF has the great advantage of using hundreds of individual images per band with extensive dithering to reduce the observational noise to very low levels and provide very high signal-to-noise ratios on thousands of sources in the images. Thus, the HUDF images are well suited to explore the uncertainties introduced into source samples through the choice of parameters in algorithms such as SExtractor, since the image noise should be negligible for the brighter objects.

\subsection{Number counts} 
The number of sources grows with increasing magnitude until the limiting magnitudes are reached. Figure 6 plots the cumulative number of objects per square minute of arc as a function of magnitude for \zband. This plot also includes the number counts taken from the GOODS survey in the same units.

%Figure 6: Cumulative number counts (surface density) in GOODS and HUDF 
\begin{figure}[ht]
\includegraphics[width=7.5 cm]{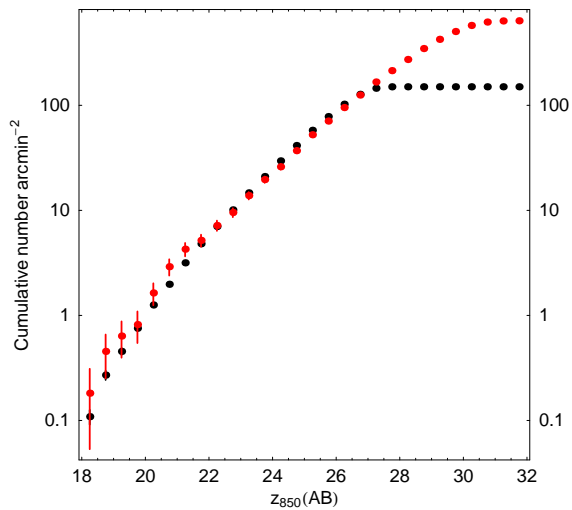}
\caption{Cumulative number counts of objects arcmin$^{-2}$ as a function of magnitude for the HUDF (red) and the GOODS (black) surveys.}
\end{figure}

The HUDF and GOODS number counts per unit area are identical in slope but offset in absolute value for magnitudes less than 26.5, the stated completeness of the GOODS data. The surface density of objects in the HUDF is smaller than in GOODS by about 10\% over the range $23 < {\rm z_{850}} < 26$. The HUDF number counts continue to rise until about magnitude 30 at which point incompleteness limits further increase. 

\subsection{Lyman break galaxies}
Objects at high redshift have little to no observable flux at wavelengths shortward of the rest-frame Ly$_\alpha$ line owing to the strong intergalactic absorption by hydrogen. When the redshift of the object is greater than about 3.5, the observational signature is a lack of flux in the \Bband\ filter but otherwise detectable emission in \Vband , \iband , and \zband . At higher redshifts, the \Vband\ and \iband\ filter fluxes also diminish relative to the longer wavelengths. The flux is said to ``dropout'' of the short wavelength filter (e.g. Steidel et al. 1996).

We identified objects as dropouts according to the following criteria similar to those used in by GOODS:

\Bband-dropouts (if all conditions met):
\begin{eqnarray}
	B_{435} -Ð V_{606} & > & (1.1 + V_{606} -Ðz_{850})  \\
	B_{435} Ð- V_{606} & >  &1.1 \\
	V_{606} Ð- z_{850}  & < & 1.6  \\
	S/N(V_{606}) &>& 5 \\ 
	S/N(i_{775}) &>& 3
\end{eqnarray}

\Vband-dropout if:
\begin{eqnarray}
	V_{606} Ð- i_{775}  &>& 1.47 + 0.89(i_{775} Ð- z_{850}) \nonumber \\
	                                   &   &  {\rm or}\ \ 2  \\ 
	V_{606} Ð- i_{775} &>& 1.2 \\
	i_{775} Ð- z_{850} &<& 1.3 \\
	S/N(i_{775})      &>& 5 \\
	S/N(B_{435})      &< & 3  \nonumber \\
           &{\rm or} & B_{435} - i_{775}  > V_{606} - i_{775} + 1 \nonumber \\
           & & 
\end{eqnarray}

\iband-dropout if:    
\begin{eqnarray}
	i_{775} Ð- z_{850} &>& 1.3 \\
	S/N(z_{850}) &>& 5 \\
	S/N(V_{606}) &< & 2  \nonumber \\
&{\rm or} & V_{606} - z_{850} > 2.8 
\end{eqnarray}

We also used the compactness index in SExtractor, sometimes called stellarity, to reject stars in the images. Through trial and error by examining different bright objects, we found that rejecting sources with a stellarity greater than 0.9 (\Vband\ and \zband ) or 0.8 (\iband ) when the magnitudes were more than 2 magnitudes below the limiting magnitudes effectively rejected stars without excluding small galaxies.

The criteria listed in equations 6--8 were used to generate three lists of dropout candidates for the HUDF using the catalog published here.  We inspected every source in these lists and rejected those that were spurious. The resulting culled lists contain 504 \Bband -dropouts, 204 \Vband -dropouts, and 54 \iband -dropouts over the 11 arcmin$^2$ HUDF field. We applied the same criteria to the published GOODS catalog, culled the list visually, and found 1606 \& 1661 \Bband -dropouts, 486 \& 370 \Vband -dropouts, and 117 \& 125 \iband -dropouts in the north (157 arcmin$^2$) \& south (161 arcmin$^2$) fields, respectively. The corresponding areal densities of sources in the two surveys show similar ratios of \Bband :\Vband :\iband -dropouts but are 8-10 times higher in the deeper HUDF image than in GOODS. This higher source density means we are going farther down the luminosity function with the additional depth. The added sensitivity partially compensates for the smaller HUDF area for discovering high-redshift objects.

Complete tables of all dropout sources are available at {\color{blue} http://www-int.stsci.edu/~svwb/hudf.html\#man}.

In the area where GOODS overlaps the HUDF, there are three \iband -dropouts in GOODS and 6 in the HUDF, two of which are common. The one GOODS source not in the HUDF subset has a color of 1.48 in GOODS and 0.83 in the HUDF. The four HUDF sources not in the GOODS catalog have \zband\  magnitudes greater than 26. These differences can be understood because the GOODS \iband\ image is shallower than the \zband\ image, and noise in \iband\ becomes a limiting factor for GOODS when selecting red \iband$ - $\zband\ sources.

Tables~6-8 list the sources in each of these samples both for the HUDF and GOODS. The electronic version of this paper contains the complete source lists; the printed tables include only the first 36 objects from each sample.

Figures 7 to 9 show examples of these dropout sources. Each figure consists of images of $64\times 64$ pixels or 1\arcsec.92 on a side, corresponding to approximately 12\,kpc in comoving coordinates. The dropout sources are always in the exact center of these images, although in some cases there are several objects that dropout in the small field. 

%Figure 7: Postage stamp images of selected B-dropouts
\begin{figure*}
\includegraphics[width=16. cm]{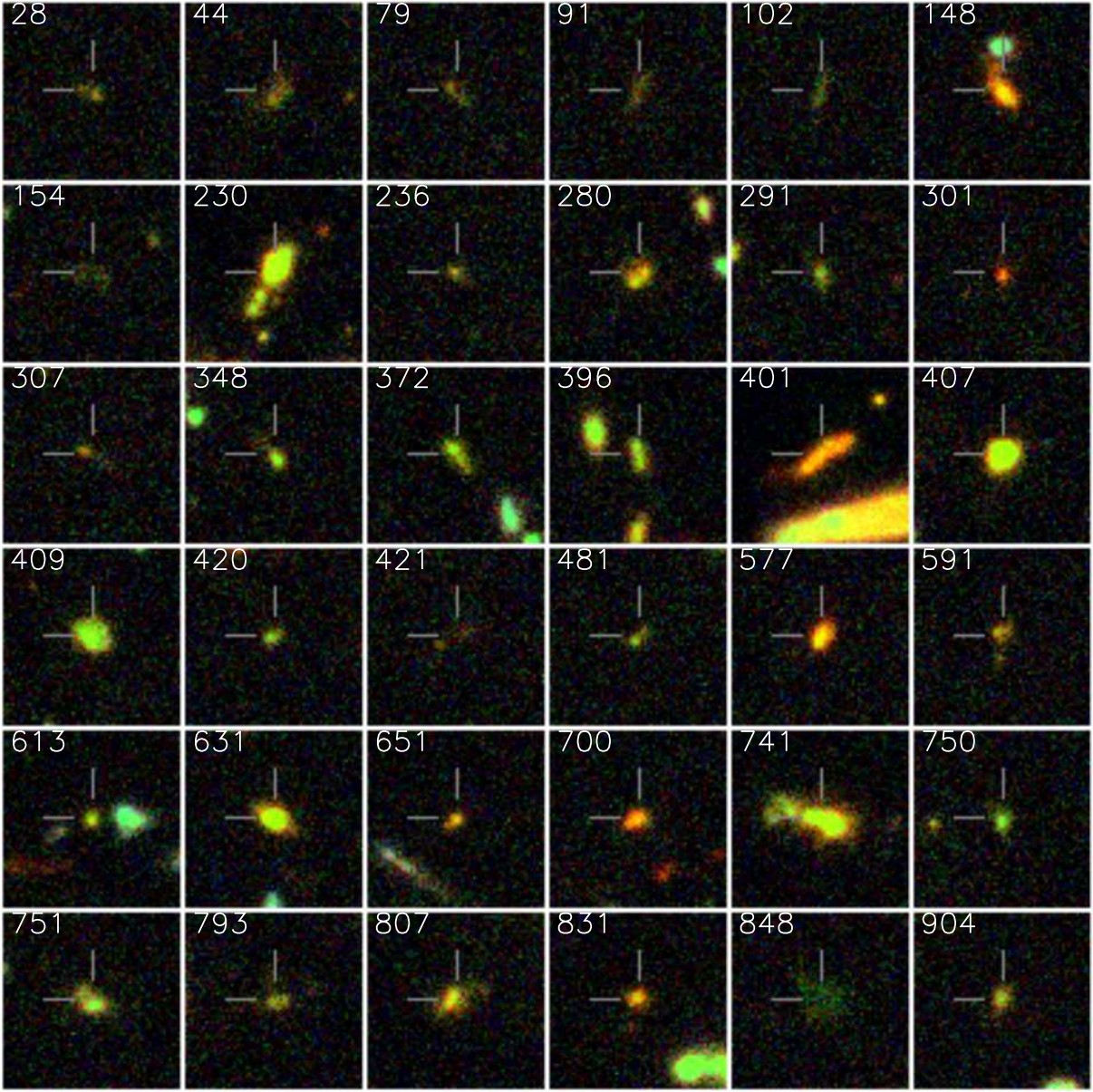} \hspace{1. cm}
\caption{Images of the first 36 \Bband-dropout sources. Each image is $64\times 64$ pixels (1\arcsec.92 square), corresponding to 13\,kpc at $z=4$. The RGB colors coincide with the  \iband, \Vband, and \Bband\ bands to represent the source colors. The small lines denote the exact center position for the objects in the table. }
\end{figure*}

%Figure 8: Postage stamp images of selected V-dropouts
\begin{figure*}
\includegraphics[width=16. cm]{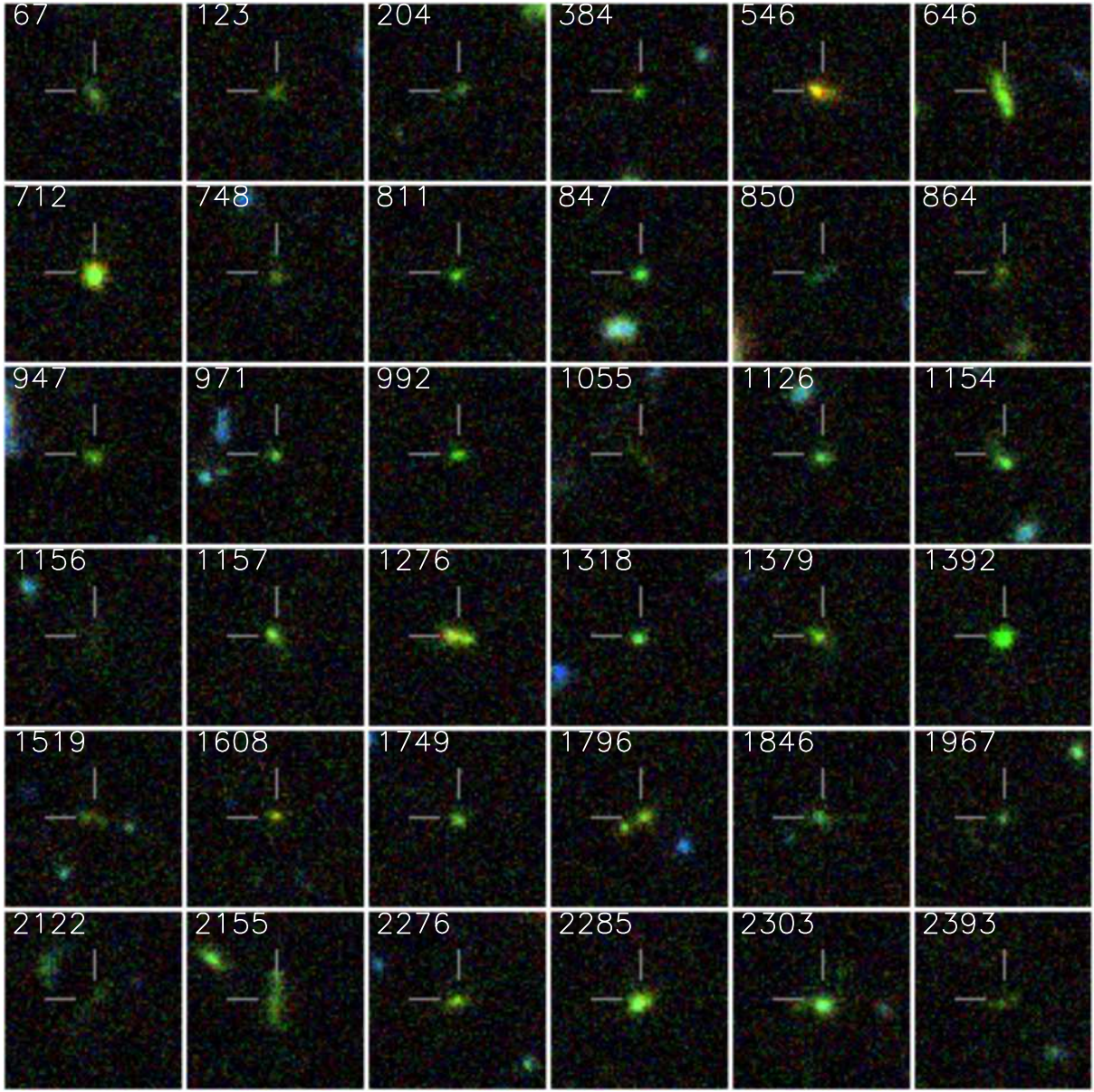} \hspace{1. cm}
\caption{Images of the first 36 \Vband-dropout sources. Each image is 1\arcsec.92 square as in Figure 7, corresponding to 12\,kpc on a side at $z=5.$ The RGB colors coincide with the \zband, \iband, and \Vband\ bands. }
\end{figure*}

%Figure 9: Postage stamp images of selected i-dropouts
\begin{figure*}
\includegraphics[width=16. cm]{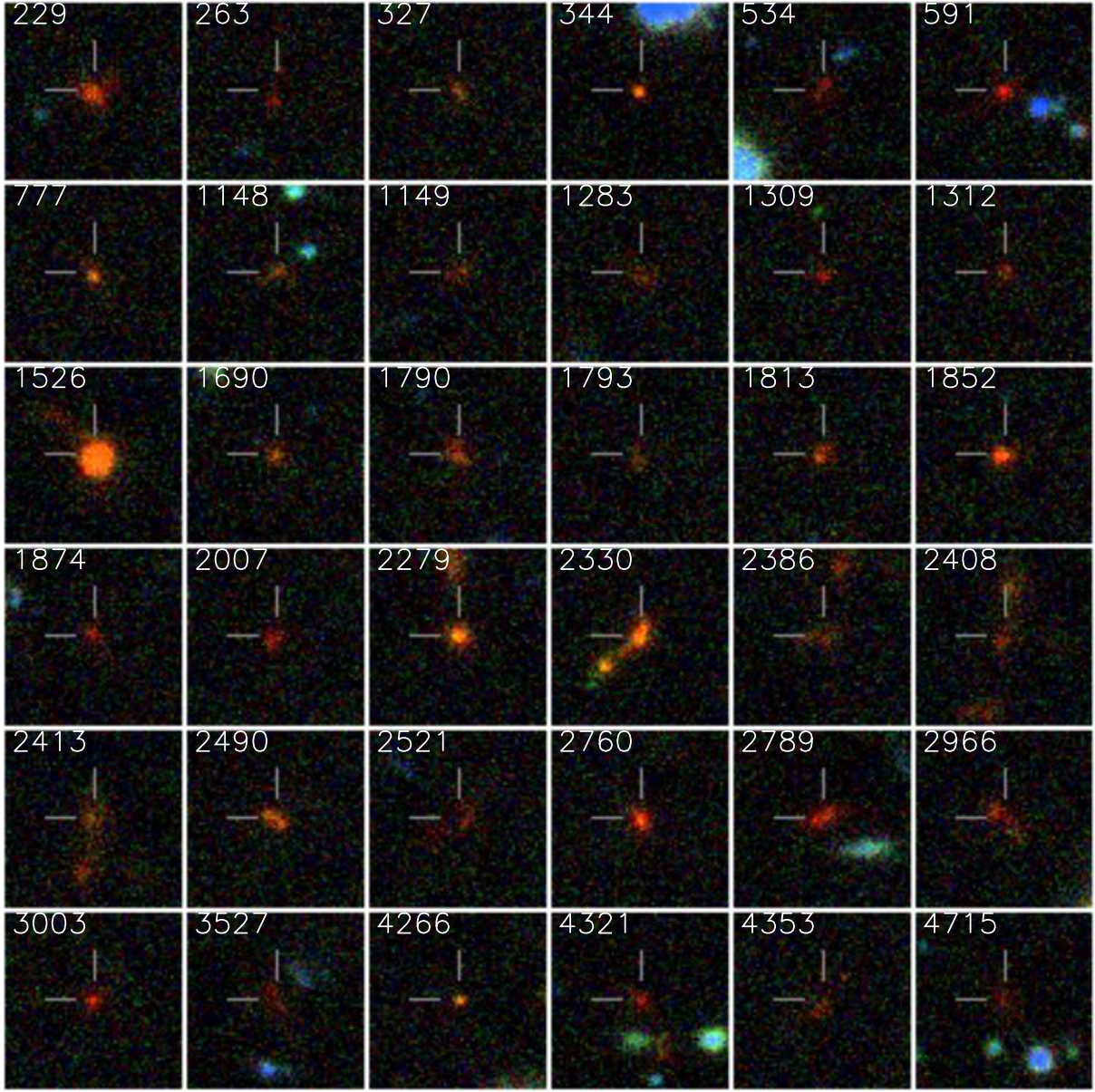} \hspace{1. cm}
\caption{Images of the first 36 \iband-dropout sources. Each image is 1\arcsec.92 square as in Figure 7, corresponding to 11\,kpc on a side at $z=6.$ The RGB colors coincide with the \zband, \iband, and \Vband\ bands. The source IDs in this figure are from the \zband-selected catalog and will not match with the \Bband - and \Vband -dropout IDs.}
\end{figure*}

%Figure 10: 24 bright B-dropouts compared to Lotz Fig. 11 simulations
\begin{figure*}
\includegraphics[width=16. cm]{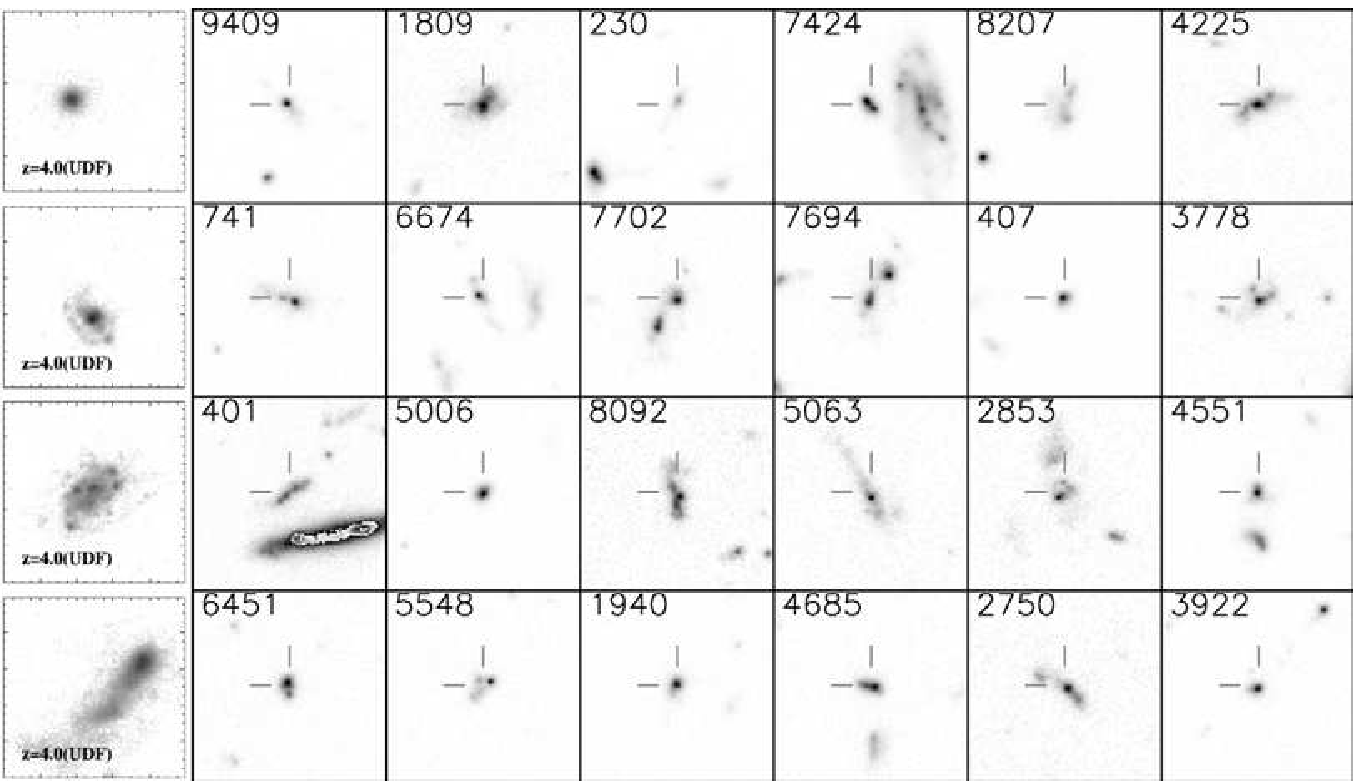}
\caption{Simulated images of nearby galaxies from Lotz et al. (2006) on the left compared with images of the 24 brightest \Bband-dropout sources on the right. All figures are 3\arcsec\ boxes corresponding to 21\,kpc at $z=4.1$. The images on the right are a sum of \Vband - and \iband -flux as assumed by the simulations. The dropout sources are denoted by the small lines in the center of each frame as in Figures 7--9.}
\end{figure*}

Visual inspection of the dropout sources in Figures 7 through 9 shows a large number of irregular structures along with many compact objects. High redshift galaxies appear much smaller and less regular than we see in the local universe, confirming similar findings from other deep surveys with the resolution of Hubble, notably the HDF. The majority of dropout sources are compact, of order 1\,kpc in extent. They often show multiple components and irregular structures. Few of the dropout sources show the regular structures of local spiral or elliptical galaxies. 

Lotz et al. (2006) simulated the appearance of four nearby galaxies as they would appear in the HUDF at a redshift of 4 using the ultraviolet images from GALEX. Figure 10 shows these simulations alongside the 24 brightest \Bband -dropout sources from this present sample, presented as combinations of the \Vband - and \iband -images. Few if any of these bright \Bband - dropout sources resemble the normal galaxies but many resemble the merger shown in the bottom panel.  Although this simple comparison ignores some of the inherent problems of observing faint nearby galaxies with little ultraviolet flux (Lotz et al. 2006), the HUDF, nevertheless, confirms the striking result of the HDF that the universe looks different at redshifts of 3 and above, corresponding to times when the universe was less than about 2 Gyr old. Elmegreen et al. (2005) reach a similar conclusion about galaxies in the HUDF by morphologically classifying all large galaxies in the field and comparing them with local distributions.

For the \iband -dropouts, there are other source lists for comparison that are useful to understand how the combination of catalog generation and subsequent selection produce reproducible samples. Bunker et al. (2004) identified 55 \iband -dropouts in the HUDF, very similar to the 54 in this paper. Close examination shows that while there are many coincidences, the source properties often differ markedly owing to the different catalog generation processes used by the separate groups. The Bunker et al. sample breaks up several of our single sources into multiple sources and find several sources near bright objects that do not appear in our catalog owing to our conservative choice of deblending. Formally, only 38 of the objects in the current catalog are coincident with Bunker et al. sources to 0\arcsec.1. This number grows to 48 when the criteria for proximity is relaxed to 1\arcsec , showing that many of the sources identified here as single are multiple objects in the Bunker et al. catalog. Of the remaining seven sources, four are close to bright galaxies and not separated from them with our choice of deblending parameter, and the other three do not pass our selection filter owing to slightly different magnitudes and uncertainties or observed flux in short wavelength filters. 

Yan and Windhorst (2004) find 108 \iband-dropouts in the HUDF to a limiting magnitude \zband $= 30$. Of these 108, 48 match STScI sources to 0\arcsec.1, a few do not appear in the STScI catalog owing to proximity to bright galaxies, and almost all the others are fainter than the limiting magnitude used in this study. Stronger deblending in the Yan and Windhorst sample means that their source sizes are typically smaller than ours, and several are sub-components of single sources in the present catalog.

Visual inspection of all sources in both samples show they are real and not image artifacts. Our inspection of the Yan and Windhorst (2004) catalog of \iband-dropout sources using similar selection criteria indicated that all those objects are also real, even though they have many objects unique to their catalog. The salient point is that different catalog parameters and different photometric magnitudes produced by the SExtractor software combine to create rather different source samples in this very simple case, introducing an important but not easily quantifiable uncertainty into subsequent interpretations of the sources properties.

\subsection{Sample redshifts and volumes}
The redshift ranges sampled by these criteria depend on the (unknown) spectra of the sources and to some extent on the source magnitudes. To assess the impact of spectrum on selection, we calculated the redshift ranges over which galaxies with different characteristic spectra may be selected. We adopted synthetic galaxy spectra from Bruzual and Charlot (2003) and convolved them with absorption by intergalactic hydrogen according to the prescription of Madau (1995) to simulate galaxies at different redshifts, adding also two very simple spectra employing a step function at Ly$_\alpha$. 

Figure 11 shows the different spectra in the rest frame (i.e. without intergalactic absorption) and how the different selection criteria act on the galaxies with different spectra as functions of redshift for the three samples. Tables~9 and 10 give the exact redshift ranges and equivalent co-moving volumes per square minute of arc in the HUDF field without and with extinction for comparison. We stress that these calculations assume noiseless observations and assume perfect photometry, so they are not equivalent to calculations of the effective volume (e.g. SAGDP99). Nevertheless, they are useful to illustrate how differences caused by spectral variations in the sources affect the sample selection. 

%Figure 11: Comparison of selection criteria
\begin{figure*}
\includegraphics[width=8. cm]{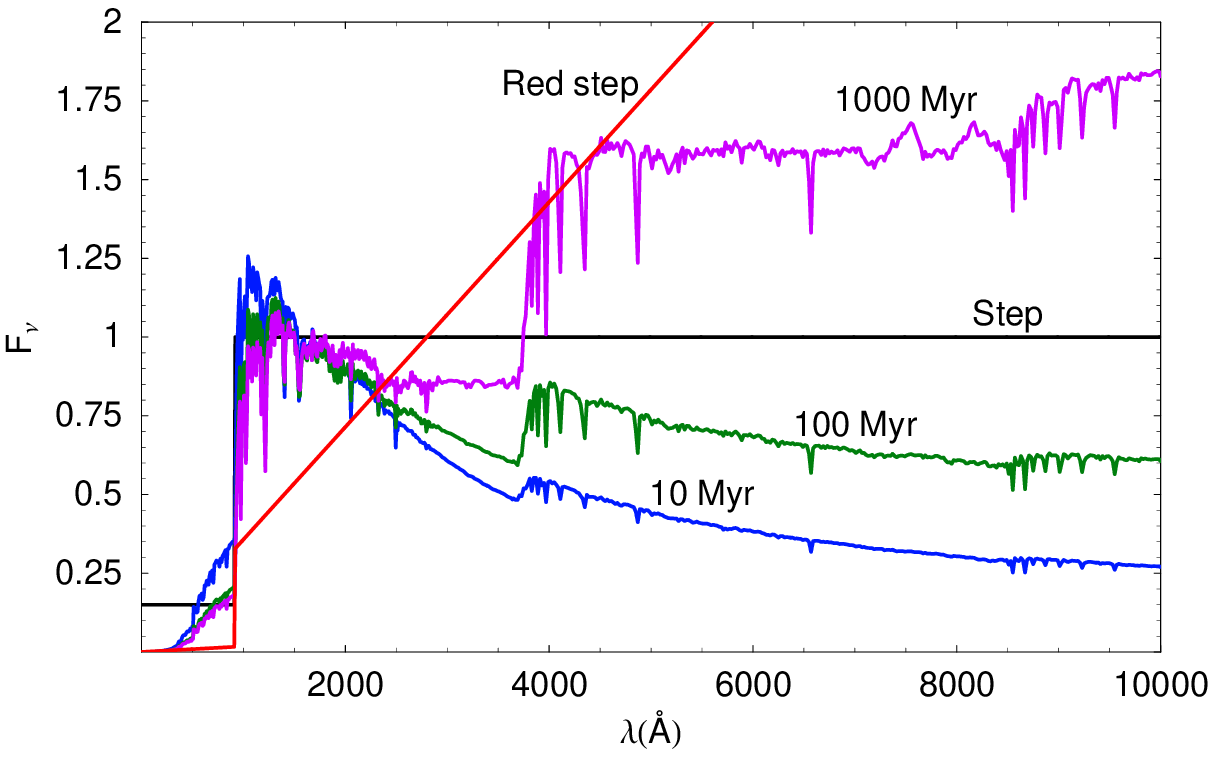} \hspace{0.2 cm}
\includegraphics[width=8. cm]{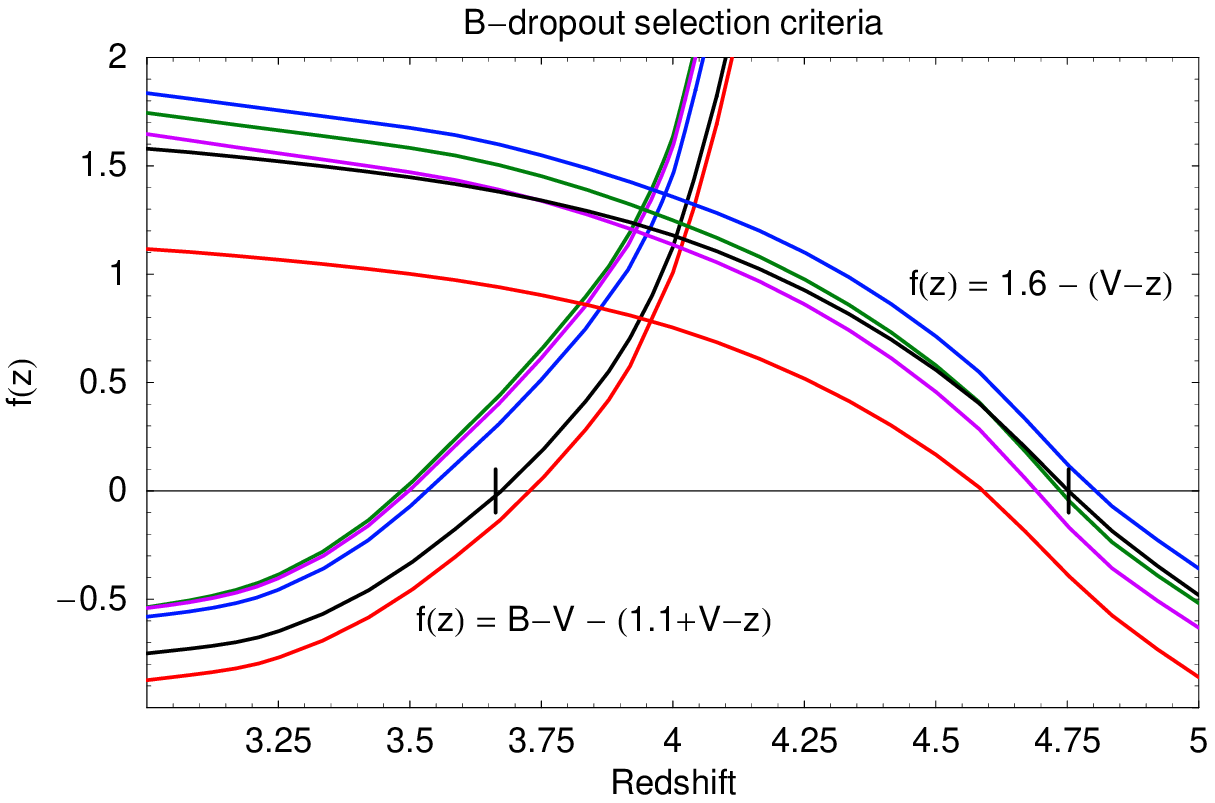} \\ \\
\includegraphics[width=8. cm]{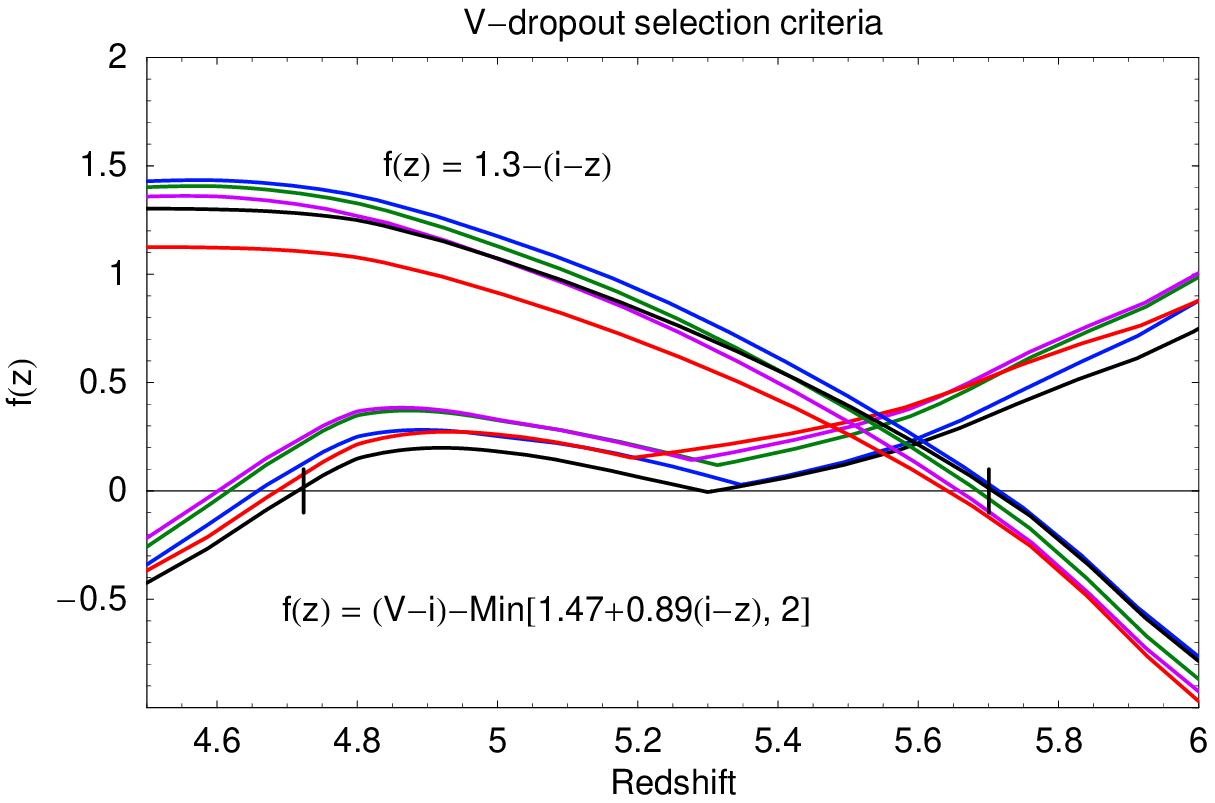} \hspace{0.2 cm}
\includegraphics[width=8. cm]{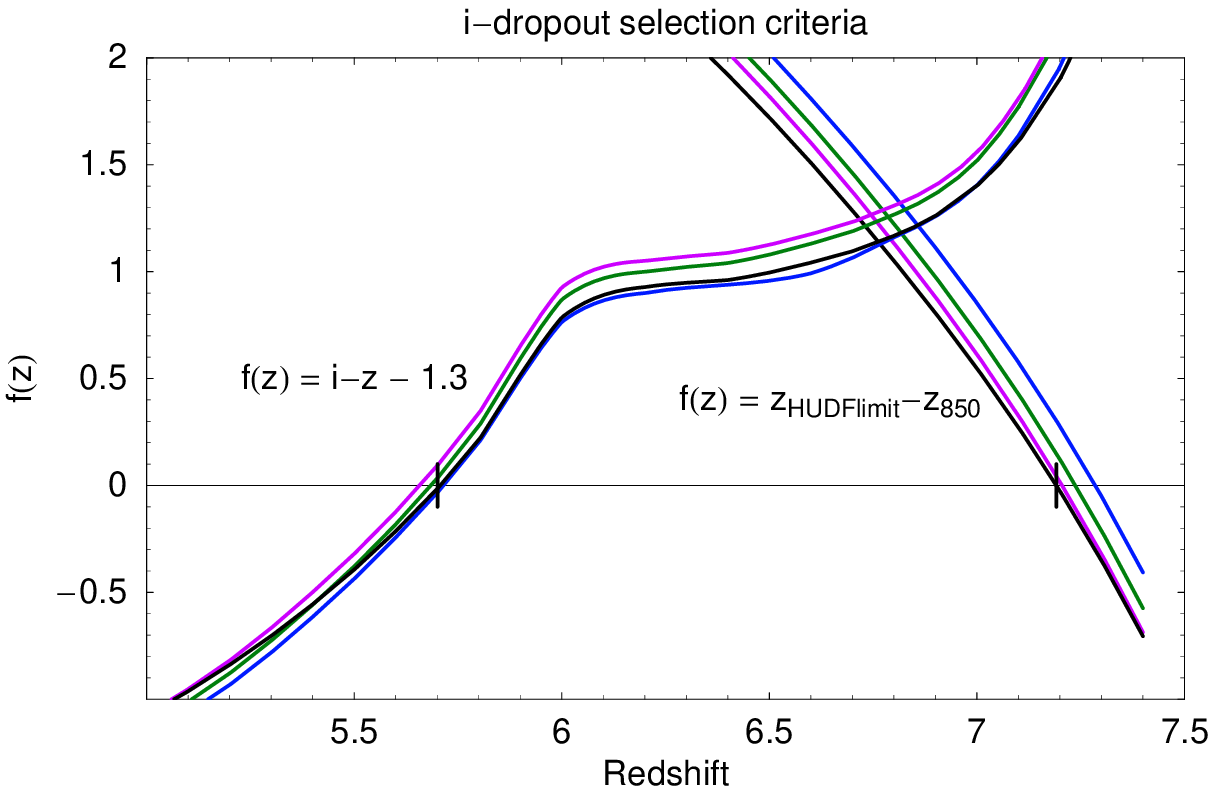}
\caption{Upper left: Synthetic spectra of galaxies to assess dropout sample criteria. Upper right, lower left, lower right: Each of the criteria in equations 6--8 are plotted as leftside-rightside, so that they satisfy the inequalities when they cross zero. The quantities that increase through zero, e.g. $B-V-(1.1+V-z)$, define the start of the selection range; the quantities that decrease through zero, e.g. $1.6-(V-z)$, define the end of the selection range. The color coding for models is the same as in the upper right: 10\,Myr CSF (blue), 100\,Myr CSF (green), 1000\,Myr CSF (magenta), Step (black), Red step (red). The vertical ``ticks'' on the $f(z) = 0$ line illustrate the selection redshifts ($z_{min}$ and $z_{max}$ )for the step function spectrum.} 
\end{figure*}

The differences in selection owing to spectral variations are generally modest, but they can affect the mean redshift and sample volumes especially if reddening by dust is important. These differences are minimized in the highest redshift samples. One additional effect for the \iband-dropouts is the importance of observational sensitivity to the most distant sources. The upper limit to the redshift range is technically around 7.4 for very bright objects ($L \gg L_*$), but in practice may be limited to a smaller value by the paucity of bright galaxies. This effect is a minor problem for the \Bband- and \Vband -dropouts where the color criteria themselves limit the upper end of the redshift range. 

Several papers have noted that photometric errors can scatter objects in and out of the color selection windows. This scattering will be most pronounced for selection criteria with shallow slopes as they approach zero in Figure 11, for example, $f(z) = (V-i) - {\rm Min}[1.47+0.89(i-z), 2]$ corresponding to the condition $V_{606} - i_{775} > \left[1.47 + 0.89(i_{775}-z_{850})\ {\rm or}\  2\right]$ for \Vband -dropout limits near $z \sim 5.3$, or for the upper limit selection of \Bband -dropouts for red spectra (cf. Tables 9 \& 10.)

These examples do not include sources with strong Ly$_\alpha$ lines and so ignore at least one known population of objects in the analysis. Only very strong line-emission will affect the broad-band colors enough to change the selection criteria, probably less than the variations in color seen in our synthetic spectra. 
The selection criteria in principle allow detection of more evolved stellar populations that are intrinsically red in redshift ranges outside those of interest. We tested the spectrum of a 1\,Gyr old population at solar metallicity, also from Bruzual and Charlot. The \Bband-dropout criteria effectively excluded the selection of this population at any redshift. Both the \Vband-dropout and \iband-dropout criteria allowed detection of some interlopers, but the expected contamination from the expected space density of such objects is small. In fact, the \iband-dropout criteria allow detection of these populations over two redshift ranges, the second being limited by the redshift at which the age of the universe is 1,Gyr. 

At these redshifts, the selection criteria are mainly sensitive to the absorption edge of the intergalactic medium for any population with detectable amounts of ultraviolet radiation. When there is little or no ultraviolet radiation present, as in a more evolved population, the criteria pick up the galaxies at redshifts where their intrinsic spectra show strong edges, complicating the interpretation. The dominant interlopers in these samples are likely to be evolved stellar populations at low redshifts. In addition, reddening by dust is likely to make some high-redshift galaxies undetectable in our samples, decreasing the completeness. Extinction by dust can be the largest single uncertainty in deriving source luminosities from the limited spectral information and can have profound consequences for understanding the total amount of light emitted by the distant populations, but it only affects source selection at redshifts below about 4---see, for example, Figure 1 of SAGDP99. 

\subsection{Size distributions}
It is possible to get a crude measure of galaxy sizes using the parameters produced by SExtractor.  Figure 12 plots the distributions of SExtractor radii enclosing 50\% of the flux for the three dropout samples in the two surveys together with log-normal functions for comparison. The plotted uncertainties are proportional to the square-root of the number of sources in each bin. Also shown in the figure is the size distribution for the 1388 sources that appear in both the HUDF and GOODS catalogs (coordinates within 0\arcsec.1) measured separately in each catalog. A log-normal distribution characterizes the GOODS data remarkably well, considering the crude measure of size used here. The sources in the HUDF sample appear smaller on average for the \Bband - and \Vband -dropouts. 

%Figure 12: Dropout size distributions
\begin{figure*}
\includegraphics[width=7.5 cm]{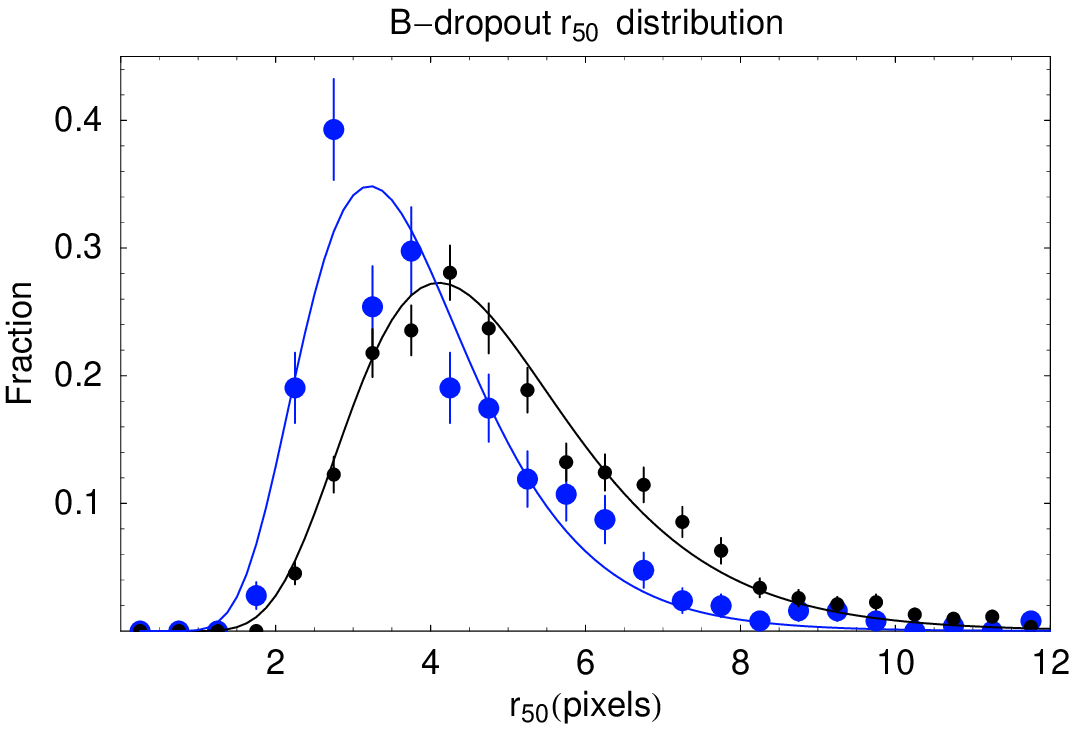} \hspace{1. cm}
\includegraphics[width=7.5 cm]{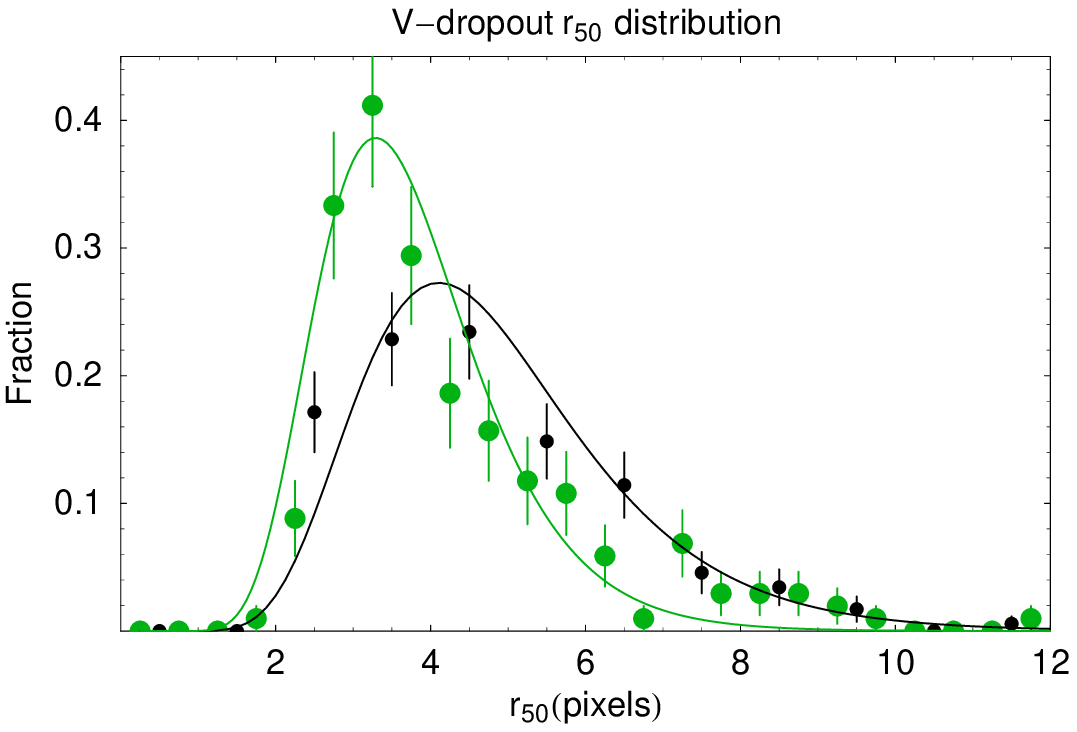} \\ \\
\includegraphics[width=7.5 cm]{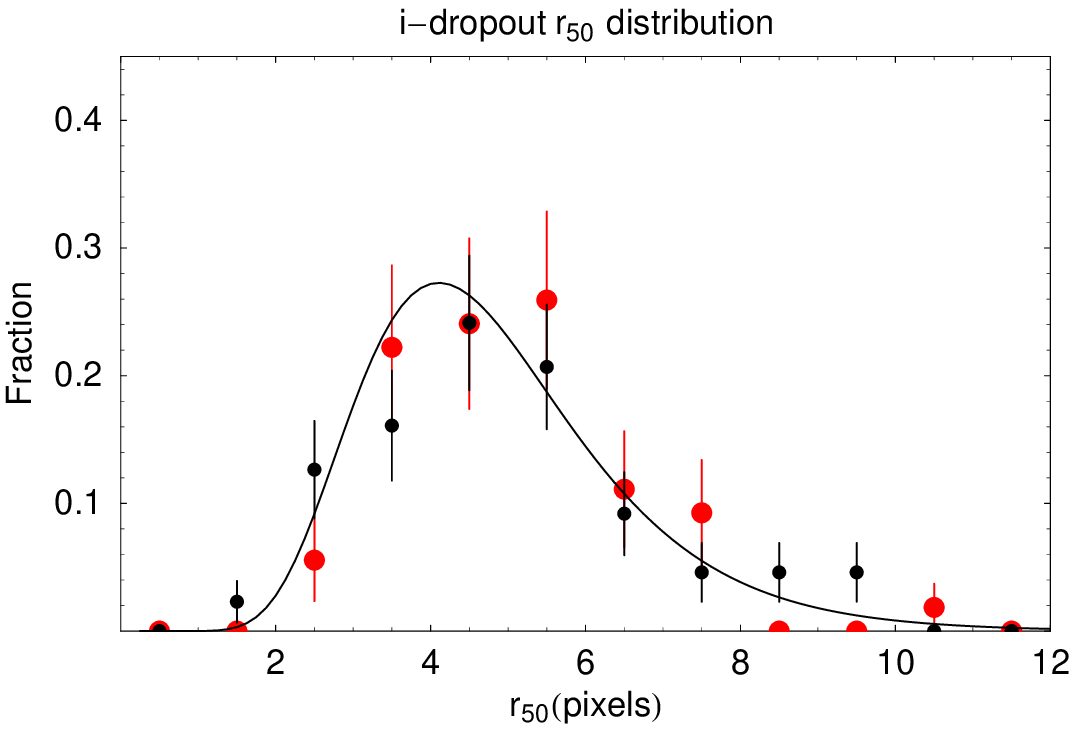} \hspace{1. cm}
\includegraphics[width=7.5 cm]{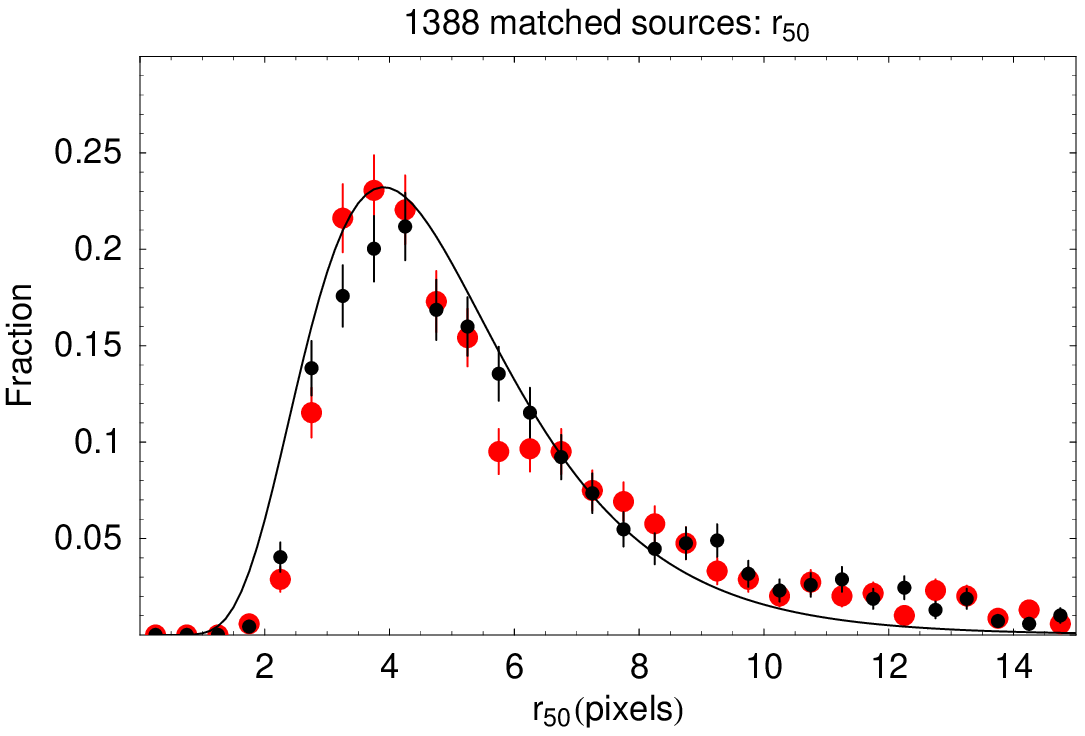}
\caption{The size distributions of the HUDF and GOODS dropout sources and for 1388 sources in the GOODS catalog that also appear in the HUDF (lower right.) The size parameter is the radius (in pixels) enclosing 50\% of the flux as returned by SExtractor. Large colored points are HUDF, small black points are GOODS in each plot. Uncertainties are Poisson noise of numbers per bin. The black lines are log-normal functions with the peak and width given as $\mu = \log(4.6)$ and $S = \log(1.5)$. The colored lines have fitting parameters: $\mu = \log(3.6)$, $S = \log(1.4)$ (blue), $\mu = \log(3.6)$ and $S = \log(1.35)$ (green).  }
\end{figure*}

%Figure 13: Probability detection functions: pUDF(V, i, z)
\begin{figure*}
\includegraphics[width=7.5 cm]{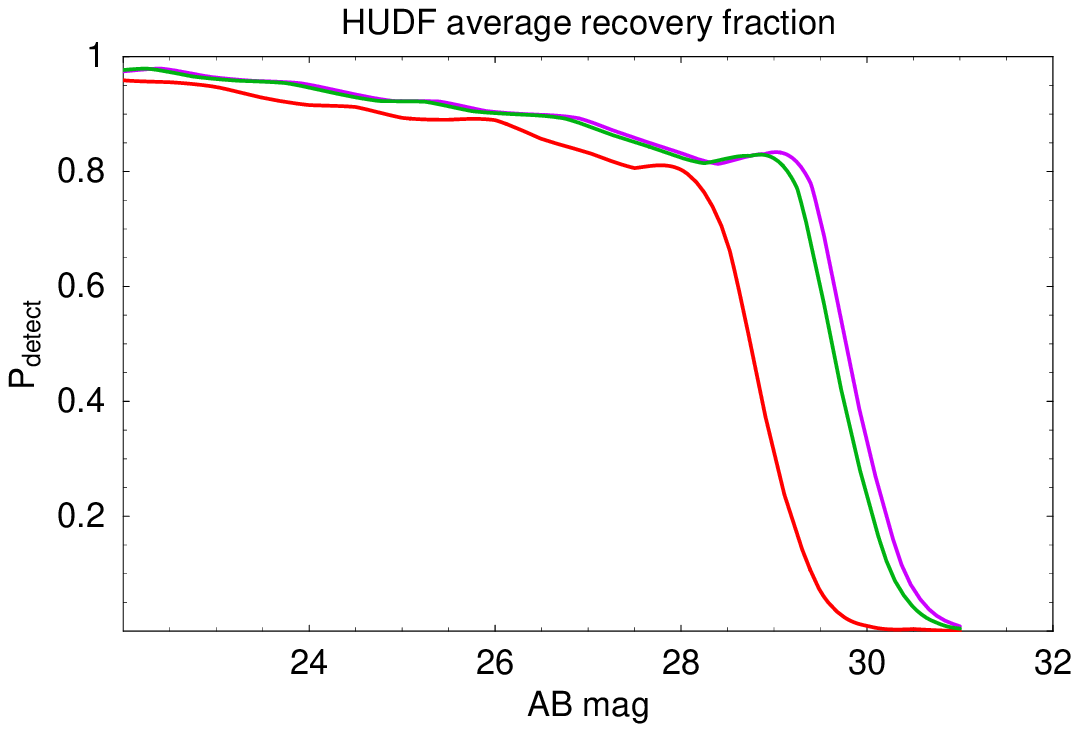} \hspace{1. cm}
\includegraphics[width=7.5 cm]{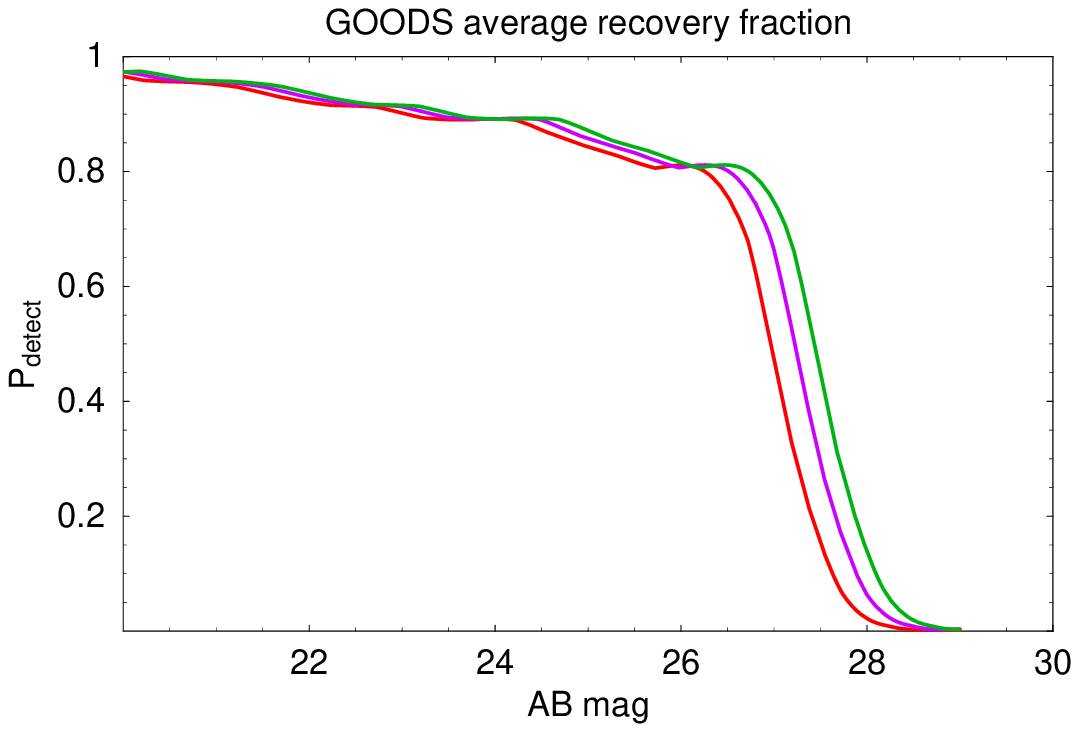}
\caption{The assumed probability of detecting dropout sources vs. magnitude for the HUDF (left) and GOODS (right) in the three bands, \Vband (green), \iband (magenta), and \zband (red) based on the convolution of a log-normal size distribution with the recovery fractions as a function of size and magnitude from Figure 3.}
\end{figure*}

The differences between the size distributions for the HUDF and GOODS samples indicate the general problem of characterizing these irregular sources with single sizes, comparing sizes for different source luminosities and survey sensitivities, and also highlights the difficulty of assessing changes with redshift. There is modest evidence for a slight increase in typical source size in the HUDF \iband -drop sample presented here, but we suspect the uncertainties in deriving these samples coupled with the difficulties of using the 50\% flux radius to characterize them preclude any conclusion that the source sizes are evolving or staying constant. As noted in the previous section, it is not always straightforward to determine when a close grouping of several sources should be treated as a single, larger object. This problem resulted in a large mismatch between the Bunker et al. (2004) \iband -dropout sample and the \iband -dropout sample presented in this paper, even though the color selection was nearly identical.

Figures 7---9 show that these distant galaxies are often close to foreground objects that would be blended in images with poorer angular resolution. A substantial fraction of these galaxies will be blended with foreground objects in images from ground-based telescopes limited to $\sim 1$\arcsec\ resolution by the Earth's atmosphere. This blending and the uncertainties in the models needed to correct for its effect introduce an additional complication in comparing samples derived from space-based and ground-based observations at faint magnitudes.

Figure 13 shows the probability of detecting dropout sources as a function of source magnitude in the HUDF and GOODS by using the log-normal distribution and convolving it with the recovery fractions shown in Figure 3. These probabilities will be used subsequently to assess the expected number of dropout sources seen in the two surveys. 

\subsection{Cosmic error}
The spatial distribution of galaxies is correlated over many scales, leading to fluctuations in the number density of galaxies. For small volumes, this correlation can dominate the errors in number counts of galaxies. The cosmic error is the square root of the field-to-field variance due to large scale structure in excess of the Poisson noise for any sample. It is an important problem for the HUDF owing to the small field of view. The relative cosmic error, $\sigma_v$, is the root variance corrected for Poisson noise, divided by the average number density (see Somerville et al. 2004, hereafter S04, equation 1).

There are several ways to estimate this error for the HUDF samples. An empirical lower limit for the bright objects can be obtained by dividing the large GOODS area into equal-sized regions of approximately the area of the HUDF and looking at the variance of sources among them. Dividing GOODS into two $3 \times 5$ grids of 10.9 square arc\-minute rectangles, the distribution of \Bband-dropouts among these is GOODS-N: (82, 76, 105, 93, 134, 128, 100, 115, 115, 95, 107, 123, 116, 110, 106), and GOODS-S: (60, 153, 140, 134, 131, 99, 133, 108, 106, 133, 93, 95, 95, 88, 93). 

The fractional root variance of this distribution is 19\% with a statistical sampling error of 9.6\% (approximately 110 sources per cell).  The relative cosmic error after removal of shot noise is then $\sigma_v = 0.16$. This measurement applies only to the brighter \Bband-dropouts detected at the GOODS depth, and gives a crude measure of the fluctuations that might be expected in the HUDF dropout samples. However, we note that the actual variations among fields is as large as a factor of 2.5, 60 vs. 153 in the first two GOODS-S sub-fields, for example.

%Figure 14: B-dropout variance in GOODS
\begin{figure*}
\includegraphics[width=6.8 cm]{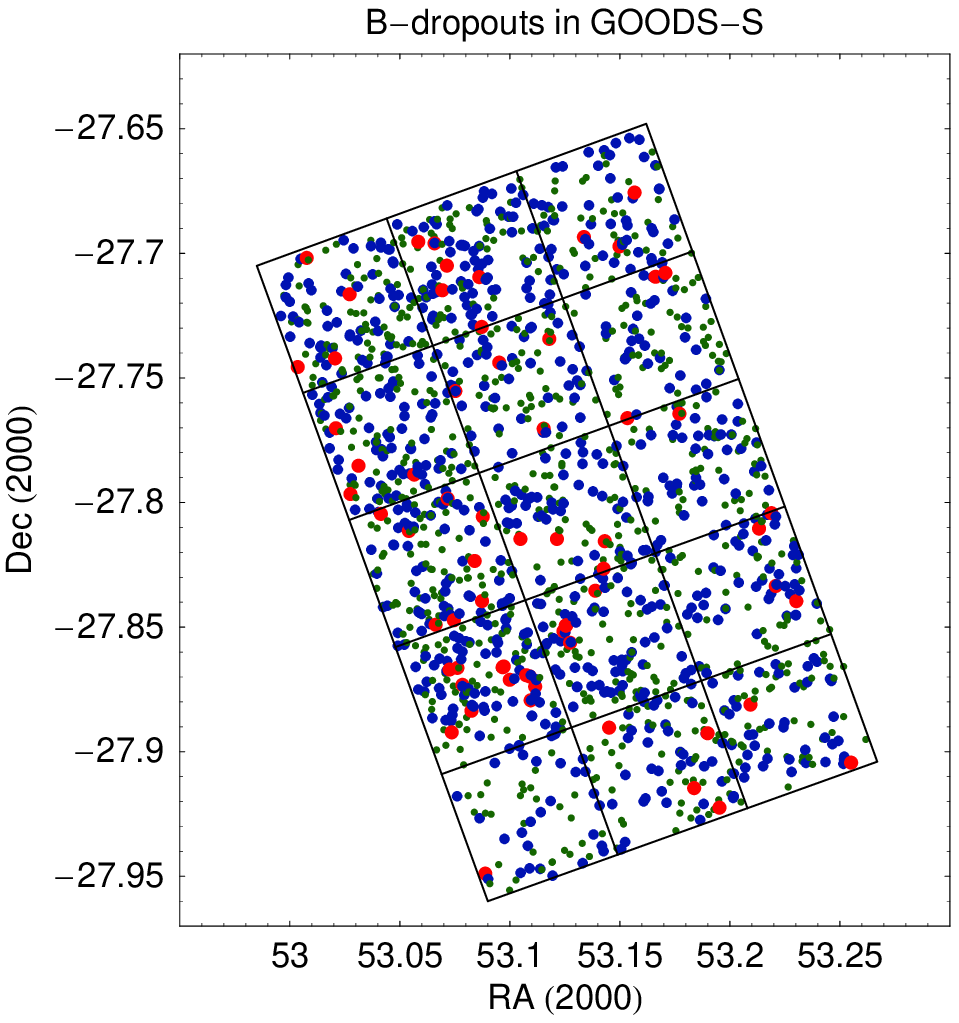} \hspace{1. cm}
\includegraphics[width=7.5 cm]{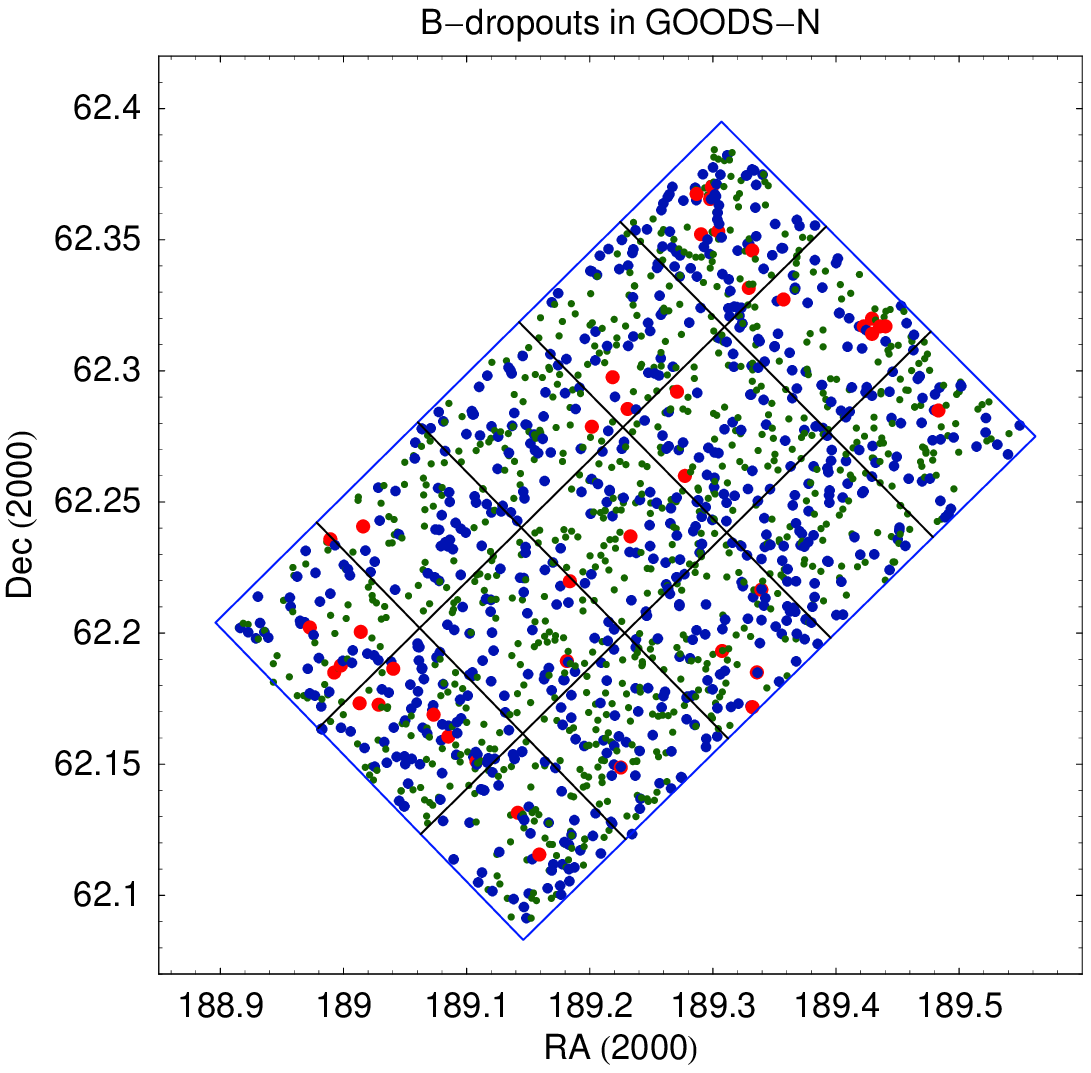}
\caption{The location of the \Bband-dropout sources are plotted on the GOODS areas. Each GOODS field is divided into a grid of $3 \times 5$ HUDF-sized areas to show the typical variance of the sample location on the sky. The size and color of the points depend on the \Vband\ magnitude as follows: large red: \Vband $< 25.5$; medium blue: $25.5 \le$\Vband $<27$; small green: $27 \le$\Vband.}
\end{figure*}

Figure 14 plots the distribution of \Bband-dropout sources on the two GOODS fields divided into a $3\times 5$ grid. It is easy to see the variations among these different HUDF-sized areas. These plots segment the samples into bright (red), moderate (blue), and faint (green) sources to allow separate estimations of source variance at different luminosities. More luminous objects are more rare and also more clustered, leading to larger field-to-field variation in their number counts. 

A similar calculation for the \Vband-dropouts yields a relative cosmic error of 16\% with a statistical sampling error of 16\%. The number of sources per HUDF-sized area is 32 and 25 in GOODS-N/S, respectively, so the field-to-field variance due to shot noise is comparable to the variance due to large scale structure. Note that the variance between GOODS-N and GOODS-S for \Vband-dropouts is about 14\% between two 160 square arcminute areas. The number of \iband-dropouts is too small to give a meaningful measure of the field-to-field variance.

This empirical approach suffers from use of the brighter and potentially more clustered objects and the use of a relatively small number of cells. A second approach is to use the measured correlation function from Lee et al. (2006) for GOODS \Bband\ and \Vband - dropouts and the analytical expression in Equation (3) of S04. For the faintest \Bband -dropout sample at limiting magnitude, \zband = 27, the correlation scale length is $r_0 = 4.14$\,Mpc. Assuming $\gamma = 1.8$ implies $\sigma_v{\rm (B_{435} drop)} = 0.31$ for spherical cells. We corrected this result for the very elongated cell geometry of the HUDF assuming scale independent bias, yielding $\sigma_v{\rm (B_{435} drop)} = 0.14$. A similar estimate for \Vband-dropouts using $r_0 = 6.3$, $\gamma = 1.8$ yields $\sigma_v{\rm (V_{606} drop)} = 0.49$ for spherical cells and $\sigma_v{\rm (V_{606} drop)} = 0.25$ for the HUDF cell geometry.

A third approach is to use the theoretical expectations for the clustering of dark matter in a hierarchical cosmology plus an estimate of the bias of the galaxy population in question, as described in S04. Here we have improved on the results presented in S04 by correcting for the pencil-beam geometry of the HUDF rather than using spherical cells. For the \Bband-dropouts and \Vband-dropouts, we can estimate the bias of the fainter HUDF sample by extrapolating to lower luminosities using the halo occupation model derived by Lee et al. (2006) based on the brighter GOODS samples. For the \Bband-dropouts, the co-moving sample volume is $4.24 \times 10^4$\,Mpc$^3$, yielding a source density of 0.011\,Mpc$^{-3}$ for 504 sources. From Figure 9 of Lee et al. (2006), we estimate the bias at this number density to be $b \sim 2.4$. Using $\sigma_{\rm DM} = 0.14$ (for spherical cells, as plotted in Figure 3 of S04), we then get $\sigma_v{\rm (B_{435} drop)} = 0.34$. After correcting for the HUDF cell geometry, this is reduced to $\sigma_v{\rm (B_{435} drop)} = 0.15$. Similarly, for \Vband -dropouts, the halo occupation model in Figure 9 of Lee et al. (2006) and the number density in the HUDF gives a bias estimate of 2.7, resulting in $\sigma_v{\rm (V_{606} drop)} = 0.17$ with corrections to the HUDF geometry.

There is no extant measure of the clustering or constraints on the halo occupation model for \iband-dropouts, but we can estimate the bias assuming one galaxy per halo as in S04. With a number density of $54 / (43,867\,{\rm Mpc}^3) = 0.0013$ at $z = 6.46$, the predicted bias is 4.7, and $\sigma_v{\rm (i_{775} drop)} = 0.23$ for the HUDF (with the geometrical correction for non-spherical cells). 

The three methods give consistent estimates of the cosmic error for the \Bband- and \Vband-dropouts, an encouraging result considering the uncertainties. In what follows, we adopt 17\% as a characteristic value for the cosmic error. 

\subsection{Luminosity functions}
The distribution of apparent source brightnesses results from binning the samples and comparing the number density of galaxies on the sky at different magnitudes. It is useful to allow the bin sizes to vary to keep the number of objects in each bin similar.

Tables~11--13 list the number of objects per square arcminute in different magnitude ranges from the samples above. The relative uncertainties are the square root of the total number of objects in each bin combined in quadrature with a fractional uncertainty of 0.17 to include at least minimally cosmic error. The actual uncertainties are certainly higher owing to selection biases especially within the small area of the HUDF; there is no attempt to remove these biases here. Each table also lists the total surface brightness of radiation in the shortest waveband for source detection (e.g. \Vband\ for \Bband--dropouts.)

Figure 15 plots these number density functions for the three redshift intervals of the samples. The GOODS N and S sources are plotted separately to show the variances in samples over the relatively large areas of GOODS. The previous section demonstrated that the empirical variance within an HUDF area is at least 17\% for GOODS \Bband-dropouts, and this variance is likely to apply to all the HUDF samples.

Figure 15 also has Schechter functions (Schech\-ter 1976) overplotted on the data. The fits to the number density distributions all assume a faint end slope parameter, $\alpha = -1.6$. The fits are done in two ways to illustrate the sensitivity to the parameters and the selection biases associated with the sensitivity of the observations. The dashed lines are fits of a Schechter functional form with two parameters, $m_*$, and $\Sigma_*$, to the apparent magnitudes to derive a characteristic magnitude and sky density. The solid lines assume Schechter function distributions with a characteristic absolute magnitude, $M_*$, and volume density, $\phi_*$, integrating the function over the sample redshift interval weighting each redshift by the co-moving volume and relative contribution of the Schechter function for an apparent magnitude bin:
\begin{equation}
	{L \over L_*}(m_b) = {D_L(z)^2 \over 1 + z} 10^{(M_*(z) + k_b(z) - m_b)/2.5}, 
\end{equation}
\begin{equation}
	f({L \over L_*}) = \Delta\Omega \int_{z_1}^{z_2} \phi_*^\prime  \left(L \over L_*\right)^{\alpha + 1} e^{-{L \over L_*}} {dV \over dz} dz,
\end{equation}
where $D_L(z)$ is the luminosity distance at redshift, $z$, in 10 pc units, $m_b$ the apparent magnitude at band, $b$, $k_b(z)$ is the k-correction for a specific spectrum, $\Delta\Omega$ is the solid angle (arcmin$^2$), $\alpha$ and $M_*$ are the usual Schechter parameters, , $\phi_*^\prime$ is the Schechter $\phi_*$ divided by $2.5 \log(e)$, and $dV \over dz$ is the co-moving volume at redshift, $z$. The redshift interval, $z_1$--$z_1$ was that appropriate to the Bruzual and Charlot 100 Myr continuous star formation model with 0.4 solar metallicity and without extinction (cf. Figure~11 and Table~9); different model spectra mostly affected $\phi_*$ owing to different sample volumes. The k-correction tends to be dominated by the position of the absorption edge in the relevant band in all cases, not the long wavelength slope of the spectrum. Indeed, a simple Schechter function assuming all galaxies are at the same redshift gives almost the same best-fit parameters as the integral over redshift with slight differences in the ratios of $\Sigma_*/\phi_*$ among the samples. To fix the magnitude scale, we normalized all galaxy spectra to make $M_*$ correspond to 1400\,\AA.

%Figure 15: Surface density luminosity functions
\begin{figure*}
\includegraphics[width=7.5 cm]{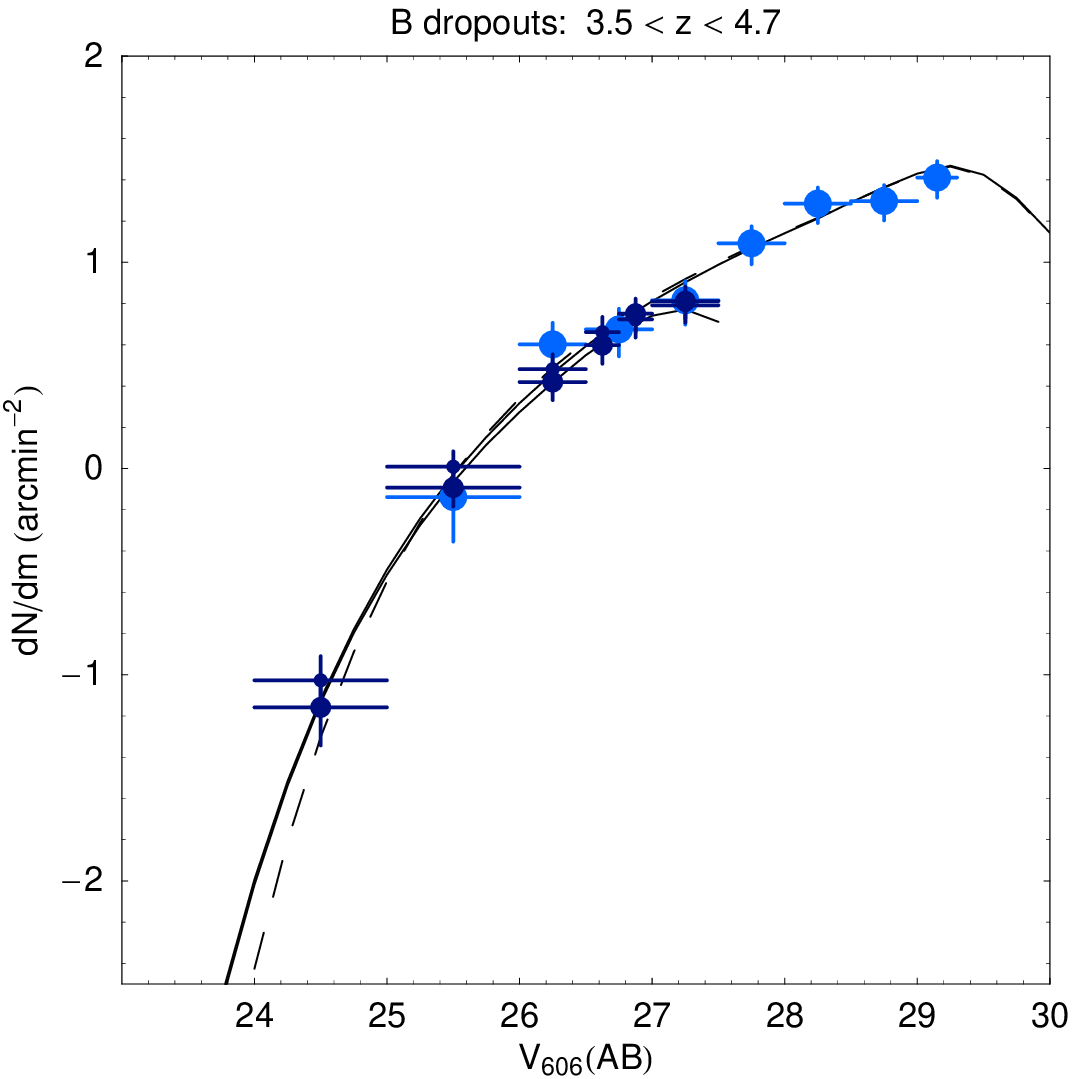} \hspace{0.5 cm}
\includegraphics[width=7.5 cm]{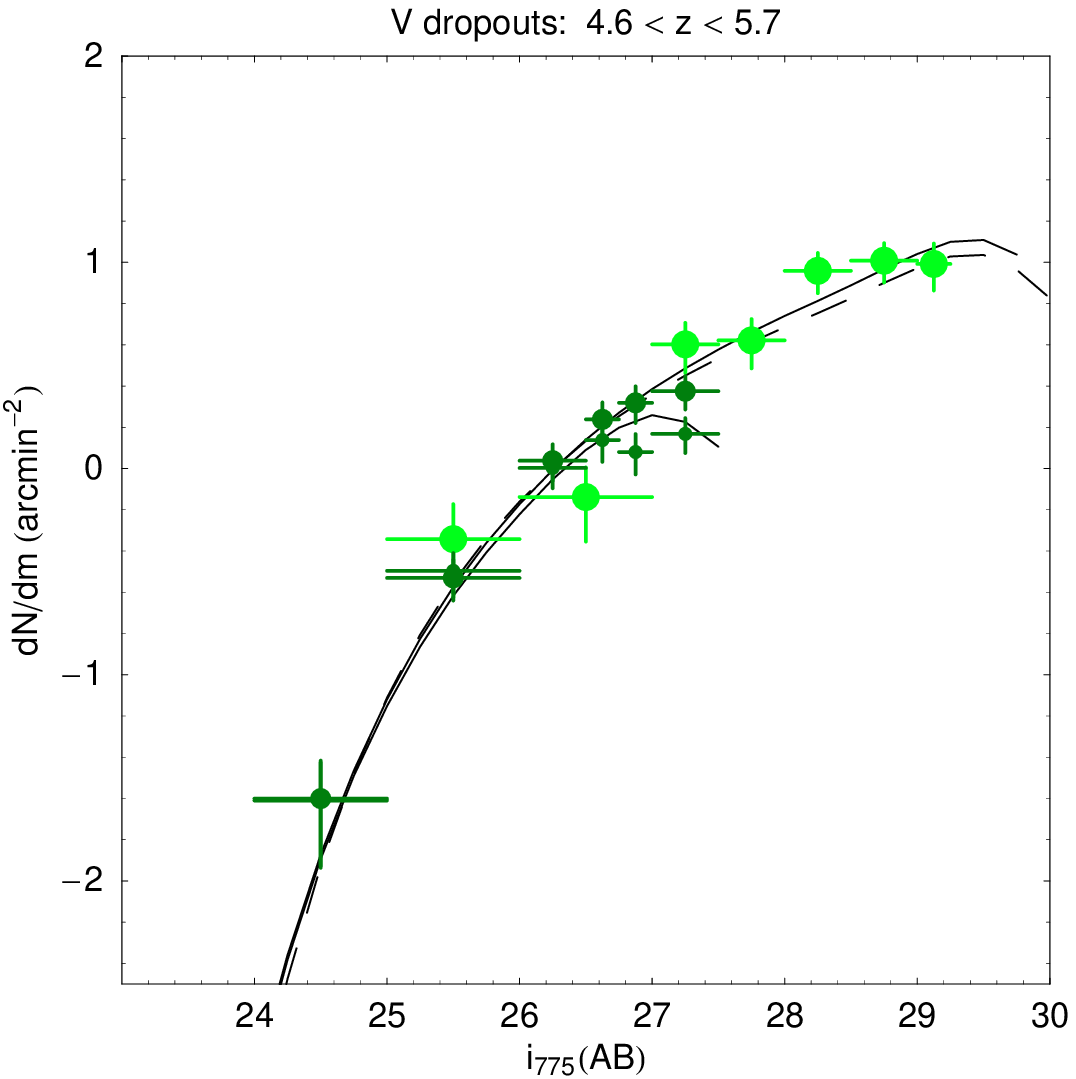} \\ \\
\includegraphics[width=7.5 cm]{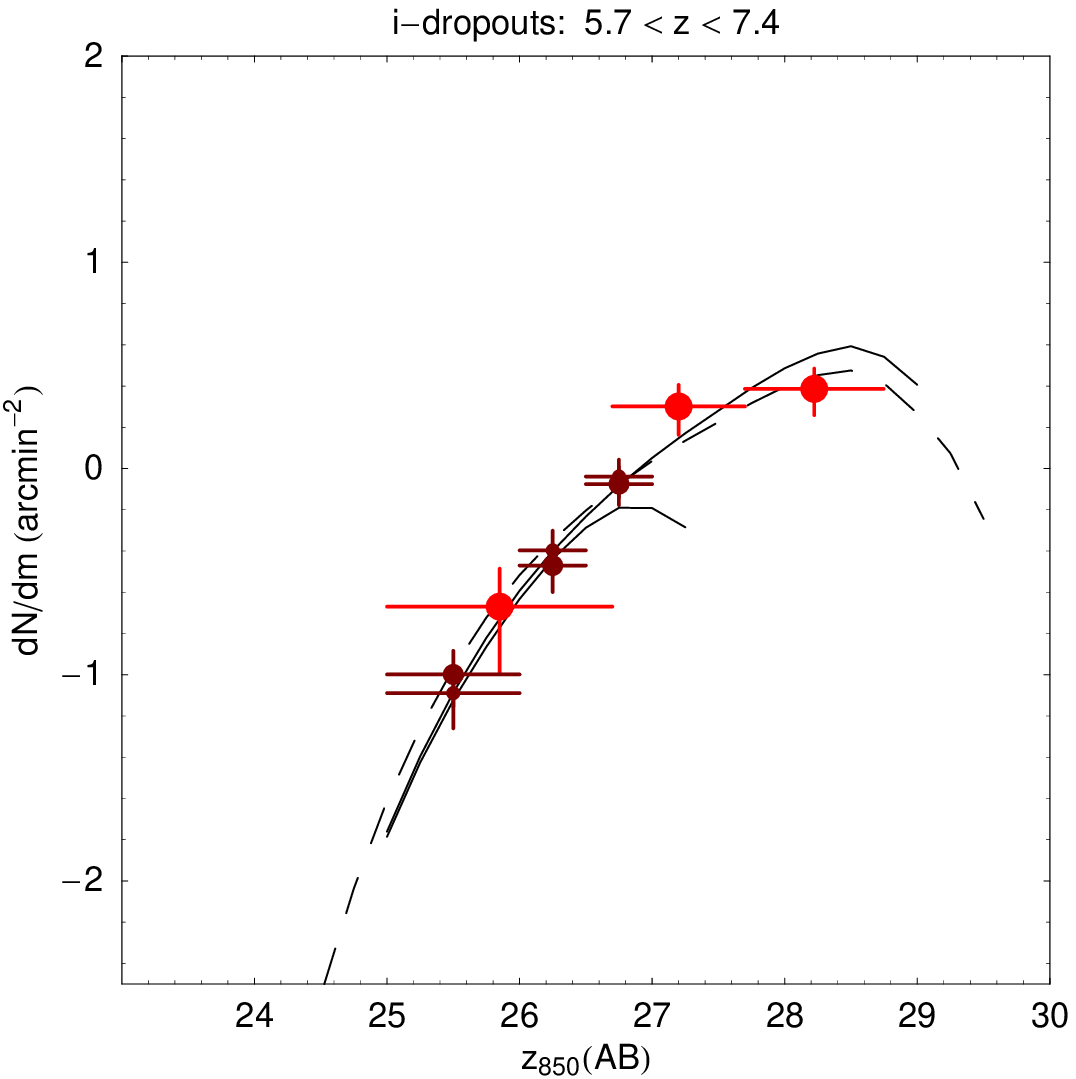} \hspace{0.5 cm}
\includegraphics[width=7.5 cm]{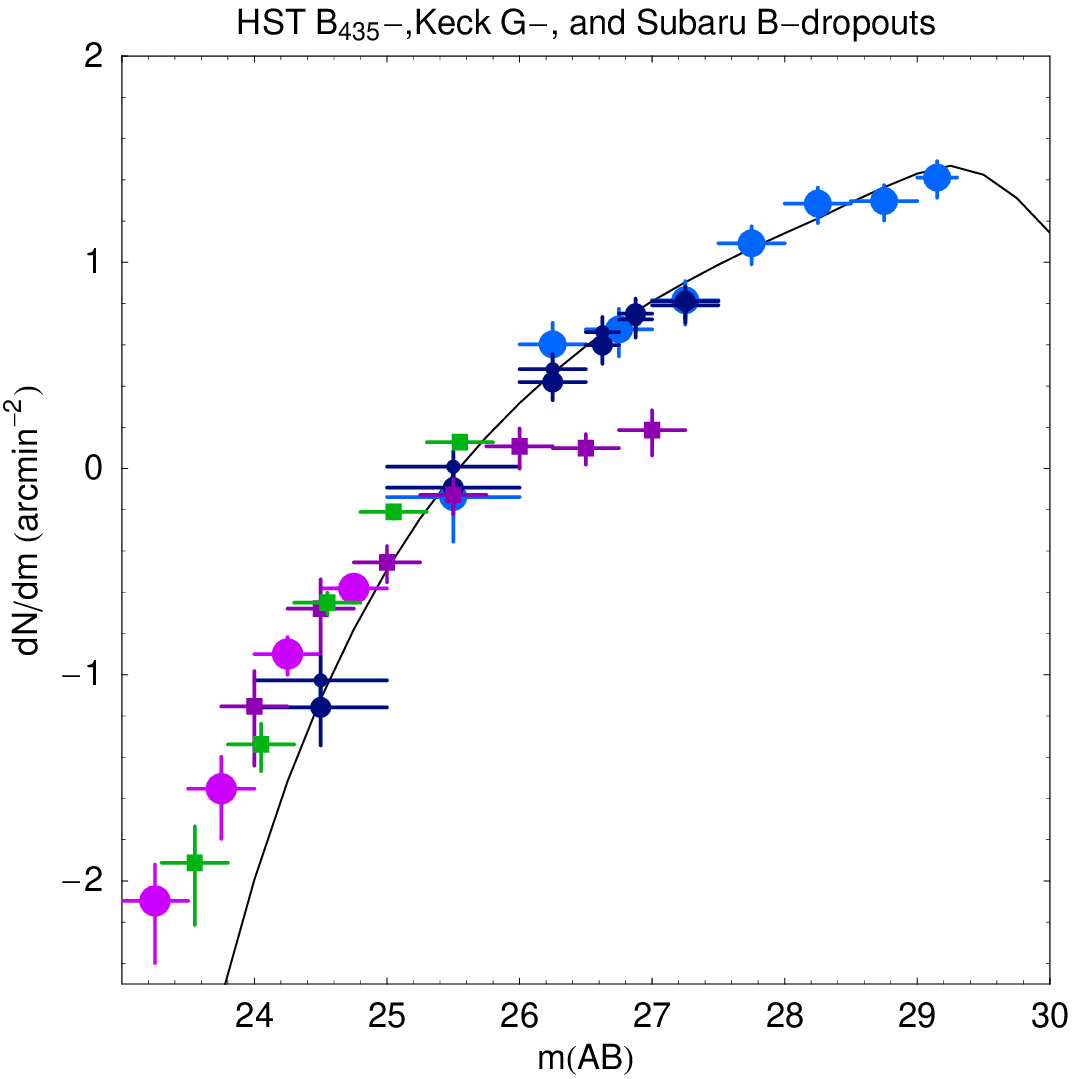}
\caption{ Number counts per square minute of arc per magnitude vs. magnitude (\Vband, \iband, and \zband ) for the dropout samples. The horizontal bars show the range of magnitudes covered for each point. The vertical bars indicate uncertainties that are proportional to the square root of the counts$^{-1}$ in that interval plus the variance of 0.17. The HUDF are the light-colored points: \Bband -dropouts (blue), \Vband -dropouts (green), and \iband -dropouts (red); GOODS are darker colored points that are large for GOODS-N or small for GOODS-S. Each sample is also compared with two Schechter function fits all with $\alpha = -1.6$. The dashed lines are Schechter functions fitted directly to the number counts with parameters:  $m_* = 26.0, \Sigma_* = 6.6$\,arcmin$^{-2}$ (\Bband -drop), $m_* = 26.1, \Sigma_* = 2.4$\,arcmin$^{-2}$ (\Vband -drop), and $m_* = 26.4, \Sigma_* = 1.7$\,arcmin$^{-2}$ (\iband -drop). The solid lines are in-situ Schechter functions integrated over the selection redshift intervals for 100\,Myr CSF galaxies with parameters: $M_* = -20.3, \phi_* = 0.0018$\,Mpc$^{-3}$ (\Bband -drop), $M_* = -20.4, \phi_* = 0.0011$\,Mpc$^{-3}$ (\Vband -drop), and $M_* = -20.5, \phi_* = 0.0007$\,Mpc$^{-3}$ (\iband -drop). The Schechter functions have been multiplied by the detection probabilities for the different surveys to show the anticipated dropoff in number counts; there have been no corrections for incompleteness applied to the data. The plot in the lower right shows the HST \Bband -dropouts compared to the G-dropouts from SAGDP99 (magenta circles) and Sawicki \& Thompson (2006) (magenta squares) and B-band dropouts from Ouchi et al. (2004) (green squares) together with the fixed-Schechter fit from the upper left plot. There are no completeness corrections applied to any points.}
\end{figure*}

The solid lines plotted in the figures are actually $p_{sb}(m_b) f(m_b)$, where $p_{sb}(m)$ is the probability of detecting an object from sample, $s$, in band, $b$, as shown in Figure 13. It is important to note is that the probability of detection depends on the object size. Our approach assumes that the distribution of object sizes is approximately log-normal, as shown in the previous sections, and to use the same size distribution (i.e. log-normal parameters) for all three samples. This assumption only affects the $\chi^2$ contribution for the faintest magnitude bins in the GOODS and HUDF samples, and the subsequent analyses use data far enough away from the detection limits to ensure that $p_{sb}(m_b)$ is greater than 0.8.

Table~14 lists the parameters for these various fits along with the formal normalized $\chi^2$ values. The fits are remarkably good considering the simplicity of the assumptions and the lack of any attempt to correct for observational biases such as variation of spectra within the samples, extinction by dust, and selection bias in the presence of noise. This latter bias, usually modeled with Monte Carlo calculations, could be considerable at the faint end of the samples.

To test for density evolution among different redshift samples, each luminosity function was fitted assuming a single Schechter function according to the prescription given above for integrating over the redshift interval and allowing only $\phi_*$ to vary. The results of these fits are also given in Table~14 in the last three columns. It was not possible to get acceptable fits by fixing $\phi_*$ and allowing $M_*$ to vary.

The conclusions of these fitting exercises can be summarized as follows:
\begin{itemize}
\item{The apparent luminosity functions may be fitted adequately by assuming simple Schechter function distributions of either the apparent magnitudes or intrinsic source luminosities.}
\item{The best fit Schechter functions have characteristic magnitudes at $\sim 1400\,$\AA\ (rest frame) of -20.3 to -20.7. The local value B-band value is -20.2 (Schechter 1976.)} 
\item{The characteristic volume densities, $\phi_*$, required to fit the apparent distributions vary between 0.002 Mpc$^{-3}$ at $z\sim 4$ to 0.0005 Mpc$^{-3}$ at $z\sim 6$. The local B-band value is 0.016 Mpc$^{-3}$.}
\item{It is possible to fit the all these luminosity functions adequately assuming no change in characteristic magnitude, $M_*$, but requiring a change in the characteristic density, $\phi_*$, dropping by a factor of three from $z \sim 4$ to $z \sim 6$. }
\item{It not possible to fit these distributions with $\phi_* \sim {\rm constant}$ and variable $M_*$.}
\item{The data are consistent with a faint end slope $\alpha = -1.6$ and do not require variations of $\alpha$ to fit the distributions.}
\item{There are strong degeneracies between $M_*$, $\phi_*$, and $\alpha$ that allow different combinations to give acceptable fits, allowing for variations in luminosity and density among samples, depending on the choice of values within these degeneracies.}
\end{itemize}

\section{Discussion}
One important goal of the HUDF was to search for galaxies at redshifts greater than 5 to assess how they have evolved in the first few billion years after the Big Bang. The increased sensitivity of these observations makes it possible to detect galaxies out to redshifts of order 7 and to compare samples at different redshifts looking for changes in characteristic luminosity, co-moving density, and morphology of typical objects. The great advantages of the HUDF are its ability to observe low surface brightness emission in these faint sources and to characterize the luminosity functions at magnitudes fainter than $M_*$. The primary disadvantages are the small area of the sky over which the observations can be made and the limited spectral information available for the faint sources, making it difficult to determine the exact redshifts of the objects and to eliminate many of the subtle biases that are unavoidable when using coarse filtering techniques to cull enormous samples.

The public release of the HUDF data stimulated considerable interest by many groups who used the data to study the early evolution of galaxies. Indeed, one of our goals was to stimulate this community interest. The subsequent research papers cover a wide-range of techniques to analyze the HUDF images and discover new properties of galaxy populations. We encourage the reader to study the resulting publications to appreciate the richness of the data and its interpretations, including Bunker et al. 2004 (the first published paper); Yan and Windhorst 2004; Stiavelli, Fall, and Panagia 2004a,b; Stanway et al. 2004; Bouwens et al. 2004a,b; Thompson et al. 2004, Stanway et al. 2005, and Straughn et al. 2006. Bouwens et al. (2005, hereafter BIBF05) summarizes most of the results of these early works and presents the most comprehensive analysis to date of \iband -dropouts in all deep ACS data from the HUDF and GOODS surveys as well as an extensive discussion of the various potential selection biases, ways to correct these, and results for large samples of sources from these data. 

Our purpose is to complement these approaches by adopting the simplest possible assumptions about the data and source samples to see if we can reach general conclusions that can be bounded by likely variations in the sample biases. In keeping with the original goals of the HUDF project, we confine ourselves only to the most salient results implied by the observations. We restrict the analyses to samples taken from the HUDF and GOODS with identical selection criteria and presumably similar observational biases.

\subsection{Evolution of galaxy luminosities}
It is of great interest to characterize the distribution of galaxy numbers, sizes, and luminosities at different epochs in the universe to understand how the present day distribution evolved from the past. One challenge to this characterization is source identification itself, because the structures seen in observations of the early universe are complex and do not necessarily correspond to the structures we see today. The choices about how sources are identified and their subsequent measurements can have profound effects on global quantities such as the characteristic luminosity, typical size, and number density of individual objects. It is therefore important to separate as much as possible observables derived from the data and model-dependent assumptions about the early universe used to analyze these observables to assess the relative impacts of the assumptions on any interpretation of how the universe evolves.

Source samples derived from space-based and ground-based observations differ in their selection biases and can be compared only with attention to these differences. It has become common practice to correct for observational biases by deriving an ``effective volume,'' $V_{\rm eff}$, for each sample by simulating different selection effects in a computer and doing Monte Carlo calculations of their average effect on model populations with different luminosities, shape, and sizes at different redshifts then using $V_{\rm eff}$ to correct the observed distributions of sources to characterize the sample through luminosity functions (SAGDP99; Ouchi et al. 2004, hereafter O04;  BIBF05; Sawicki \& Thompson 2006) This method necessarily includes assumptions about the underlying populations of galaxies that are then difficult to separate from the observations in the final results. It is especially problematic when comparing samples derived using two different observational techniques, such as ground-based telescopes limited by seeing and sky brightness with space-based telescopes, because the biases in each method are quite different and may not be equally well corrected by the effective volumes.

In this paper, we present number counts per solid angle, $\Sigma(m)$, as a function of apparent magnitude as a purely observational quantity derived from the data (Section 4.7.) The model-dependent assumptions needed to fit these functions are embedded in the fitting routines explicitly. The biases introduced by different spectral distributions of the underlying populations can be calculated for the redshift ranges allowed by the color-selection criteria. Although incomplete, when applied consistently to data sets gathered using the same method, it should be a good way to compare samples at different redshifts without a priori knowledge of the underlying source population to discover significant evolution of the source populations.

This approach suggests that the number counts at redshifts between 4 and 7 may be characterized by Schechter functions as seen in Figure 15. The HUDF is sufficiently deep to detect galaxies somewhat fainter than the characteristic magnitude, $M_*$, at all redshifts $\le 7$ , a key goal of the observations. It is therefore an excellent complement to shallower surveys such as GOODS that sample the bright end of the luminosity distribution. Furthermore, it is possible to fit the distributions well by adopting a faint end slope, $\alpha \approx -1.6$, and a value for the characteristic absolute magnitude, $M_* \approx -20.3$, similar to the local value, although the wavelengths are quite different (Johnson B-magnitude for the local value, $M_{AB} \sim 1400\,$\AA\ for the analysis here.)

The two salient problems in comparing the space-based to the ground-based data sets are that they sample different ends of the luminosity function, and the comparisons are normally made after using different model assumptions to compute the effective volumes to correct observed surface densities to in situ volume densities. There are potential biases created by different model assumptions in $V_{\rm eff}$. SAGDP99 pointed out that small differences in the color selection criteria used for dropout samples can create large differences in the resulting absolute luminosity functions, although the differences are minimized when making relative comparisons between samples observed and analyzed the same way. The \Vband -dropouts are especially sensitive to the color of the underlying galaxies owing to the shallow slopes of the main selection function, $ V_{606} Ð- i_{775}  > 1.47 + 0.89(i_{775} Ð- z_{850})$, near the selection boundary; see Figure 11c.  Thus, when comparing different samples, it is important to use the same assumptions for the analysis of $V_{\rm eff}$.

Figure 15 includes a comparison between the space-based and ground-based samples at $z \sim 4$ from four samples: SAGDP99, O04, Sawicki and Thompson (2006), and the samples in this paper. The observed surface density data from two of the ground-based samples match the space-based samples reasonably well in the regions of overlap, although the faint end of the Sawicki and Thompson data are significantly under the space-based and O04 counts. It is evident from this figure that the shape of the apparent luminosity functions between the ground-based and space-based samples look different without normalization to the sample volumes.

We derived rudimentary sample volumes for three of the samples plotted in Figure 15 adopting the entire co-moving volumes within the color-selection redshifts. The absolute magnitudes are calculated from the apparent magnitudes using distance moduli and k-corrections from a Bruzual and Charlot continuous star formation model over 100\,Myr with 0.4 solar metallicity and no extinction convolved with the different filter response functions. For the G-dropouts of SAGDP99, we used the spectroscopic redshift distribution for the volume and the ratios of $V_{\rm eff}$ given in their Table 4 to correct for incompleteness in the two faintest magnitude bins of their study.  For the O04 data, only color-selection sample volumes were used. The same technique was applied to the \Bband -dropouts presented in this paper using the same assumptions and limiting the magnitude range to make completeness corrections less than 20\% (cf. Table 9 and Figure 13.) We could not correct the Sawicki \& Thompson (2006) data without knowledge of their completeness corrections, although it is evident from inspection of their figures that the faint end luminosity functions will not match the space-based counts unless the corrections for incompleteness are factors of 3 or more.

Figure 16 shows the resulting \Bband- , B-, and G-dropout luminosity functions along with Schechter function fits to the data. The data are easy to fit with a single Schechter function with a characteristic luminosity and faint end slope similar to those derived for various samples using both ground and space-based observations. Figure 8 of SAGDP99 gives best fit parameters to the $z\sim4$ sample of $m_* = 24.97$, $\phi_* = 0.013$\,Mpc$^{-3}$ for an Einstein-de Sitter universe with $h=1$. For a concordance model and k-corrections assumed in this paper, we calculate the equivalent $M_* =  -20.9$ and $\phi_* = 0.0011$\,Mpc$^{-3}$. The values derived in Figure 16, $M_* = -20.7$ and $\phi_* = 0.0013$\,Mpc$^{-3}$, are entirely consistent with the SAGDP99 values, considering the degeneracy between $M_*$ and $\phi_*$ in the fits and different methods of calculating volumes between the two papers. The absolute luminosities depend on the k-corrections applied to the different filter sets used for the SAGDP99, O04, and Hubble observations. 

%Figure 16: Volume luminosity functions
\begin{figure*}
\includegraphics[width=7.5 cm]{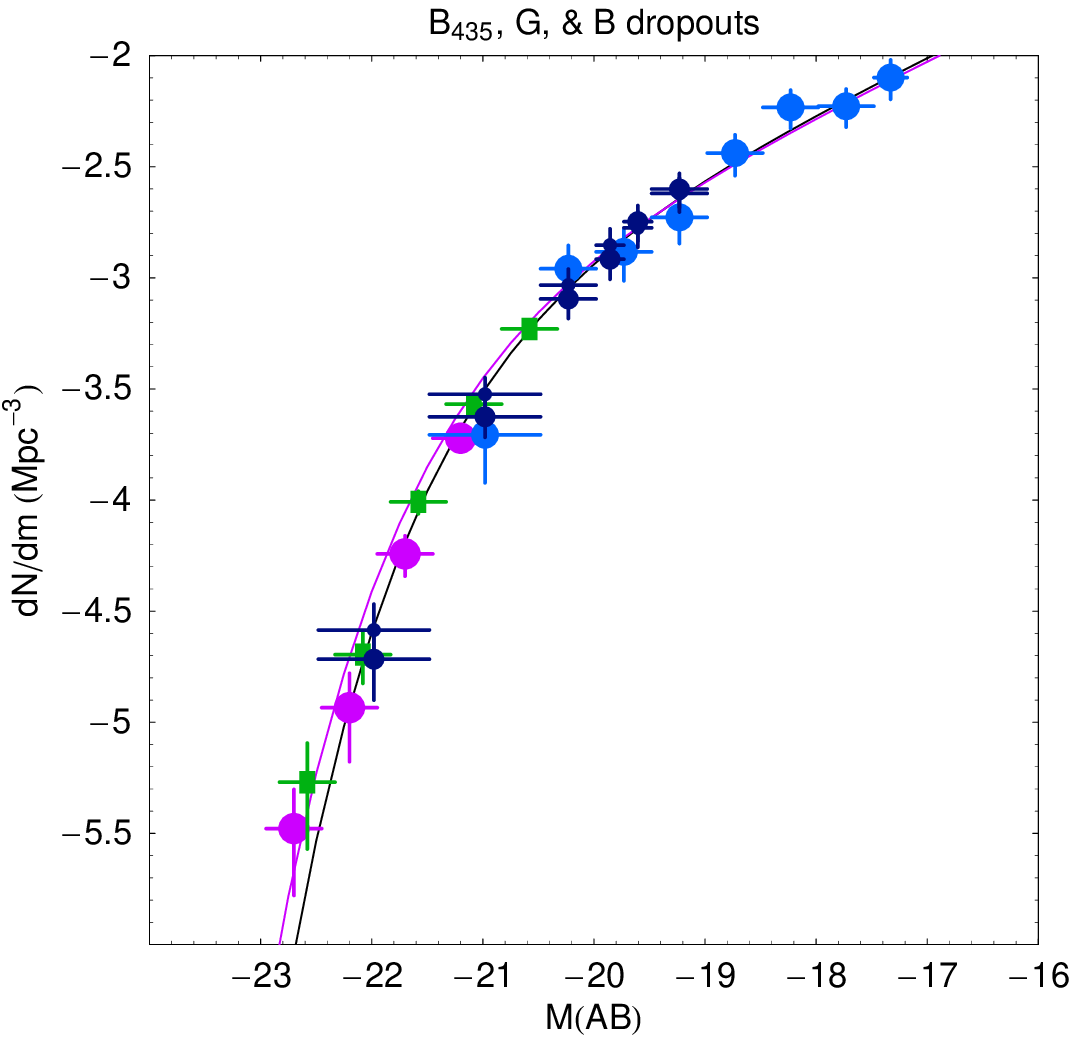} \hspace{1. cm}
\includegraphics[width=7.5 cm]{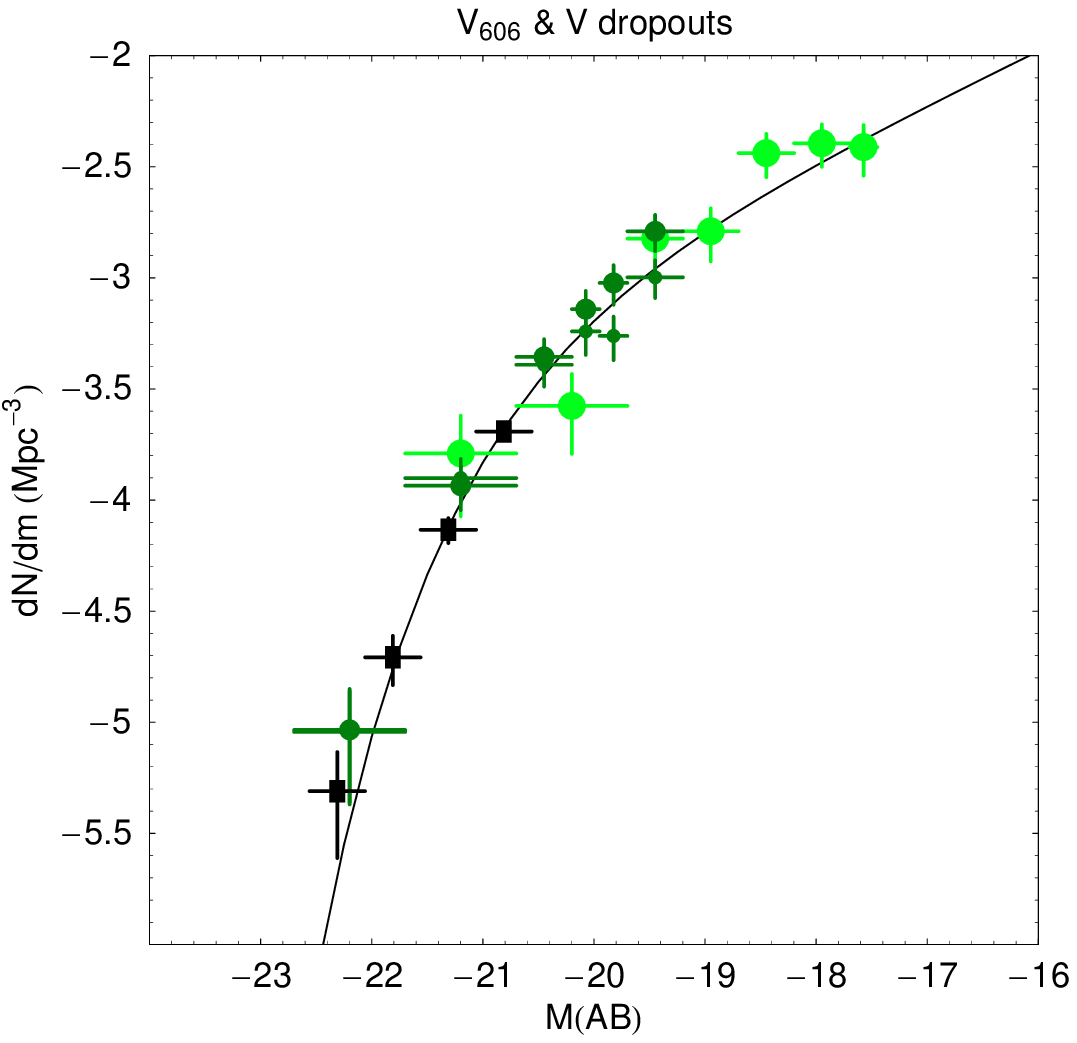} \\ \\
\includegraphics[width=7.5 cm]{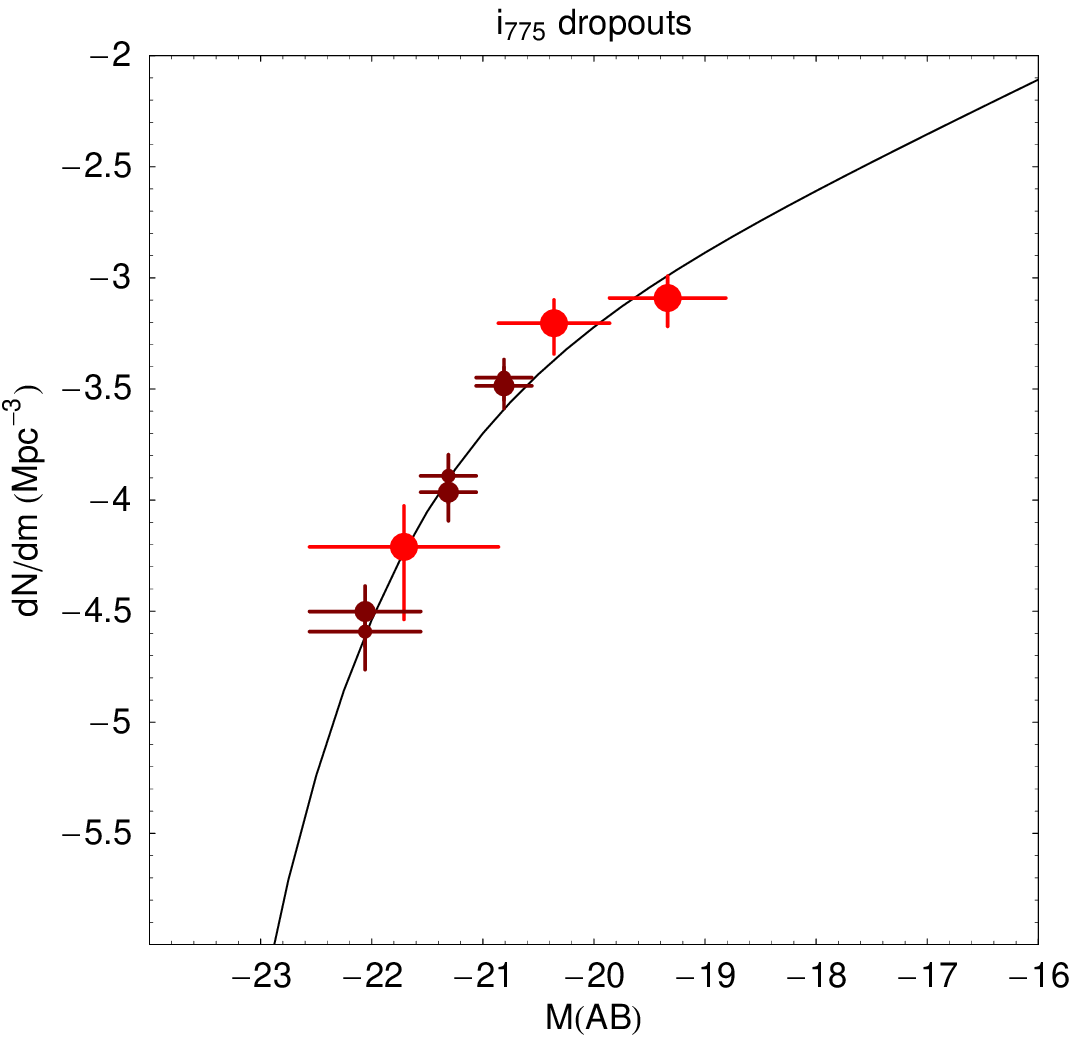}
\caption{The in situ luminosity functions of the dropout sources from the HUDF, GOODS, SAGDP99, and O04  surveys corrected using the model-dependent $V_{s}$ and distance moduli as discussed in Section 5. Color and shape codes are the same as in Fig. 15.  Upper left: \Bband-dropouts. The black line is the best-fit Schechter function for this combined sample: $M_* = -20.7, \phi_* = 0.0013$\,Mpc$^{-3}$ ($\chi^2 = 0.6$. The magenta line is our translation of the original fit from Figure 8 of SAGDP99 using $M_* = -20.9$ and $\phi_* = 0.0011$\,Mpc$^{-3}$ for the concordance cosmology and k-corrections for a 100\,Myr CSF Bruzual spectrum using our model calculations. We were unable to include the Sawicki \& Thompson (2006) data in this plot without knowledge of their completeness corrections. Upper right: \Vband -dropouts from the HST samples together with \Vband-dropouts from O04 (black squares).  Fit parameters are: $M_* -20.55, \phi_* = 0.0009$\,Mpc$^{-3}$ ($\chi^2 = 1.1$.) Lower left: HST \iband -dropouts. Fit parameters are: $M_* = -21.1, \phi_* = 0.0005$\,Mpc$^{-3}$ ($\chi^2 = 0.6$.)  }
\end{figure*}

This method allows us to tie the data sets together under a single set of assumptions. However, the absolute values of $M_*$, $\phi_*$, and $\alpha$ depend both on the model assumptions and the completeness corrections, which we have estimated using relatively simple assumptions. It is possible to fit all the distributions in this paper well by assuming constant $M_*$ and varying $\phi_*$. It is not possible to fit all the distributions in this paper well by assuming a constant density of galaxies and varying $M_*$ and $\alpha$. In all cases, the average density of galaxies at redshifts above 6 appears smaller than at redshifts $\le 4$ by a factor of order three for fixed $M_*$. 

Our treatment of galaxy evolution using samples all derived with similar techniques and assumptions calls into question one of the strongest conclusions of BIBF05, who state ``Scenarios, such as density evolution ($\phi^*$), which do not include this evolution in $M^*$ or $\alpha$ are ruled out at 99.9999\% confidence -- demonstrating quite convincingly that galaxies at $z \sim 6$ are lower in luminosity than galaxies at $z \sim 3$.'' A simple way to test this statement using the samples presented here is to examine the ratio of luminosity density above a certain magnitude limit to that below the limit and see if this ratio changes with redshift. Table~16 presents the results of this test. Each dropout sample was divided into a bright and faint part with the boundary set as the apparent magnitude of $M_* = -20.8$ galaxy at the mean redshift of the sample using k-corrections appropriate to a Bruzual \& Charlot 100\,Myr continuous star formation model. The faint end limit was set 2 magnitudes above the boundary to ensure the inclusion of similar bright/faint magnitude ranges among the three samples. Table~16 gives the resulting surface brightness and in-situ energy densities as well as faint/bright ratios. It demonstrates that at least for these samples there is no evidence for an evolution of the faint/bright ratio within the uncertainties.

There are certainly biases in our technique that could affect this conclusion. If there is less absorption by the intergalactic medium than assumed by the Madau (1995) model for low redshifts (see, e.g., Fontanot et al. 2006), we could miss bright galaxies that have enough residual \Bband\ flux to be excluded as \Vband -dropouts, say, even though the colors might otherwise qualify: there is an extra filter in equation (15) to deal with this problem but it was determined by trial and error and may not be robust for large samples. BIBF05 discovered many more faint galaxies in their \iband -dropout sample that will boost the faint/bright ratio in Table~16, an effect that would be offset by including the additional faint galaxies in the \Bband - and \Vband -samples by extending the faint magnitude limit below $M = -18.8$. 

The biases should apply in a similar fashion to each of the samples presented here, however. The ground-based data have a different set of biases, and it appears that even when great care is taken to model instrumental and observational effects it is difficult to calculate large corrections accurately. The faint end luminosity functions of Sawicki \& Thompson (2006) in  Figure 15 clearly show incompleteness in the last few points, even though their paper uses these points to establish a flatter faint end slope than seen in any of the space-based samples or in the Subaru data. Examination of Figure 7 shows how much crowding exists on scales of $\sim 1$\arcsec, typical of ground-based seeing, and how it would be difficult to correct for loss of faint galaxies owing to blending with other objects.

It is encouraging that Figure 16 shows a consistency between the \Bband-, B- and G-dropouts when analyzed with similar assumptions. It remains to be seen if this consistency will hold up under a more detailed analysis that assesses the impact of observational noise on the selection of these separate samples. Observations at longer wavelengths for all these samples might also be used to constrain the spectral variations---to assess the impact of dust, for  example---and better refine the model assumptions using the data themselves. Spectra of the high redshift samples derived here will be essential to remove some of the largest uncertainties.

The actual luminosity density decreases modestly from low to high redshifts, as shown in Figure 17 which is an update of Figure 1 from Giavalisco et al. (2004.) Giavalisco et al. (2004) were the first to note the difficulty of assessing effective volumes especially for the \iband -dropout sample, where the sample redshift range is determined more by limiting sensitivity than pure color selection---note the comparison between the open and filled circles in Figure 17 that use the same data but different assumptions about the effective volumes of the samples. The HUDF greatly ameliorates this problem with additional sensitivity that allows detection of $L_*$ galaxies to redshifts very close to that for which the Ly$_\alpha$ edge moves out of the \zband\ filter. Adding the faint \iband -dropout sources discovered by BIBF05 below our cutoff will certainly boost the high redshift point in this figure, suggesting little if any evolution in the luminosity density between redshifts 6 and 3.

%Figure 17: Energy density vs. Redshift
\begin{figure}[ht]
\includegraphics[width=7.5 cm]{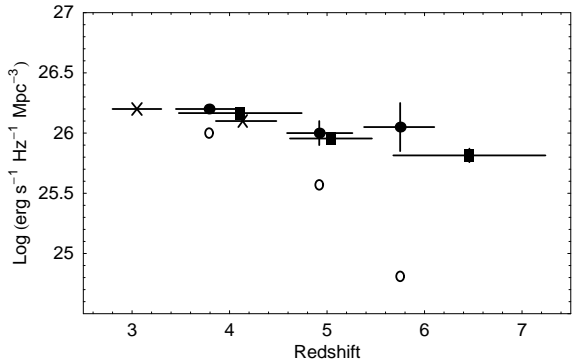}
\caption{A comparison of the total energy densities from different surveys after Figure 1 of Giavalisco et al. (2004): crosses: SAGDP99, filled and open circles: Giavalisco et al. for two different $V_{\rm eff}$, rectangles: total light from HUDF sources using k-corrections and mean dropout distances for HUDF only.}
\end{figure}

The important point is that {\it some derived properties of the galaxy population such as the luminosity distributions often depend more strongly on the assumptions used to correct biases than they do on the data themselves}. In many cases, the correction factors are large, as noted by SAGDP99, meaning the absolute densities and absolute magnitudes can depend almost entirely on the corrections. Without separate consistency checks to establish ground truth---high resolution spectroscopy of every source in a field, say---it is dangerous to take the model results too seriously. This problem is exacerbated in differential comparisons between derived quantities from data sets taken with different instruments and using assumptions about the underlying populations for their analysis.

\subsection{Cosmological dimming of galaxies}
For the individual objects seen in the images, a comparison of sources seen in both the HUDF and GOODS suggests that the Hubble observations adequately measure the total luminosity of galaxies brighter than the completeness limits. Figure 5 shows that there are few if any cases of low-surface brightness emission at the periphery of galaxies picked up in the deeper HUDF survey that are not seen in the shallower GOODS images. The populations at high redshift appear to be much more compact than local galaxies, making them easy to see with the high resolution of the Hubble images. The colors of the bright and faint galaxies in each dropout sample are similar as is the emission in the \iband\ and \zband\ for the \Bband -dropouts (cf. Figure 18), suggesting that obscuration by dust is not large for the galaxies seen here. Figure~18 also indicates that the long wavelength spectra of galaxies at redshifts near 4 derived using the dropout technique are flat, and there is not a strong indication of changes in reddening by dust at different redshifts in the two lowest redshift populations.

The apparent surface brightness of radiation from the entire population of high-redshift galaxies is increased by the addition of the HUDF compared to the GOODS samples, typically by a factor of about two. This higher surface brightness results from the contribution to the faint end of the luminosity functions that dominate the total luminosity in each sample. There is no obvious turnover in these functions (Figure 15) before reaching the limiting magnitudes of the HUDF (see also the discussion by Yan and Windhorst 2004.) Certainly, the observed surface brightness is a lower limit that would be improved only by much deeper observations than even the HUDF.

%Figure 18: Average spectra of dropout sources
\begin{figure*}
\includegraphics[width=8. cm]{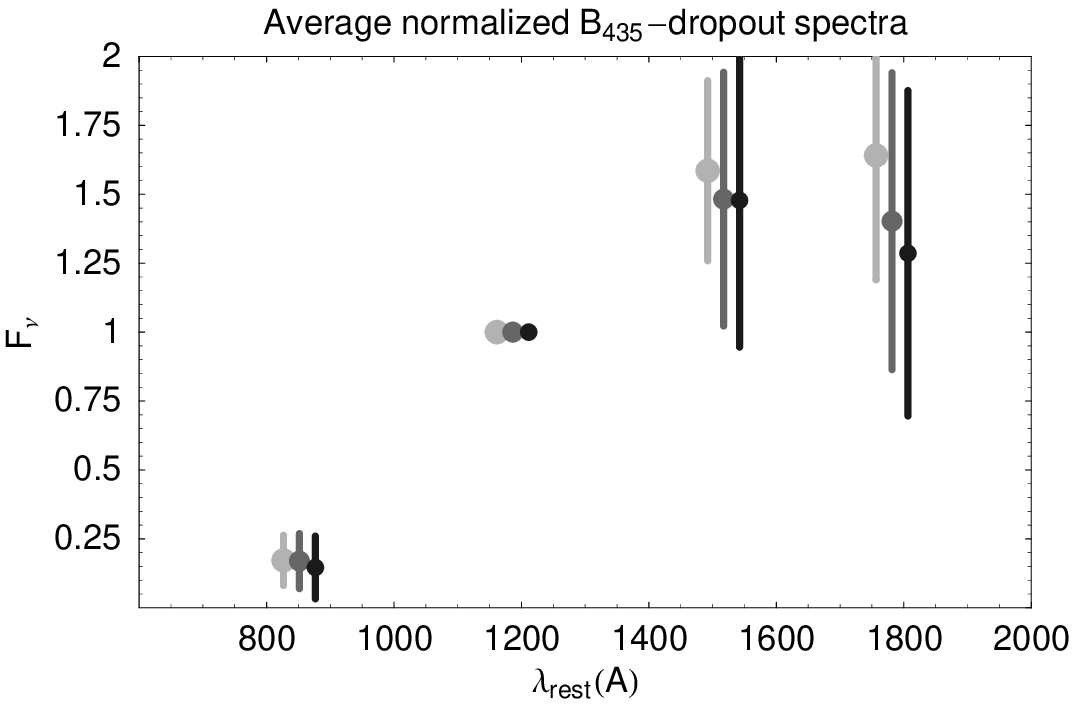} \hspace{0.4 cm} 
\includegraphics[width=8. cm]{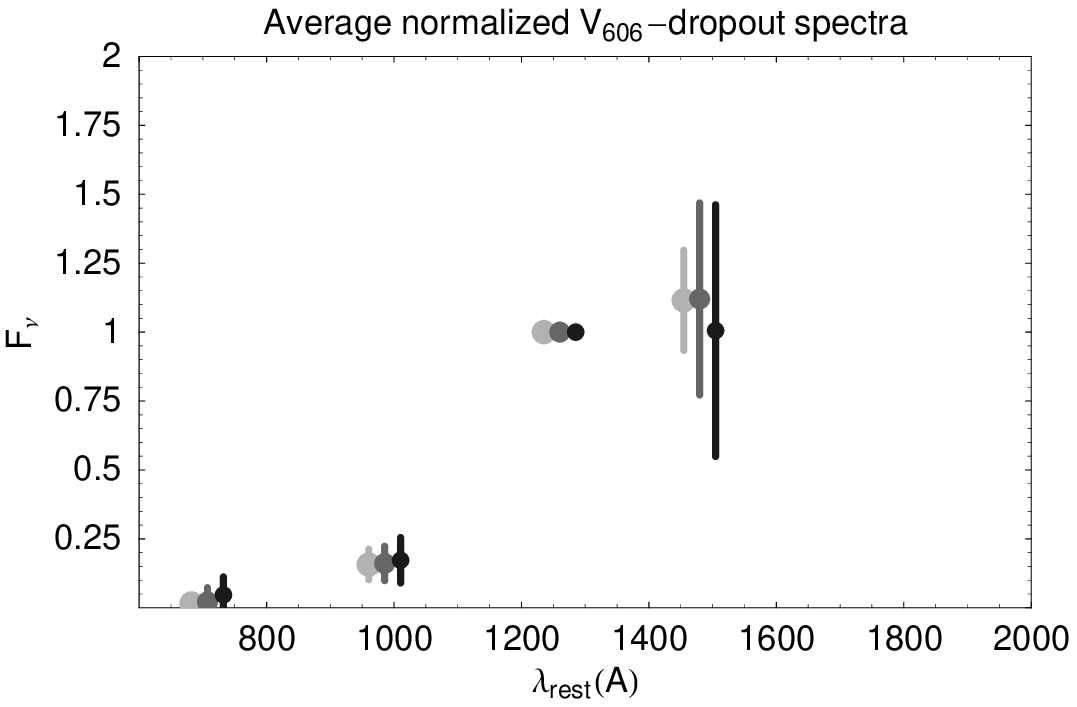}
\caption{The average spectrum for the B- and \Vband-dropout samples vs. rest-frame wavelength. All sources are assumed to be at the mean redshift in each sample: 4.1 (B-drops) and 5.0 (V-drops). The flux densities through each filter have been averaged after normalizing to the selection filter (\Vband\ or \iband ) flux density. The error bars are the standard deviations of the samples. In each plot, the averages take place in three bins of selection magnitude. For \Bband -dropouts, these are: ${\rm V < 26.5}$, ${\rm 26.5 \le V < 28}$, ${\rm V \ge 28}$. The ranges are the same for \Vband -dropouts except the magnitude is \iband . The three sub-samples are plotted from left to right in the figures with decreasing symbol sizes and offset -25, 0, and 25\,\AA\ from the mean rest frame wavelength.}
\end{figure*}

\section{Conclusions}

The Hubble Ultra Deep Field reached limiting magnitudes within a few tenths of the zodiacal-light limited values and approximately 1.5 magnitudes deeper than the next deepest observations. It uncovered more than 10,000 new galaxies, the vast majority of which are below the sensitivity of other surveys. Many of these sources have redshifts greater than three corresponding to the epoch less than two billion years after the Big Bang.

Three samples of galaxies in different redshift intervals were identified in the HUDF from objects with missing flux in the shortest wavelength bands, those that ``dropped out'' of the \Bband, \Vband, or \iband\ filters, corresponding approximately to redshift ranges $3.5 < z < 4.7$, $4.6 < z < 5.5$, and $5.7 < z < 7$, respectively. The same technique was applied to the GOODS survey to get samples of similar objects at the bright end of the luminosity function but over a larger region of the sky. Comparison of these samples showed that:
\begin{itemize}
\item{As found in previous observations with the Hubble Space Telescope, galaxies at these high redshifts are smaller and less symmetric in shape than galaxies at lower redshifts. These results confirm the conclusion of the first Hubble Deep Field observations that typical galaxies in the early universe look markedly different than galaxies today, showing that galaxies evolved rapidly in the first few billion years after the Big Bang.} 
\item{The faint sources detected by the HUDF are smaller on average than the brighter objects seen in shallower surveys such as GOODS or the HDF, with typical sizes (radius enclosing 50\% of the flux) about 0\arcsec.15. There is not strong evidence for evolution of source size with redshift among these samples.}
\item{The HUDF samples the luminosity function approximately 2 magnitudes below the characteristic magnitude at redshifts as large as 6, thus fulfilling a key goal of the observations.}
\item{It is possible to characterize the apparent luminosity functions of all samples with Schechter functions assuming only a variation in source density ($\phi_*$) and not luminosity ($M_*$) or faint-end slope ($\alpha$). It is not possible to fit the luminosity functions assuming evolution of the characteristic luminosity alone. The uncertainties in deriving and fitting the luminosity functions are large enough to accommodate luminosity evolution for comparison with theory, for example, but it is not required by the data.}
\item{The co-moving volume density of galaxies needed to fit these high-redshift samples is at least 10 times smaller than the local value of $0.016\,{\rm Mpc}^{-3}$ using the simplest assumptions about source selection and the entire co-moving volumes allowed by color selection. Use of an effective volume approach may raise these densities, and it is quite likely that the Lyman-break dropout technique picks up only a fraction of the galaxies at high redshift, meaning the absolute value of $\phi_*$ derived here is a lower limit.}
\item{The shape of the bright-end luminosity function at $z\sim 4$ estimated from \Bband -dropouts closely matches that of ground-based samples using a similar dropout technique, when the samples are treated with similar (albeit uncertain) assumptions.}
\item{Although the luminosity density of galaxies at wavelengths $\sim 1400\,$\AA\ decreases modestly from redshifts of a few out to redshifts greater than 6, vigorous star formation was already underway when the universe was less then 1 billion years old.}
\end{itemize}

Our purpose in carrying out the HUDF observations was to provide a deep dataset as a legacy to science for study of the early universe. The many publications arising from the original data release testify to the utility of the HUDF for this purpose. The images provide a rich hunting ground for those who wish to understand the nature of the universe at the time when stars and galaxies first came into existence after the Big Bang.

\begin{acknowledgments}
These observations would not have been possible without the help of many people at the Space Telescope Science Institute and in the community. We are especially indebted to Stefano Casertano, Ian Jordan, Max Mutchler, and Bill Workman. We profited from discussions with Mark Dickinson, Richard Ellis, Michael Fall, Andy Fruchter, Mauro Giavalisco, Hans-Walter Rix, Dan Stark, Charles Steidel, Rodger Thompson, and especially Robert Williams. For advice on the choice of observational parameters as well as occasional discussions about the scientific directions, we thank the group of distinguished scientists who served as the scientific advisory committee to this project: Ron Ekers, Catherine Cesarsky, Guenther Hasinger, Garth Illingworth, Jeremy Mould, Matt Mountain, Anneila Sargent, Tom Soifer, Harvey Tananbaum, Robert Williams, and Rogier Windhorst. We also acknowledge the many good suggestions that came out of the ACS High Latitude Survey Workshop in March 2001 and the Hubble Treasury workshop in Baltimore in November, 2001. SB is grateful to the Max-Planck-Institut f\"ur Astronomie for providing research facilities and hosting several visits to enable much of the analysis presented here and to a number of important suggestions from the MPIA staff for that analysis. We are also grateful to the California Institute for Technology for hosting one of us (SB) and generously sharing information about the dropout techniques used there. The referee was most helpful in pointing out weaknesses in the arguments and stimulating analysis of the data. Of course, none of these observations could have taken place without the dedication of the hundreds of engineers, managers, and scientists at NASA and its contractors who gave us the Hubble Space Telescope as a lasting legacy to the world.
\end{acknowledgments}

\clearpage

%--------------------------------------------------   Tables ------------------------------------------------------------------------
%Table 1: Characteristic of the HUDF Field
\begin{deluxetable}{ll}
\tabletypesize{\scriptsize}
\tablecaption{HUDF Field Characteristics}
\tablewidth{0pt}
\startdata
Field Center: &   RA(J2000) = 3\,32\,39.0 \\
                         &  Dec(J2000) = -27\,47\,29.1\\  
                         & $l = 223.6^\circ$, $b = -54.4^\circ$ \\
                         & Ecliptic latitude: $-45.2^\circ$ \\
Size:                & 200\arcsec$\times$200\arcsec \\
Area:               & 11.0 arcmin$^2$ \\
Position Angle:  &   $\sim 42^\circ$ \\
E(B-V):  & 0.008 \\
HI column density:     & $7.9\times 10^{19}$\,cm$^{-2}$ \\
Radio sources: & two \@ 300 mJy (365 MHz) at $\sim 10$\arcmin \\
                            & one \@ 1 Jy (408 MHz) at 51\arcmin \\
\hline
\enddata
\end{deluxetable}

%Table 2: predicted number of objects vs. redshift  
\begin{deluxetable}{crrrr}
\tabletypesize{\scriptsize}
\tablecaption{Potential number of sources in HUDF}
\tablewidth{0pt}
\tablehead{
\colhead{Redshift} & \colhead{Total $dN \over dz$} & \colhead{\Vband\ $dN \over dz$} & \colhead{\iband\ $dN \over dz$} & \colhead{\zband\ $dN \over dz$}
}
\startdata
1 & 26,800 & 23,600 & 23,900 & 13,900 \\
2 & 17,000 & 17,500 & 15,800 & 7,700 \\
3 & 10,700 & 11,500 & 11,000 & 5,300 \\
4 & 7,000 & 5,900 & 7,700 & 3,700 \\
5 & 4,900 & 930 & 4,800 & 2,600 \\
6 & 3,600 & 0 & 490 & 1,600 \\
7 & 2,700 & 0 & 0 & 120 \\
\hline
\enddata
\\
Differential numbers are rounded for clarity; m$_{\rm AB}$ limits: 29, 29, 29, 28.\\
Columns 3--5 represent estimates using an unreddened model spectrum \\
starburst galaxy with intergalactic hydrogen absorption; \\ 
$\phi_* = 0.016$\,Mpc$^{-3}$, $M_* = -20.2$, $\alpha = -1.6$. \\
\end{deluxetable}
  
%Table 3: HUDF Observations
\begin{deluxetable}{lrrrrr}
\tabletypesize{\scriptsize}
\tablecaption{HUDF Orbits}
\tablewidth{0pt}
\tablehead{
\colhead{Date range} & \colhead{PA} & \colhead{\Bband} & \colhead{\Vband} & \colhead{\iband} & \colhead{\zband} 
}
\startdata
2003.09.24 Ð 2003.10.28 & 310 & 6 & 8 & 18 & 18 \\
2003.09.24 Ð 2003.10.28 & 314 & 22 & 20 & 48 & 56 \\
2003.12.04 Ð 2004.01.15 & 40 & 6 & 8 & 18 & 20 \\
2003.12.04 Ð 2004.01.15 & 44 & 22 & 20 & 50 & 50 \\
\hline
Total orbits & &  56 & 56 & 144 & 144 \\
Total ksec & & 134.9 & 135.3 & 347.1& 346.6 \\
\enddata
\end{deluxetable}
  
%Table 4: Dither pattern
\begin{deluxetable}{ccccc}
\tabletypesize{\scriptsize}
\tablecaption{Dither pattern}
\tablewidth{0pt}
\tablehead{
\colhead{x\arcsec } & \colhead{y\arcsec } & \colhead{x (pix)} & \colhead{y (pix)} & \colhead{Exp \#} 
}
\startdata
0.0000 & 0.00000 & 0.0 & 0.0 & 1\\
0.1488 & 0.08595 & 3.0 & 1.5 & 2\\
0.2232 & 0.24075 & 4.5 & 4.5 & 3\\
0.0744 & 0.15480 & 1.5 & 3.0 & 4\\
\enddata
\end{deluxetable}

%Table 5: Sensitivities achieved
\begin{deluxetable}{lccccccc}
\tabletypesize{\scriptsize}
\tablecaption{HUDF Sensitivities}
\tablewidth{0pt}
\tablehead{
\colhead{Filter} & \colhead{Exposure} & \colhead{ZP} & \colhead{$<$Sky$>$} &  \colhead{Predict} & \colhead{10$\sigma$/0\arcsec.2} &  \colhead{50\% recovery} \\
& \colhead{ksec} & & \colhead{s$^{-1}$ pix$^{-1}$} & \colhead{m$_{AB}$} & \colhead{m$_{AB}$} &  \colhead{m$_{AB}$}
}
\startdata
F435W & 134.9 & 25.67 & 0.0284 & 29.09 & 29.08 &    \\
F606W & 135.3 & 26.49 & 0.1358 & 29.29 & 29.45 &   \\
F775W & 347.1 & 25.65 & 0.0780 & 29.21 & 29.35 &   \\
F850LP & 346.6 & 24.86 & 0.0415 & 28.67 & 28.70 &  29.4 \\
\enddata
\\ 
The ZP \& $<$Sky$>$ from the ACS manual. Read noise $\sigma_{\rm read} = 5$.\\
50\% recovery magnitude is the average disk/bulge for point sources.
\end{deluxetable}

%Table 6: B dropouts; first 36 sources each from HUDF samples
\begin{deluxetable}{rrrlrrrrrrrr}
\tabletypesize{\scriptsize}
\tablecaption{First 36 \Bband-dropout sources}
\tablewidth{0pt}
\tablehead{
\colhead{STScI ID} & \colhead{RA (J2000)} &\colhead{Dec (J2000)} & \colhead{r50(\arcsec)} & \colhead{B} & \colhead{$\sigma_B$} & \colhead{V} & \colhead{$\sigma_V$} & \colhead{i} & \colhead{$\sigma_i$} & \colhead{z} & \colhead{$\sigma_z$}
}
\startdata
28 & 53.16485 & -27.82724 & 0.089 & 30.84 & 0.54 & 29.08 & 0.09 & 29.00 & 0.07 & 29.50 & 0.20 \\ 
44 & 53.16797 & -27.82653 & 0.187 & 29.67 & 0.38 & 27.92 & 0.08 & 27.52 & 0.04 & 27.71 & 0.08 \\ 
79 & 53.16769 & -27.82542 & 0.115 & 31.14 & 0.86 & 28.93 & 0.09 & 28.67 & 0.07 & 28.69 & 0.12 \\ 
91 & 53.17054 & -27.82467 & 0.133 & 31.33 & 1.13 & 29.02 & 0.21 & 28.50 & 0.06 & 28.71 & 0.13 \\ 
102 & 53.16347 & -27.82437 & 0.116 & 30.39 & 0.33 & 29.2 & 0.11 & 29.31 & 0.09 & 29.28 & 0.16 \\ 
148 & 53.16454 & -27.82316 & 0.133 & 30.21 & 0.53 & 27.53 & 0.04 & 26.66 & 0.02 & 26.65 & 0.03 \\ 
154 & 53.16586 & -27.82344 & 0.167 & 30.10 & 0.34 & 28.91 & 0.16 & 28.85 & 0.08 & 29.60 & 0.28 \\ 
230 & 53.15400 & -27.82118 & 0.176 & 27.44 & 0.07 & 25.67 & 0.01 & 25.23 & 0.01 & 25.23 & 0.01 \\ 
236 & 53.15455 & -27.82214 & 0.077 & 31.27 & 0.74 & 29.30 & 0.13 & 29.03 & 0.07 & 29.14 & 0.14 \\ 
280 & 53.15610 & -27.82122 & 0.121 & --        & --       & 27.71 & 0.04 & 27.37 & 0.03 & 27.50 & 0.05 \\ 
291 & 53.17409 & -27.82107 & 0.107 & 30.86 & 0.72 & 28.44 & 0.07 & 28.31 & 0.05 & 28.43 & 0.10 \\ 
301 & 53.16346 & -27.82102 & 0.093 & --         & --      & 29.19 & 0.13 & 28.09 & 0.05 & 28.22 & 0.09 \\ 
307 & 53.16511 & -27.82100 & 0.113 & 32.12 & 1.74 & 29.32 & 0.13 & 29.02 & 0.08 & 29.11 & 0.15 \\ 
348 & 53.16972 & -27.82032 & 0.111 & 29.78 & 0.25 & 28.32 & 0.06 & 28.09 & 0.04 & 28.13 & 0.07 \\ 
372 & 53.17023 & -27.81995 & 0.119 & 30.57 & 0.50 & 27.91 & 0.04 & 27.87 & 0.03 & 28.12 & 0.07 \\ 
396 & 53.17545 & -27.81957 & 0.140 & 28.99 & 0.17 & 27.57 & 0.04 & 27.31 & 0.03 & 27.24 & 0.05 \\ 
401 & 53.16197 & -27.81909 & 0.245 & 29.23 & 0.39 & 26.20 & 0.02 & 25.28 & 0.01 & 25.14 & 0.01 \\ 
407 & 53.16069 & -27.81918 & 0.097 & 28.32 & 0.10 & 26.11 & 0.01 & 25.65 & 0.01 & 25.70 & 0.01 \\ 
409 & 53.17234 & -27.81933 & 0.139 & 28.31 & 0.09 & 26.99 & 0.02 & 26.83 & 0.02 & 26.95 & 0.03 \\ 
420 & 53.15847 & -27.81947 & 0.094 & 30.75 & 0.50 & 28.81 & 0.08 & 28.66 & 0.06 & 28.67 & 0.10 \\ 
421 & 53.15600 & -27.81939 & 0.180 & 31.44 & 1.26 & 29.31 & 0.16 & 28.69 & 0.08 & 29.26 & 0.22 \\ 
481 & 53.15569 & -27.81870 & 0.089 & 30.28 & 0.29 & 29.06 & 0.09 & 28.99 & 0.07 & 29.11 & 0.13 \\ 
577 & 53.16145 & -27.81744 & 0.110 & 30.73 & 0.71 & 28.05 & 0.05 & 27.09 & 0.02 & 27.05 & 0.03 \\ 
591 & 53.17494 & -27.81738 & 0.108 & --         & --      & 28.64 & 0.08 & 28.28 & 0.05 & 28.45 & 0.09 \\ 
613 & 53.16039 & -27.81736 & 0.193 & 29.62 & 0.30 & 28.04 & 0.07 & 27.78 & 0.04 & 27.78 & 0.07 \\ 
631 & 53.16702 & -27.81700 & 0.100 & 29.84 & 0.36 & 26.90 & 0.02 & 26.43 & 0.01 & 26.50 & 0.02 \\ 
651 & 53.14933 & -27.81696 & 0.084 & --         & --      & 29.11 & 0.10 & 28.35 & 0.04 & 28.49 & 0.08 \\ 
700 & 53.16573 & -27.81646 & 0.085 & 30.70 & 0.60 & 28.27 & 0.05 & 27.37 & 0.02 & 27.41 & 0.04 \\ 
741 & 53.16223 & -27.81577 & 0.209 & 27.49 & 0.07 & 25.89 & 0.01 & 25.45 & 0.01 & 25.42 & 0.01 \\ 
750 & 53.17981 & -27.81593 & 0.120 & 29.73 & 0.23 & 28.34 & 0.07 & 28.26 & 0.05 & 28.37 & 0.09 \\ 
751 & 53.16381 & -27.81591 & 0.135 & 28.59 & 0.11 & 27.48 & 0.04 & 27.29 & 0.03 & 27.48 & 0.05 \\ 
793 & 53.14770 & -27.81547 & 0.172 & 29.71 & 0.36 & 28.03 & 0.07 & 27.72 & 0.04 & 27.97 & 0.10 \\ 
807 & 53.17183 & -27.81518 & 0.139 & 30.12 & 0.48 & 27.57 & 0.05 & 27.15 & 0.03 & 27.19 & 0.05 \\ 
831 & 53.14756 & -27.81495 & 0.107 & --         & --      & 28.39 & 0.07 & 27.76 & 0.03 & 27.97 & 0.07 \\ 
848 & 53.16470 & -27.81486 & 0.194 & 31.65 & 2.06 & 28.05 & 0.11 & 28.50 & 0.09 & 28.90 & 0.23 \\ 
904 & 53.16409 & -27.81418 & 0.116 & 30.40 & 0.38 & 28.81 & 0.12 & 28.33 & 0.04 & 28.40 & 0.08 \\ 
\enddata
\end{deluxetable}

%Table 7: V-dropouts; first 36 sources from HUDF sample
\begin{deluxetable}{rrrlrrrrrrrr}
\tabletypesize{\scriptsize}
\tablecaption{First 36 \Vband-dropout sources}
\tablewidth{0pt}
\tablehead{
\colhead{STScI ID} & \colhead{RA (J2000)} &\colhead{Dec (J2000)} & \colhead{r50(\arcsec)} & \colhead{B} & \colhead{$\sigma_B$} & \colhead{V} & \colhead{$\sigma_V$} & \colhead{i} & \colhead{$\sigma_i$} & \colhead{z} & \colhead{$\sigma_z$}
}
\startdata
67 & 53.16471 & -27.82572 & 0.104 & 31.08 & 0.69 & 30.62 & 0.31 & 28.80 & 0.15 & 29.00 & 0.13 \\ 
123 & 53.16771 & -27.82393 & 0.112 & --        & --          & 31.21 & 0.62 & 28.86 & 0.09 & 29. & 0.16 \\ 
204 & 53.16884 & -27.82266 & 0.101 & --        & --          & 31.47 & 0.61 & 29.24 & 0.12 & 29.61 & 0.21 \\ 
384 & 53.15979 & -27.82000 & 0.066 & 34.59 & 12.11 & 32.27 & 0.99 & 29.38 & 0.11 & 29.99 & 0.24 \\ 
546 & 53.15804 & -27.81794 & 0.145 & --        & --          & 30.28 & 0.35 & 27.77 & 0.06 & 27.04 & 0.04 \\ 
646 & 53.16606 & -27.81678 & 0.148 & --        & --          & 29.39 & 0.14 & 27.41 & 0.04 & 27.32 & 0.04 \\ 
712 & 53.17836 & -27.81625 & 0.095 & 31.81 & 1.87 & 29.74 & 0.19 & 27.26 & 0.03 & 27.11 & 0.03 \\ 
748 & 53.17549 & -27.81601 & 0.138 & 31.94 & 1.93 & 31.06 & 0.60 & 28.82 & 0.15 & 28.92 & 0.16 \\ 
811 & 53.14928 & -27.81537 & 0.078 & --        & --        & 31.29 & 0.47 & 29.00 & 0.09 & 29.62 & 0.20 \\ 
847 & 53.16759 & -27.81499 & 0.092 & --        & --        & 29.77 & 0.15 & 28.48 & 0.07 & 28.87 & 0.13 \\ 
850 & 53.16398 & -27.81503 & 0.102 & --        & --        & 30.64 & 0.25 & 29.38 & 0.14 & 29.96 & 0.27 \\ 
864 & 53.16253 & -27.81490 & 0.118 & 31.26 & 0.80 & 31.26 & 0.55 & 28.95 & 0.09 & 28.93 & 0.13 \\ 
947 & 53.16205 & -27.81389 & 0.082 & --        & --       & 31.76 & 0.78 & 28.91 & 0.12 & 28.97 & 0.12 \\ 
971 & 53.16449 & -27.81373 & 0.064 & 31.37 & 0.66 & 31.17 & 0.38 & 29.32 & 0.19 & 29.53 & 0.17 \\ 
992 & 53.17062 & -27.81350 & 0.066 & 34.33 & 9.19 & 32.06 & 0.79 & 29.14 & 0.10 & 29.71 & 0.18 \\ 
1055 & 53.16123 & -27.81267 & 0.231 & 32.57 & 5.95 & 31.93 & 2.28 & 28.40 & 0.14 & 28.61 & 0.21 \\ 
1126 & 53.17548 & -27.81208 & 0.097 & --        & --       & 29.96 & 0.17 & 28.57 & 0.07 & 29.11 & 0.16 \\ 
1154 & 53.17563 & -27.81167 & 0.125 & 31.76 & 1.58 & 29.50 & 0.14 & 28.07 & 0.07 & 28.24 & 0.09 \\ 
1156 & 53.16736 & -27.81171 & 0.351 & --        & --      & 29.73 & 0.39 & 28.16 & 0.11 & 29.00 & 0.39 \\ 
1157 & 53.15931 & -27.81169 & 0.110 & --        & --      & 30.32 & 0.25 & 28.39 & 0.06 & 28.53 & 0.10 \\ 
1276 & 53.14984 & -27.81067 & 0.127 & --        & --       & 29.80 & 0.19 & 27.77 & 0.04 & 27.63 & 0.05 \\ 
1318 & 53.16177 & -27.81035 & 0.059 & 31.34 & 0.62 & 30.25 & 0.16 & 28.94 & 0.08 & 29.21 & 0.12 \\ 
1379 & 53.14944 & -27.80973 & 0.118 & 30.81 & 0.65 & 30.46 & 0.32 & 28.32 & 0.08 & 28.47 & 0.10 \\ 
1392 & 53.15636 & -27.80959 & 0.078 & --         & --      & 30.86 & 0.45 & 27.63 & 0.03 & 28.27 & 0.08 \\ 
1519 & 53.15031 & -27.80882 & 0.106 & --         & --      & 31.62 & 0.67 & 29.17 & 0.12 & 29.12 & 0.13 \\ 
1608 & 53.14378 & -27.80828 & 0.091 & 31.62 & 1.19 & 33.56 & 4.87 & 29.01 & 0.10 & 28.72 & 0.11 \\ 
1749 & 53.15279 & -27.80729 & 0.071 & --         & --      & 30.82 & 0.27 & 29.09 & 0.09 & 29.48 & 0.16 \\ 
1796 & 53.15324 & -27.80697 & 0.145 & 30.95 & 0.70 & 30.12 & 0.23 & 28.21 & 0.06 & 27.99 & 0.06 \\ 
1846 & 53.17214 & -27.80674 & 0.109 & --         & --      & 30.23 & 0.23 & 28.71 & 0.09 & 29.00 & 0.15 \\ 
1967 & 53.17512 & -27.80612 & 0.084 & 31.78 & 1.04 & 30.52 & 0.23 & 29.29 & 0.15 & 30.05 & 0.29 \\ 
2122 & 53.13990 & -27.80500 & 0.180 & 31.07 & 0.91 & 30.18 & 0.28 & 28.90 & 0.17 & 29.32 & 0.24 \\ 
2155 & 53.17774 & -27.80466 & 0.225 & --         & --      & 29.14 & 0.15 & 27.49 & 0.06 & 27.53 & 0.07 \\ 
2276 & 53.16717 & -27.80429 & 0.126 & --         & --      & 30.39 & 0.34 & 28.36 & 0.09 & 28.21 & 0.09 \\ 
2285 & 53.16834 & -27.80413 & 0.105 & --         & --      & 29.54 & 0.15 & 27.71 & 0.04 & 27.72 & 0.05 \\ 
2303 & 53.18131 & -27.80394 & 0.159 & --         & --      & 28.91 & 0.11 & 27.41 & 0.04 & 27.44 & 0.06 \\ 
2393 & 53.13875 & -27.80375 & 0.128 & --         & --      & 34.19 & 8.45 & 29.12 & 0.13 & 29.02 & 0.14 \\ 
\enddata
\end{deluxetable}

%Table 8: i dropouts; first 36 sources from HUDF sample
\begin{deluxetable}{rrrlrrrrrrrr}
\tabletypesize{\scriptsize}
\tablecaption{First 36 \iband-dropout sources}
\tablewidth{0pt}
\tablehead{
\colhead{STScI ID} & \colhead{RA (J2000)} &\colhead{Dec (J2000)} & \colhead{r50(\arcsec)} & \colhead{B} & \colhead{$\sigma_B$} & \colhead{V} & \colhead{$\sigma_V$} & \colhead{i} & \colhead{$\sigma_i$} & \colhead{z} & \colhead{$\sigma_z$}
}
\startdata
229 & 53.17161 & -27.82079 & 0.160 & --      & --    & 31.83 & 1.47 & 29.04 & 0.13 & 26.94 & 0.04 \\ 
263 & 53.17585 & -27.81997 & 0.239 & 32.05 & 4.24 & --         & --    & 29.47 & 0.30 & 27.51 & 0.12 \\ 
327 & 53.16271 & -27.81897 & 0.120 & --         & --      & 32.82 & 2.63 & 29.54 & 0.14 & 28.14 & 0.09 \\ 
344 & 53.16691 & -27.81875 & 0.066 & 31.99 & 1.34 & 31.24 & 0.46 & 29.90 & 0.15 & 28.13 & 0.07 \\ 
534 & 53.16040 & -27.81606 & 0.230 & 31.99 & 3.35 & 33.36 & 8.11 & 29.76 & 0.33 & 27.30 & 0.09 \\ 
591 & 53.15532 & -27.81516 & 0.150 & 33.20 & 7.34 & 31.18 & 0.79 & 32.08 & 2.03 & 27.31 & 0.07 \\ 
777 & 53.17048 & -27.81243 & 0.124 & 30.84 & 0.65 & --         & --    & 29.14 & 0.11 & 27.75 & 0.08 \\ 
1148 & 53.15203 & -27.80822 & 0.156 & --      & --       & 31.42 & 0.85 & 29.32 & 0.14 & 27.89 & 0.10 \\ 
1149 & 53.17985 & -27.80821 & 0.112 & --      & --       & 32.14 & 1.28 & 29.91 & 0.19 & 28.38 & 0.11 \\ 
1283 & 53.14169 & -27.80694 & 0.173 & --        & --     & --         & --    & 29.29 & 0.16 & 27.85 & 0.11 \\ 
1309 & 53.14225 & -27.80677 & 0.098 & 31.65 & 1.15 & --         & --    & 30.45 & 0.30 & 28.43 & 0.10 \\ 
1312 & 53.13888 & -27.80671 & 0.108 & 33.67 & 8.61 & --         & --    & 30.53 & 0.37 & 28.31 & 0.14 \\ 
1526 & 53.16672 & -27.80416 & 0.157 & --         & --    & 29.63 & 0.33 & 26.68 & 0.02 & 25.11 & 0.01 \\ 
1690 & 53.16574 & -27.80338 & 0.128 & 32.39 & 2.81 & 31.13 & 0.61 & 29.24 & 0.12 & 27.84 & 0.08 \\ 
1790 & 53.13043 & -27.80231 & 0.167 & 30.83 & 0.91 & 31.81 & 1.53 & 28.90 & 0.12 & 27.26 & 0.07 \\ 
1793 & 53.14623 & -27.80253 & 0.112 & --         & --    & --         & --    & 30.02 & 0.18 & 28.60 & 0.12 \\ 
1813 & 53.17748 & -27.80245 & 0.126 & --         & --    & --         & --    & 29.56 & 0.17 & 27.62 & 0.08 \\ 
1852 & 53.14075 & -27.80211 & 0.098 & --         & --    & 33.24 & 3.95 & 29.20 & 0.11 & 27.31 & 0.04 \\ 
1874 & 53.14617 & -27.80190 & 0.137 & --         & --    & 30.90 & 0.53 & 30.01 & 0.26 & 27.91 & 0.10 \\ 
2007 & 53.17832 & -27.80090 & 0.106 & --         & --    & --         & --    & 31.94 & 1.27 & 28.02 & 0.08 \\ 
2279 & 53.14394 & -27.79889 & 0.118 & --         & --    & 31.84 & 1.34 & 28.64 & 0.08 & 27.00 & 0.04 \\ 
2330 & 53.13587 & -27.79833 & 0.145 & 31.51 & 1.54 & 31.2 & 0.81 & 28.13 & 0.05 & 26.72 & 0.03 \\ 
2386 & 53.14294 & -27.79821 & 0.166 & --         & --    & 30.95 & 0.59 & 29.16 & 0.13 & 27.71 & 0.08 \\ 
2408 & 53.14287 & -27.79800 & 0.223 & --         & --    & --         & --    & 29.05 & 0.16 & 27.24 & 0.08 \\ 
2413 & 53.14285 & -27.79785 & 0.192 & 32.14 & 3.01 & 32.78 & 3.81 & 28.63 & 0.09 & 27.28 & 0.06 \\ 
2490 & 53.15950 & -27.79758 & 0.127 & --         & --    & 32.28 & 1.92 & 29.07 & 0.11 & 27.46 & 0.06 \\ 
2521 & 53.15261 & -27.79724 & 0.314 & --         & --    & --         & --    & 29.38 & 0.27 & 26.99 & 0.08 \\ 
2760 & 53.13928 & -27.79580 & 0.112 & 31.25 & 0.97 & 31.33 & 0.74 & 29.48 & 0.15 & 27.38 & 0.05 \\ 
2789 & 53.17322 & -27.79562 & 0.162 & --         & --    & --         & --    & 29.52 & 0.20 & 27.01 & 0.05 \\ 
2966 & 53.12271 & -27.79459 & 0.200 & 30.65 & 0.79 & 30.5 & 0.47 & 29.13 & 0.15 & 27.22 & 0.08 \\ 
3003 & 53.14607 & -27.79449 & 0.120 & --         & --    & --         & --    & 31.43 & 0.85 & 27.81 & 0.08 \\ 
3527 & 53.12559 & -27.79124 & 0.236 & --         & --    & 31.05 & 0.95 & 29.13 & 0.18 & 27.42 & 0.09 \\ 
4266 & 53.16193 & -27.78699 & 0.076 & 31.09 & 0.61 & 31.80 & 0.82 & 30.09 & 0.18 & 28.50 & 0.11 \\ 
4321 & 53.18625 & -27.78655 & 0.185 & 29.90 & 0.38 & 30.68 & 0.56 & 29.63 & 0.24 & 27.43 & 0.09 \\ 
4353 & 53.16161 & -27.78633 & 0.186 & --         & --     & --         & --    & 29.66 & 0.21 & 27.86 & 0.09 \\ 
4715 & 53.13482 & -27.78412 & 0.180 & 31.66 & 1.85 & --         & --    & 29.84 & 0.27 & 27.73 & 0.11 \\ 
\enddata
\end{deluxetable}

%Table 9: redshift ranges & volumes of dropout samples
\begin{deluxetable}{rccccccccccc}
\tabletypesize{\scriptsize}
\tablecaption{Hubble dropout selection volumes}
\tablewidth{0pt}
\tablehead{
\colhead{} & \multicolumn{3}{c}{\Bband-dropouts} & \multicolumn{3}{c}{\Vband-dropouts} & \multicolumn{5}{c}{\iband-dropouts} \\ 
\colhead{Spectrum} & \colhead{$z_{min}$} & \colhead{$z_{max}$} & \colhead{V(Mpc$^3$)} & \colhead{$z_{min}$} & \colhead{$z_{max}$} & \colhead{V(Mpc$^3$)} & \colhead{$z_{min}$} & \colhead{$z_{Hmax}$} & \colhead{$V_H$(Mpc$^3$)} & \colhead{$z_{Gmax}$} & \colhead{$V_G$(Mpc$^3$)}
}
\startdata
10 Myr      & 3.53 & 4.80 & 3981 & 4.67 & 5.71 & 2990 & 5.71 & 7.29 & 3924 & 6.62 & 2335 \\
100 Myr   & 3.48 & 4.74 & 4046 & 4.62 & 5.68 & 3051 & 5.68 & 7.24 & 3897 & 6.56 & 2279 \\
1000 Myr & 3.49 & 4.69 & 3866 & 4.60 & 5.65 & 3037 & 5.65 & 7.21 & 3893 & 6.52 & 2253 \\
Flat Step  & 3.66 & 4.75 & 3466 & 4.72 & 5.70 & 2797 & 5.70 & 7.19 & 3734 & 6.47 & 2007 \\
Red Step & 3.72 & 4.59 & 2787 & 4.69 & 5.64 & 2717 & 5.64 & 7.14 & 3785 & 6.42 & 2056 \\
\enddata
\\ 
V is the co-moving volume (Mpc$^3$ arcmin$^{-2}$) within the redshift range: $V \equiv  {\rm arcmin}^2 \int_{z_{min}}^{z_{max}} {dV \over dz} dz$.\\
$V_G$ and $V_H$ are GOODS and HUDF, respectively. $z_{Hmax}$ \& $z_{Gmax}$ illustrate $M_*=-20.8$ at the \zband\ limits.
\end{deluxetable}

%Table 10: redshift ranges & volumes of dropout samples with E(B-V)=0.15
\begin{deluxetable}{rccccccccccc}
\tabletypesize{\scriptsize}
\tablecaption{Hubble dropout selection volumes: E$_{B-V}=0.15$}
\tablewidth{0pt}
\tablehead{
\colhead{} & \multicolumn{3}{c}{\Bband-dropouts} & \multicolumn{3}{c}{\Vband-dropouts} & \multicolumn{5}{c}{\iband-dropouts} \\ 
\colhead{Spectrum} & \colhead{$z_{min}$} & \colhead{$z_{max}$} & \colhead{V(Mpc$^3$)} & \colhead{$z_{min}$} & \colhead{$z_{max}$} & \colhead{V(Mpc$^3$)} & \colhead{$z_{min}$} & \colhead{$z_{Hmax}$} & \colhead{$V_H$(Mpc$^3$)} & \colhead{$z_{Gmax}$} & \colhead{$V_G$(Mpc$^3$)}
}
\startdata
10 Myr      & 3.35 & 4.46 & 3614 & 4.53 & 5.58 & 3046 & 5.58 & 7.17 & 4031 & 6.58 & 2456 \\
100 Myr   & 3.30 & 4.40 & 3622 & 4.50 & 5.55 & 3061 & 5.55 & 7.13 & 4011 & 6.55 & 2404 \\
1000 Myr & 3.31 & 4.33 & 3385 & 4.48 & 5.52 & 3031 & 5.52 & 7.10 & 4022 & 6.43 & 2398 \\
Flat Step  & 3.46 & 4.36 & 2956 & 4.56 & 5.55 & 2869 & 5.55 & 7.08 & 3870 & 6.37 & 2161 \\
Red Step & 3.51 & 4.05 & 1804 & 4.55 & 5.48 & 2719 & 5.48 & 7.03 & 3972 & 6.33 & 2258 \\
\enddata
\\ 
V is the co-moving volume (Mpc$^3$ arcmin$^{-2}$) within the redshift range: $V \equiv  {\rm arcmin}^2 \int_{z_{min}}^{z_{max}} {dV \over dz} dz$.\\
$V_G$ and $V_H$ are GOODS and HUDF, respectively. $z_{Hmax}$ \& $z_{Gmax}$ illustrate $M_*=-20.8$ at the \zband\ limits.
\end{deluxetable}

%Table 11: Surface densities of B-dropout samples
\begin{deluxetable}{lrrr}
\tabletypesize{\scriptsize}
\tablecaption{Surface densities of \Bband-dropouts vs. magnitude}
\tablewidth{0pt}
\tablehead{
\colhead{\Vband\ range}  & \colhead{$\Sigma$(HUDF)} & \colhead{$\Sigma$(GOODS-N)} & \colhead{$\Sigma$(GOODS-S)} \\
         & \multicolumn{3}{c}{mag$^{-1}$ arcmin$^{-2}$} 
}
\startdata
24--25 & & 0.069$\pm$0.024 & 0.094$\pm$0.029 \\
25--26 & & 0.808$\pm$0.155 & 1.021$\pm$0.191 \\
26--26.5 & & 2.627$\pm$0.482 & 3.031$\pm$0.551 \\
26.5--26.75 & & 3.966$\pm$0.745 & 4.585$\pm$0.85 \\
26.75--27 & & 5.633$\pm$1.029 & 5.286$\pm$0.97 \\
27--27.5 & & 6.479$\pm$1.138 & 6.188$\pm$1.088 \\
25--26 & 0.727$\pm$0.285&& \\ 
26--26.5 & 4.$\pm$1.091&& \\ 
26.5--27 & 4.727$\pm$1.227&& \\ 
27--27.5 & 6.545$\pm$1.558&& \\ 
27.5--28 & 12.364$\pm$2.582&& \\ 
28--28.5 & 19.273$\pm$3.773&& \\ 
28.5--29 & 19.818$\pm$3.867&& \\ 
29--29.3 & 25.758$\pm$5.194&& \\ 
29.3--29.624 & 12.065$\pm$2.755&& \\ 
\enddata
\\
Total surface brightness derived from HUDF only: \\
23.7 mag arcmin$^{-2}$, at $\lambda_{\rm rest} \sim 1400\,$\AA.
\end{deluxetable}

%Table 12: Surface densities of V-dropout samples
\begin{deluxetable}{lrrr}
\tabletypesize{\scriptsize}
\tablecaption{Surface densities of \Vband-dropouts vs. magnitude}
\tablewidth{0pt}
\tablehead{
\colhead{\iband\ range}  & \colhead{$\Sigma$(HUDF)} & \colhead{$\Sigma$(GOODS-N)} & \colhead{$\Sigma$(GOODS-S)} \\
 & \multicolumn{3}{c}{mag$^{-1}$ arcmin$^{-2}$} 
}
\startdata
24--25 & & 0.025$\pm$0.013 & 0.025$\pm$0.013 \\ 
25--26 & & 0.295$\pm$0.066 & 0.319$\pm$0.07 \\ 
26--26.5 & & 1.091$\pm$0.219 & 1.006$\pm$0.204 \\ 
26.5--26.75 & & 1.731$\pm$0.361 & 1.374$\pm$0.297 \\ 
26.75--27 & & 2.083$\pm$0.421 & 1.202$\pm$0.267 \\ 
27--27.5 & & 2.371$\pm$0.438 & 1.472$\pm$0.284 \\ 
25--26 & 0.455$\pm$0.217&& \\ 
26--27 & 0.727$\pm$0.285&& \\ 
27--27.5 & 4.$\pm$1.091&& \\ 
27.5--28 & 4.182$\pm$1.125&& \\ 
28--28.5 & 9.091$\pm$2.01&& \\ 
28.5--29 & 10.182$\pm$2.202&& \\ 
29--29.25 & 9.818$\pm$2.521&& \\ 
29.25--29.524 & 4.313$\pm$1.403&& \\ 
\enddata
\\
Total surface brightness derived from HUDF only: \\
24.8 mag arcmin$^{-2}$, at $\lambda_{\rm rest} \sim 1400\,$\AA. 
\end{deluxetable}

%Table 13: Surface densities of i-dropout samples
\begin{deluxetable}{llrrr}
\tabletypesize{\scriptsize}
\tablecaption{Surface densities of \iband-dropouts vs. magnitude}
\tablewidth{0pt}
\tablehead{
\colhead{\zband\ range} & \colhead{$\Sigma$(HUDF)} & \colhead{$\Sigma$(GOODS-N)} & \colhead{$\Sigma$(GOODS-S)} \\
 & \multicolumn{3}{c}{mag$^{-1}$ arcmin$^{-2}$} 
}
\startdata
25--26 & & 0.1$\pm$0.03 & 0.081$\pm$0.026 \\ 
26--26.5 & & 0.339$\pm$0.087 & 0.401$\pm$0.098 \\ 
26.5--27 & & 0.84$\pm$0.176 & 0.915$\pm$0.189 \\ 
25--26.7 & 0.214$\pm$0.113 && \\ 
26.7--27.7 & 2.$\pm$0.545 && \\ 
27.7--28.746 & 2.434$\pm$0.619 && \\ 
\enddata
\\
Total surface brightness derived from HUDF only: \\
25.6 mag arcmin$^{-2}$, at $\lambda_{\rm rest} \sim 1400\,$\AA.
\end{deluxetable}

%Table 14: Schechter fits to dropout samples
\begin{deluxetable}{lcccccccccc}
\tabletypesize{\scriptsize}
\tablecaption{Schechter function fits}
\tablewidth{0pt}
\tablehead{
  & \multicolumn{3}{c}{Simple Schechter} & \multicolumn{2}{c}{$M_*$, $\phi_*$ from $m_*$, $\Sigma_*$} & \multicolumn{3}{c}{$\int \phi[m(z)] {dV\over dz} dz$} & \multicolumn{2}{c}{$M_*=-20.3$}  \\ 
\colhead{Sample} & \colhead{$m_*$} & \colhead{$\Sigma_*$(min$^{-2}$)} & \colhead{$\chi^2$}  & \colhead{$M_*$}(equiv) & \colhead{$\phi_*$(equiv)} & \colhead{$M_*$} & \colhead{$\phi_*$(Mpc$^{-3}$)} & \colhead{$\chi^2$} &   \colhead{$\phi_*$(Mpc$^{-3}$)} & \colhead{$\chi^2$}
}
\startdata 
\Bband-dropouts & 26.0 & 6.6 & 0.5  & -20.5 & 0.0016 &  -20.3 & 0.0020 &  0.4 & 0.0020  & 0.4 \\
\Vband-dropouts & 26.1 & 2.4 & 1.2 & -20.6 & 0.0008 &  -20.5 & 0.0010 &  1.1 & 0.0012  & 1.3  \\
\iband-dropouts  & 26.3 & 1.5 & 0.8 & -21.3 & 0.0004 &  -20.5 & 0.0008 &  0.9 & 0.0010  & 1.1  \\
\enddata
\\ 
Columns 5 \& 6 translate the simple Schechter function fits from columns 2 and 3 into equivalent $M_*$ and $\phi_*$ assuming \\
spectra from 100\,Myr CSF galaxy as in other cases. All $\chi^2$ contours are highly elliptical in the $m_*$, $\Sigma_*$ or $M_*$,$\phi_*$
 planes, \\
indicating strong degeneracy in the fitting parameters. All fits assume $\alpha=-1.6$. The uncertainties in $m_*$, $\Sigma_*$ are typically: \\
$\sigma$(\Bband-drop):  $\Delta(m(M)_*- \ln[\Sigma_*(\phi_*)] \approx 0.6$,  $\Delta(m(M)_*+ \ln[\Sigma_*(\phi_*)] \approx 0.1$; \\
$\sigma$(\Vband-drop): $\Delta(m(M)_*- \ln[\Sigma_*(\phi_*)] \approx 1.4$, $\Delta(m(M)_*+ \ln[\Sigma_*(\phi_*)] \approx 0.3$; \\
$\sigma$(\iband-drop):  $\Delta(m(M)_*- \ln[\Sigma_*(\phi_*)] \approx 1.5$, $\Delta(m(M)_*+ \ln[\Sigma_*(\phi_*)] \approx 0.5$ 
\end{deluxetable}

%Table 15: Combined luminosities G & B-drop samples
\begin{deluxetable}{lcccc}
\tabletypesize{\scriptsize}
\tablecaption{Combined luminosity functions G- and \Bband -dropouts}
\tablewidth{0pt}
\tablehead{
\colhead{V mag} & \multicolumn{4}{c}{$\log(\phi)$\,(mag$^{-1}$\,Mpc$^{-3}$)}  \\
& \colhead{SAGDP99} & \colhead{HUDF} & \colhead{GOODS-N} & \colhead{GOODS-S} 
}
\startdata
-22.95 to -22.45 & -5.48$\pm$0.18 &&& \\ 
-22.45 to -21.95 & -4.93$\pm$0.15 &&& \\ 
-21.95 to -21.45 & -4.24$\pm$0.08 &&& \\ 
-21.45 to -20.95 & -3.72$\pm$0.05 &&& \\ 
-22.48 to -21.48 &                              &                               & -4.72$\pm$0.13 & -4.58$\pm$0.12 \\ 
-21.48 to -20.48 &                              & -3.71$\pm$0.14 & -3.63$\pm$0.08 & -3.52$\pm$0.07 \\ 
-20.48 to -19.98 &                              & -2.96$\pm$0.10 & -3.10$\pm$0.07 & -3.03$\pm$0.07 \\ 
-19.98 to -19.73 &                              &                               & -2.92$\pm$0.07 & -2.85$\pm$0.07 \\ 
-19.73 to -19.48 &                              &                               & -2.75$\pm$0.07 & -2.77$\pm$0.07 \\ 
-19.98 to -19.48 &                              & -2.88$\pm$0.10 && \\ 
-19.48 to -18.98 &                              & -2.73$\pm$0.09 & -2.60$\pm$0.07 & -2.62$\pm$0.07 \\ 
-18.98 to -18.48 &                              & -2.44$\pm$0.08 && \\ 
-18.48 to -17.98 &                              & -2.23$\pm$0.08 && \\ 
-17.98 to -17.48 &                              & -2.23$\pm$0.08 && \\ 
-17.48 to -17.18 &                              & -2.10$\pm$0.08 && \\ 
-17.18 to -16.86 &                              & -2.32$\pm$0.09 && \\ 
\enddata
\\
Sample redshifts:  4.13 (3.75--4.5) Caltech, 4.11 (3.49--4.69) Hubble \\
Sample volumes (Mpc$^3$ arcmin$^{-2}$): 1999 (Caltech), 4045 (Hubble)\\
100\,Myr Bruzual model distance moduli: 45.95 Caltech, 46.48 Hubble \\
\end{deluxetable}

%Table 16: Luminosity density: bright and faint
\begin{deluxetable}{lccccccc}
\tabletypesize{\scriptsize}
\tablecaption{Luminosity Density Comparison}
\tablewidth{0pt}
\tablehead{
  & & & \multicolumn{2}{c}{$\Sigma$(mag arcmin$^{-2}$)} & \multicolumn{2}{c}{$\log \Phi$(erg s$^{-1}$ Hz$^{-1}$ Mpc$^{-3}$)} & \colhead{$\Phi_{faint} / \Phi_{bright} $} \\
\colhead{Sample} & \colhead{{\it \={z}}} & \colhead{$m_*$} & \colhead{GOODS $m < m_*$} & \colhead{HUDF $m \ge m_*$} & \colhead{GOODS $m < m_*$} & \colhead{HUDF $m \ge m_*$} & 
}
\startdata 
\Bband - dropout & 4.11 & 25.31 & 26.53$\pm$0.12 & 24.49$\pm$0.05 & 25.18$\pm$0.05 & 26.$\pm$0.02 & 6.6$\pm$13\%  \\ 
\Vband -dropout & 5.04 & 25.66 & 27.25$\pm$0.15 & 25.57$\pm$0.07 & 25.08$\pm$0.06 & 25.75$\pm$0.03 & 4.7$\pm$16\%  \\ 
\iband -dropout & 6.46 & 26.06 & 28.15$\pm$0.18 & 25.88$\pm$0.14 & 24.85$\pm$0.07 & 25.76$\pm$0.06 & 8.1$\pm$22\%  \\ 
\enddata \\ 
Faint/bright divide at $M = -20.8$. Faint end limit at $M = -18.8$. Mean redshifts, volumes from the 100\,Myr model in Table 9.\\ 
Listed uncertainties are $\propto 1/\sqrt{{\rm nobj}-1}$ and do not include cosmic error or model-dependent variations. \\  
\end{deluxetable}


\clearpage

\begin{thebibliography}{}
\bibitem[Becker et al. 2001]{beck01}
Becker, R. H., Fan, X., White, R. L., Strauss, M. A., Narayanan, V. K., Lupton, R. H.; Gunn, J. E., Annis, J., Bahcall, N. A., Brinkmann, J., and 20 coauthors 2001, \aj, 122, 2850.
\bibitem[Bertin \& Arnouts (1996)]{bert96}
Bertin, E. and Arnouts, S. 1996, \aap, 117, 393. 
\bibitem[Bouwens et al. 2004a]{bouw04a}
Bouwens, R. J., Illingworth, G. D., Thompson, R. I., Blakeslee, J. P., Dickinson, M. E., Broadhurst, T. J., Eisenstein, D. J., Fan, X., Franx, M., Meurer, G., and van Dokkum, P. 2004, \apj, 606, L25.
\bibitem[Bouwens et al. 2004b]{bouw04b}
Bouwens, R. J., Illingworth, G. D., Blakeslee, J. P., Broadhurst, T. J., and Franx, M. 2004, \apj, 611, L1. 
\bibitem[BIBF05]{bouw05}
Bouwens, R. J., Illingworth, G. D., Blakeslee, J. P., and Franx, M. 2006, \apj, in press (astro-ph/0509641) (BIBF05).
\bibitem[Bruzual \& Charlot 2003]{bruz03}
Bruzual, G., \& Charlot, S. 2003, \mnras, 344, 1000.
\bibitem[Bunker et al. 2004]{bunk04}
Bunker, A. J., Stanway, E. R., Ellis, R. S., and McMahon, R. G. 2004, MNRAS, 355, 374.
\bibitem[Davis \& Wilkinson 1974]{davi74}
Davis, M. and Wilkinson, D. T. 1974, \apj, 192, 251.
\bibitem[Dickinson 1995]{dick95}
Dickinson, M. 1995, in {\it Fresh views of elliptical galaxies}, eds. A. Buzzoni, A. Renzini, A. Serrano, ASP conf. ser. 86, 283
\bibitem[Driver, Windhorts, \& Griffiths 1995]{driv95}
Driver, S. P., Windhorst, R. A., and Griffiths, R. E. 1995, \apj, 453, 48.
\bibitem[Eggen, Lynden-Bell, \& Sandage 1962]{egge62}
Eggen, O. J., Lynden-Bell, D., and Sandage, A. R. 1962, \apj, 136, 748.
\bibitem[Ellis 1998]{elli98}
Ellis, R. 1998, \nat, 395, 3.
\bibitem[Elmegreen et al. 2005]{elme05}
Elmegreen, D. M., Elmegreen, B., Rubin, D. S., and Schaffer, M. A. 2005, \apj, 631, 85.
\bibitem[Fan et al. 2002]{fan02}
Fan, X., Narayanan, V. K., Strauss, M. A., White, R. L., Becker, R. H., Pentericci, L., and Rix, H.-W. 2002, \aj, 123, 1247.
\bibitem[Ford et al. 2003]{ford03}
Ford, H. C., Clampin, M., Hartig, G. F., Illingworth, G. D., Sirianni, M., Martel, A. R., Meurer, G. R., McCann, W. J., Sullivan, P. C., Bartko, F.; and 26 coauthors  2003, SPIE, 4854, 581.
\bibitem[Fontanot et al. 2006]{font06}
Fontanot, F., Cristiani, S.,Monaco, P., Nonino, M., Vanzella,  E., Grazian, A., Mao, J., Stern, D. and the GOODS Team  2006, (in preparation.)
\bibitem[Frenk et al. 1985]{frenk85}
Frenk, C. S., White, S. D. M., Efstathiou, G., and Davis, M. 1985, \nat, 317, 595.
\bibitem[Fruchter \& Hook 2002]{fruc02}
Fruchter, A. S. and Hook, R. N. 2002, \pasp, 114, 144.
\bibitem[Gabash et al. 2004a]{gaba04a}
Gabasch, A., Bender, R., Seitz, S., Hopp, U., Saglia, R. P., Feulner, G., Snigula, J., Drory, N., Appenzeller, I., Heidt, J., Mehlert, D., Noll, S., B\"ohm, A., J\"ager, K., Ziegler, B., and Fricke, K. J. 2004a, \aap, 421, 41.
\bibitem[Gabash et al. 2004b]{gaba04b}
Gabash, A.; Salvato,ÊM., Saglia,ÊR.ÊP., Bender,ÊR., Hopp,ÊU., Seitz,ÊS., Feulner,ÊG., Pannella,ÊM., Drory,ÊN., Schirmer,ÊM., and Erben,ÊT. 2004b, \apj, 616, 83.
\bibitem[Giavalisco 2002]{giav02}
Giavalisco, M. 2002, \araa, 40, 579.
\bibitem[Giavalisco et al. 2004]{giav04}
Giavalisco,ÊM., Ferguson,ÊH.ÊC., Koekemoer,ÊA.ÊM., Dickinson,ÊM., Alexander,ÊD.ÊM., Bauer,ÊF.ÊE., Bergeron,ÊJ., Biagetti,ÊC., Brandt,ÊW.ÊN., Casertano,ÊS., and 47 coauthors 2004, \apjl, 600, 93L.
\bibitem[King et al. 2005]{king05}
King, I. R., Bedin, L. R., Piotto, G., Cassisi, S., and Anderson, J. 2005, \aj, 130, 626.
\bibitem[Koekemoer 2002]{koek02}
Koekemoer, A. M. 2002, in {\it The 2002 HST Calibration Workshop:  Hubble after the Installation of the ACS and the NICMOS Cooling System}, eds. S. Arribas, A. Koekemoer, B. Whitmore, {Space Telescope Science Institute:Baltimore}, 293.
\bibitem[Kogut et al. 2003]{kogu03}
Kogut, A., Spergel, D. N., Barnes, C., Bennett, C. L., Halpern, M., Hinshaw, G., Jarosik, N., Limon, M., Meyer, S. S., Page, L., et al. 2003, \apjs, 148, 161. 
\bibitem[Koo \& Kron 1980]{koo80}
Koo, D. C., and Kron, R. T. 1980, \pasp, 92, 537.
\bibitem[Lee et al. 2006]{lee06}
Lee, K. S., Giavalisco, M., Gnedin, O. Y., Somerville, R. S., Ferguson, H. C., Dickinson, M., and Ouchi, M. 2006, \apjl, 642, 63L.
\bibitem[Lotz et al. 2006]{lotz06}
Lotz, J. M., Madau, P., Giavalisco, M., Primack, J., and Ferguson, H. C. 2006, \apj, 636, 592.
\bibitem[Madau 1995]{mada95}
Madau,ÊP. 1995, \apj, 441, 18.
\bibitem[Ouchi et al. 2004]{ouch04}
Ouchi, M., Shimasaku, K., Okamura, S., Furusawa, H., Kashikawa, N., Ota, Kl, Doi, M., Hamabe, M., Kimura, M., Komiyama, Y., Miyazaki, M., Miyazaki, S., Nakata, F., Sekiguchi, M., Yagi, M., and Yasuda, N. 2004, \apj, 611, 660 (O04).
\bibitem[Partridge 1974]{part74}
Partridge, R. B. 1974, \apj, 192, 241.
\bibitem[Partridge \& Peebles 1967a]{part67a}
Partridge, R. B., and Peebles, P. J. E. 1967a, \apj, 147, 868.
\bibitem[Partridge \& Peebles 1967b]{part67b}
Partridge, R. B., and Peebles, P. J. E. 1967b, \apj, 148, 377.
\bibitem[Sawicki \& Thompson]{sawi06}
Sawicki, M. and Thompson, D. 2006, \apj, in press.
\bibitem[Schechter 1976]{sche76}
Schechter, P. 1976, \apj, 203, 297.
\bibitem[Schlegel et al. 1998]{schl98}
Schlegel, D. J., Finkbeiner, D. P., and Davis, M. 1998, \apj, 500, 525. 
\bibitem[Somerville et al. 2004]{some04}
Somerville, R. S., Lee, K., Ferguson, H. C., Gardner, J. P., Moustakas, L. A., and Giavalisco, M. 2004, \apjl, 600, 171L (S04.)
\bibitem[Spergel et al. 2003]{sper03}
Spergel, ÊD.ÊN., Verde,ÊL., Peiris,ÊH.ÊV., Komatsu,ÊE., Nolta,ÊM.ÊR., Bennett,ÊC.ÊL., Halpern,ÊM., Hinshaw,ÊG., Jarosik,ÊN., Kogut,ÊA., and 7 coauthors 2003, \apjs, 148, 175.
\bibitem[Spergel et al. 2006]{sper06}
Spergel, D. N., et al. 2006, astro-ph/0603449.
\bibitem[Stanway et al. 2004a]{stan04a}
Stanway, E. R. et al. 2004a, \apj, 604, L13.
\bibitem[Stanway, McMahon, \& Bunker 2005]{stan05}
Stanway, E. R., McMahon, R. G., and Bunker, A. J. 2005, MNRAS, 359, 1184.
\bibitem[SAGDP99]{stei99}
Steidel, C. C., Adelberger, K. L., Giavalisco, M., Dickinson, M., and Pettini, M. 1999, \apj., 519, 1 (SAGDP99).
\bibitem[Steidel et al. 2003]{stei03}
Steidel,ÊC.ÊC., Adelberger,ÊK.ÊL., Shapley, A. E., Pettini,ÊM., Dickinson,ÊM., and Giavalisco,ÊM. 2003, \apj, 592, 728.
\bibitem[Steidel et al. 1996a]{stei96a}
Steidel,ÊC.ÊC., Giavalisco,ÊM., Pettini,ÊM., Dickinson,ÊM., and Adelberger,ÊK.ÊL. 1996a, \apjl, 462, L17.
\bibitem[Steidel et al. 1996b]{stei96b}
Steidel, C. C., Giavalisco, M., Dickinson, M., and Adelberger, K. L. 1996b, \aj, 112, 352.
\bibitem[Steidel \& Hamilton 1992]{stei92}
Steidel, C. and Hamilton, D. 1992, \aj, 104, 941.
\bibitem[Steidel et al. 2002]{stei02}
Steidel, C. C., Hunt, M. P., Shapley, A. E., Adelberger, K. L., Pettini, M., Dickinson, M., and Giavalisco, M. 2002, \apj, 576, 653.
\bibitem[Stiavelli, Fall, \& Panagia 2004a]{stia04a}
Stiavelli, M., Fall, S. M., and Panagia, N. 2004a, \apj, 600, 508.
\bibitem[Stiavelli, Fall, \& Panagia 2004b]{stia04b}
Stiavelli, M., Fall, S. M., and Panagia, N. 2004b, \apj, 610, L1.
\bibitem[Straughn et al. 2006]{yan06}
Straughn, A. N. , Cohen, S. H., Ryan, R. E., Jr., Hathi, N. P., Windhorst, R. A.,  and Jansen, R. A. 2006, \apj, in press (astro-ph/0511423.)
\bibitem[Thompson et al. 2005]{thom05}
Thompson, R. I. et al. 2005, \aj, 130, 1.
\bibitem[Tinsley 1972a]{tins72a}
Tinsley, B. M. 1972a, \apjl, 178, L39.
\bibitem[Tinsley 1972b]{tins72b}
Tinsley, B. M. 1972b, \apj, 178, 319.
\bibitem[Wherry et al. 2004]{wher04}
Wherry, N., Blanton, M. R., and Hogg, D. W. 2004, astro-ph/0406274. 
\bibitem[Williams et al. 1996]{will96}
Williams, R.ÊE., Blacker,ÊB., Dickinson,ÊM., Dixon,ÊW.ÊVanÊDyke, Ferguson,ÊH.ÊC., Fruchter,ÊA.ÊS., Giavalisco,ÊM., Gilliland,ÊR.ÊL., Heyer,ÊI., Katsanis,ÊR., and 7 coauthors 1996, \aj, 112, 1335. 
\bibitem[Williams et al. 2000]{will00}
Williams, R. E., Baum, S., Bergeron, L. E., Bernstein, N., Blacker, B. S., Boyle, B. J., Brown, T. M., Carollo, C. M., Casertano, S., Covarrubias, R., and 44 coauthors 2000, \aj, 120, 2735.
\bibitem[Yan and Windhorst 2004]{yan04}
Yan, H. and Windhorst, R. A. 2004, \apjl, 612, L93.
\end{thebibliography}
\end{document}